\documentclass[journal,onecolumn,romanappendices]{IEEEtran}

\usepackage[utf8]{inputenc} 
\usepackage[T1]{fontenc}
\usepackage{url}
\usepackage{ifthen}
\usepackage{cite}
\usepackage[cmex10]{amsmath} 
\usepackage{longtable}
\usepackage{makecell}
\usepackage{algorithm,algorithmic}
                             
\usepackage{amssymb}
\usepackage{xifthen,xparse}
\usepackage{hyperref}  % remove when submitting to IEEE Trans.
\usepackage{float}
\usepackage{caption}

\usepackage{tikz,pgfplots,tkz-graph,color}
\pgfplotsset{compat=newest}
\pgfplotsset{plot coordinates/math parser=false}
\usetikzlibrary{decorations.markings,calc,positioning,matrix}

\usepackage{textcomp}

\allowdisplaybreaks

\newif\iffull
\fulltrue

\usepackage{listings}

\definecolor{dkgreen}{rgb}{0,0.6,0}
\definecolor{gray}{rgb}{0.5,0.5,0.5}
\definecolor{mauve}{rgb}{0.58,0,0.82}

\definecolor{myblue}{RGB}{0,0,255}
\definecolor{mygreen}{RGB}{28,172,0}
\definecolor{mylilas}{RGB}{170,55,241}
\definecolor{bitcolor}{rgb}{1,0.84314,0}
\definecolor{checkcolor}{rgb}{0.52941,0.80784,1}

\newtheorem{theorem}{Theorem}
\newtheorem{corollary}[theorem]{Corollary}
\newtheorem{remark}[theorem]{Remark}
\newtheorem{lemma}[theorem]{Lemma}
\newtheorem{definition}[theorem]{Definition}

\newtheorem{fact}[theorem]{Fact}
\newcommand{\ket}[1]{\left\lvert #1 \right\rangle}

\NewDocumentCommand\ketbra{+m+g}{%
  \IfNoValueTF{#2}
    {\left\lvert #1 \right\rangle \left\langle #1 \right\vert}
  {\left\lvert #1 \right\rangle \left\langle #2 \right\rvert}%
}
\NewDocumentCommand\braket{+m+g}{%
  \IfNoValueTF{#2}
    {\left\langle #1 \vert #1 \right\rangle}
  {\left\langle #1 \vert #2 \right\rangle}%
}

\newcommand{\MCC}{\mathcal{C}}
\newcommand{\MCCd}{\mathcal{C}^{\perp}}
\newcommand{\MCQ}{\mathcal{Q}}
\newcommand{\vecnot}[1]{\underline{#1}}
\newcommand{\bg}{\boldsymbol{g}}
\newcommand{\llbr}{[\![}
\newcommand{\rrbr}{]\!]}
\newcommand{\cnot}[2]{{\rm CNOT}_{#1 \rightarrow #2}}
\newcommand{\lcnot}[2]{\overline{{\rm CNOT}}_{#1 \rightarrow #2}}
\newcommand{\cz}[2]{{\rm CZ}_{#1#2}}
\newcommand{\lcz}[2]{\overline{{\rm CZ}}_{#1#2}}
\newcommand{\lX}{\bar{X}}
\newcommand{\lZ}{\bar{Z}}
\newcommand{\lY}{\bar{Y}}
\newcommand{\lH}{\bar{H}}
\newcommand{\lP}{\bar{P}}

\newcommand{\syminn}[2]{\langle #1, #2 \rangle_{\text{s}}}

\begin{document}
\title{Synthesis of Logical Clifford Operators \\ via Symplectic Geometry}

 \author{%
   \IEEEauthorblockN{Narayanan Rengaswamy,
                     Robert Calderbank,
                     Swanand Kadhe,
                     and Henry D. Pfister}%
   \thanks{N. Rengaswamy, R. Calderbank and H.D. Pfister are with the
                     Department of Electrical and Computer Engineering,
                     Duke University,
                     Durham, North Carolina 27708, USA.
                     Email: \{narayanan.rengaswamy, robert.calderbank, henry.pfister\}@duke.edu}%
   \thanks{S. Kadhe is currently with the
                     Department of Electrical Engineering and Computer Science, University of California, Berkeley, CA 94720, USA.
            Most of this work was conducted while he was with the
                     Department of Electrical and Computer Engineering,
                     Texas A\&M University,
                     College Station, Texas 77843, USA, 
           and was visiting Duke University during Fall 2017.
                     Email: swanand.kadhe@berkeley.edu}
 }

\maketitle

\begin{abstract}
Quantum error-correcting codes can be used to protect qubits involved in quantum computation. 
This requires that logical operators acting on protected qubits be translated to physical operators (circuits) acting on physical quantum states. 
We propose a mathematical framework for synthesizing physical circuits that implement logical Clifford operators for stabilizer codes. 
Circuit synthesis is enabled by representing the desired physical Clifford operator in $\mathbb{C}^{N \times N}$ as a partial $2m \times 2m$ binary symplectic matrix, where $N = 2^m$. 
We state and prove two theorems that use symplectic transvections to efficiently enumerate all binary symplectic matrices that satisfy a system of linear equations.
As an important corollary of these results, we prove that for an $\llbr m,m-k \rrbr$ stabilizer code every logical Clifford operator has $2^{k(k+1)/2}$ symplectic solutions. 
The desired physical circuits are then obtained by decomposing each solution as a product of elementary symplectic matrices, each corresponding to an elementary circuit. 
%This enumeration of physical realizations enables optimization over them with respect to a suitable metric.
Our assembly of the possible physical realizations enables optimization over the ensemble with respect to a suitable metric.
Furthermore, we show that any circuit that normalizes the stabilizer of the code can be transformed into a circuit that centralizes the stabilizer, while realizing the same logical operation. 
However, the optimal circuit for a given metric may not correspond to a centralizing solution. 
Our method of circuit synthesis can be applied to any stabilizer code, and this paper provides a proof of concept synthesis of universal Clifford gates for the $\llbr 6,4,2 \rrbr$ CSS code.
We conclude with a classical coding-theoretic perspective for constructing logical Pauli operators for CSS codes. 
Since our circuit synthesis algorithm builds on the logical Pauli operators for the code, this paper provides a complete framework for constructing all logical Clifford operators for CSS codes.
Programs implementing the algorithms in this paper, which includes routines to solve for binary symplectic solutions of general linear systems and our overall circuit synthesis algorithm, can be found at \url{https://github.com/nrenga/symplectic-arxiv18a}.
\end{abstract}

\begin{IEEEkeywords}
Heisenberg-Weyl group, symplectic geometry, transvections, Clifford group, stabilizer codes, logical operators
\end{IEEEkeywords}

\section{Introduction}
\label{sec:intro}

The first quantum error-correcting code (QECC) was discovered by Shor~\cite{Shor-physreva95}, and CSS codes were introduced by Calderbank and Shor~\cite{Calderbank-physreva96}, and Steane~\cite{Steane-physreva96}. 
The general class of stabilizer codes was introduced by Calderbank, Rains, Shor and Sloane~\cite{Calderbank-it98*2}, and by Gottesman~\cite{Gottesman-phd97}.
A QECC protects $m-k$ logical qubits by embedding them into a physical system comprising $m$ physical qubits.
QECCs can be used both for the reliable transmission of quantum information and for the realization of fault-tolerant quantum computation~\cite{Nielsen-2010,Wilde-2013}.
For computation, any desired operation on the $m-k$ logical qubits must be implemented as a physical operation on the $m$ physical qubits that preserves the code space.

For computation with a specific QECC, the first task is defining the logical Pauli operators for the code.
For stabilizer codes, the first algorithm for this task was introduced by Gottesman in~\cite[Sec.~4]{Gottesman-phd97}.
This algorithm reduces the stabilizer matrix of the code into a standard form in order to determine the logical Pauli operators.
A similar algorithm, based on symplectic geometry, was proposed by Wilde in~\cite{Wilde-physreva09}, which applies the Symplectic Gram-Schmidt Orthogonalization Procedure (SGSOP) to the normalizer of the code and then extracts the stabilizer generators and logical Pauli operators for the code.

Given the logical Pauli operators for a stabilizer QECC, physical realizations of Clifford operators on the logical qubits can be represented by $2m \times 2m$ binary symplectic matrices, reducing the complexity \emph{dramatically} from $2^m$ complex variables to $2m$ binary variables (see~\cite{Calderbank-physrevlett97,Gottesman-arxiv09} and Section~\ref{sec:operators}).
We exploit this fact to propose an algorithm that efficiently assembles \emph{all} symplectic matrices representing physical operators (circuits) that realize a given logical operator on the protected qubits.
This makes it possible to optimize the choice of circuit with respect to a suitable metric, that might be a function of the quantum hardware.
This paper provides a proof of concept demonstration using the well-known $\llbr 6,4,2 \rrbr$ QECC \cite{Gottesman-arxiv97,Chao-arxiv17b}, where we reduce the depth of the circuit (see Def.~\ref{def:depth}) for each operator.
The primary contributions of this paper are the four theorems that we state and prove in Section~\ref{sec:discuss}, and the main synthesis algorithm which builds on the results of these theorems (see Algorithm~\ref{alg:log_ops}).
These results form part of a larger program for fault-tolerant quantum computation, where the goal is to achieve reliability by using classical computers to track and control physical quantum systems and perform error correction only as needed.

We note that there are several works that focus on exactly decomposing, or approximating, an arbitrary unitary operator as a sequence of operators from a fixed \emph{instruction set}.
For example, in~\cite{Kliuchnikov-prl13} the authors demonstrate an algorithm that can approximate a random unitary with precision $\epsilon$ using $O(\log(1/\epsilon))$ Clifford and T gates (which forms their instruction set), and employing up to two ancillary qubits. 
They show that this algorithm saturates the information-theoretic lower bound for the problem and guarantees asymptotic optimality.
In~\cite{Amy-tcad13}, the authors present an algorithm for computing depth-optimal decompositions of arbitrary $m$-qubit unitary operators, and in~\cite{Maslov-arxiv17}, the authors use the Bruhat decomposition of the symplectic group to generate shorter Clifford circuits (referred to as stabilizer circuits therein).
However, these works do not consider circuit synthesis or optimization of unitary operators in the \emph{encoded} space.
Our work enables one to accomplish this task for Clifford operators on the logical qubits encoded by a stabilizer code, and obtain physical operators and a circuit decomposition for the same (using the result of Can in~\cite{Can-2017a}.
We note that there exists some works in the literature that study this problem for specific codes and operations, e.g., see~\cite{Grassl-isit13,Chao-arxiv17b}. 
However, we believe our work is the first to propose a framework to address this problem for general stabilizer codes, and hence enable automated circuit synthesis for encoded Clifford operators.

Subsequently, we also propose a general method of constructing logical Pauli operators for CSS codes.
Although this construction is closely related to the above two algorithms by Gottesman and Wilde, we provide a completely classical coding-theoretic perspective for this task which, to the best of our knowledge, has not appeared before in the literature.
Since our circuit synthesis for logical Clifford operations relies on the existence of physical implementations for the logical Pauli operators, this paper provides a complete framework for constructing all logical Clifford operators for CSS codes.

The primary content of this paper is organized as follows.
Section~\ref{sec:operators} discusses the connection between quantum computation and symplectic geometry, which forms the foundation for this work.
In Section~\ref{sec:css642}, the process of finding universal logical Clifford gates is demonstrated for the well-known $\llbr 6,4,2 \rrbr$ CSS code\cite{Gottesman-arxiv97,Chao-arxiv17b}. 
Then in Section~\ref{sec:discuss}, the general case of stabilizer codes is discussed rigorously via four theorems and our synthesis algorithm.
Subsequently, in Section~\ref{sec:css_operators} the general construction of stabilizers and logical Pauli operators for CSS codes is described from a classical coding-theoretic perspective.
Finally, Section~\ref{sec:conclusion} concludes the paper and discusses future work.

The appendices complementing the main content are organized as follows.
Appendix~\ref{sec:elem_symp} discusses elementary symplectic transformations, their circuits and the proof of Theorem~\ref{thm:Trung}, which states that any symplectic matrix can be decomposed as a product of these elementary symplectic transformations.
Appendix~\ref{sec:alg2_matlab} provides source code for Algorithm~\ref{alg:symp_lineq_all} with extensive comments.
Appendix~\ref{sec:css642_all_phy_ops} enumerates all 8 symplectic solutions for each of the logical Clifford operators discussed in Section~\ref{sec:css642} for the $\llbr 6,4,2 \rrbr$ CSS code.
Appendix~\ref{sec:css_proofs} contains the proofs for the results discussed in Section~\ref{sec:css_operators}.
Finally, Appendix~\ref{sec:circuit_ids} provides a table of some useful quantum circuit identities, pertaining to conjugation relations of Clifford gates with Pauli operators, which are frequently used in this paper as well as in the literature.
Two examples for explicitly calculating these relations algebraically is also provided.

\section{Physical and Logical Operators}
\label{sec:operators}

This section discusses the mathematical framework for quantum error correction introduced in~\cite{Calderbank-it98*2,Calderbank-physreva96,Gottesman-phd97} and described in detail in~\cite{Gottesman-arxiv97}.
Throughout this paper, binary vectors are row vectors and quantum states (i.e., kets) are column vectors.

\subsection{The Heisenberg-Weyl Group and Symplectic Geometry}
\label{sec:heisenberg_weyl}

The quantum states of a single qubit system are expressed as $\ket{\psi} = \alpha \ket{0} + \beta \ket{1} \in \mathbb{C}^2$, where kets $\ket{0} \triangleq \begin{bmatrix} 1 \\ 0 \end{bmatrix}$ and $\ket{1} \triangleq \begin{bmatrix} 0 \\ 1 \end{bmatrix}$ are called the \emph{computational basis} states, and $\alpha, \beta \in \mathbb{C}$ satisfy $|\alpha|^2 + |\beta|^2 = 1$ as per the \emph{Born rule}\cite[Chap. 3]{Wilde-2013}.
Any single qubit error can be expanded in terms of flip, phase and flip-phase errors (on a state $\ket{\psi}$) described by the Pauli matrices
\[ 
X \triangleq 
\begin{bmatrix}
0 & 1 \\
1 & 0
\end{bmatrix} , \ 
Z \triangleq 
{\renewcommand\arraycolsep{1.65pt}
\left[
\begin{array}{cr}
1 & 0 \\
0 & -1
\end{array}\right]} \ {\rm and}\ 
Y \triangleq \iota XZ = 
{\renewcommand\arraycolsep{1.65pt}
\left[
\begin{array}{cr}
0 & -\iota \\
\iota & 0
\end{array}\right]}
\]
respectively\cite[Chap. 10]{Nielsen-2010}, where $\iota \triangleq \sqrt{-1}$.
These operators are both unitary and Hermitian, and hence have eigenvalues $\pm 1$.
The states $\ket{0}$ and $\ket{1}$ are the eigenstates of $Z$, the \emph{conjugate basis} states $\ket{+} \triangleq \frac{1}{\sqrt{2}} (\ket{0} + \ket{1})$ and $\ket{-} \triangleq \frac{1}{\sqrt{2}} (\ket{0} - \ket{1})$ are the eigenstates of $X$, and the states $\frac{1}{\sqrt{2}} (\ket{0} + \iota \ket{1})$ and $\frac{1}{\sqrt{2}} (\ket{0} - \iota \ket{1})$ are the eigenstates of $Y$.
The states of an $m$-qubit system are described by (linear combinations of) Kronecker products of single-qubit states, and the corresponding (Pauli) errors are expressed as Kronecker products 
$\iota^{\kappa} \ E_1 \otimes E_2 \otimes \cdots \otimes E_m$, 
where $\kappa \in \{0,1,2,3\}, E_i \in \mathcal{P} \triangleq \{ I_2, X, Z, Y \}$ is the error on the $i$-th qubit and $I_2$ is the $2 \times 2$ identity matrix.
The set $\mathcal{P}$ forms an orthonormal basis for the space of unitary operators on $\mathbb{C}^2$ with respect to the trace inner product.
It is now well-known that any code that corrects these types of quantum errors will be able to correct errors in arbitrary models, assuming that the errors are not correlated among large number of qubits, and that the error rate is small~\cite{Ekert-physrevlett96}.

\begin{definition}
Given row vectors $a = [a_1,\ldots,a_m],\ b = [b_1,\ldots,b_m] \in \mathbb{F}_2^m$ we define the $m$-qubit operator
\begin{align}
\label{eq:d_ab}
D(a,b) \triangleq X^{a_1} Z^{b_1} \otimes \cdots \otimes X^{a_m} Z^{b_m} .
\end{align}
For $N = 2^m$, the \emph{Heisenberg-Weyl} group $HW_N$ of order $4N^2$ is defined as $HW_N \triangleq \{ \iota^{\kappa} D(a,b) \mid a,b \in \mathbb{F}_2^m, \kappa \in \{0,1,2,3\} \}$.
This is also called the Pauli group on $m$ qubits.
\end{definition}

Note that the elements of $HW_N$ can be interpreted either as errors, i.e., $HW_N = \{ \iota^{\kappa} \ E_1 \otimes E_2 \otimes \cdots \otimes E_m \mid E_i \in \mathcal{P} \}$, or, in general, as $m$-qubit operators.
Since $X$ and $Z$ anti-commute, i.e., $XZ = -ZX$, multiplication in $HW_N$ satisfies
\begin{align}
D(a, b) D(a', b') & = \left( \bigotimes_{j=1}^{m} X^{a_j} Z^{b_j} \right) \left( \bigotimes_{j=1}^{m} X^{a_j'} Z^{b_j'} \right) \nonumber \\
  & = \bigotimes_{j=1}^{m} X^{a_j} \left( Z^{b_j} X^{a_j'} \right) Z^{b_j'} \nonumber \\
%  & = \bigotimes_{j=1}^{m} X^{a_j} (-1)^{b_j a_j'} X^{a_j'} Z^{b_j} Z^{b_j'} \nonumber \\
  & = \bigotimes_{j=1}^{m} (-1)^{b_j a_j'} X^{a_j'} \left( X^{a_j} Z^{b_j'} \right) Z^{b_j}\ \left( \because Z^{b_j} X^{a_j'} = (-1)^{b_j a_j'} X^{a_j'} Z^{b_j} \right) \nonumber \\
%  & = \bigotimes_{j=1}^{m} (-1)^{b_j a_j'} X^{a_j'} (-1)^{a_j b_j'} Z^{b_j'} X^{a_j} Z^{b_j} \nonumber \\
  & = \prod_{j=1}^{m} (-1)^{a_j b_j'} (-1)^{b_j a_j'} \bigotimes_{j=1}^{m} X^{a_j'} Z^{b_j'} X^{a_j} Z^{b_j}\ \left(\because X^{a_j} Z^{b_j'} = (-1)^{a_j b_j'} Z^{b_j'} X^{a_j} \right) \nonumber \\
  & = (-1)^{a' b^T + b' a^T} \left( \bigotimes_{j=1}^{m} X^{a_j'} Z^{b_j'} \right) \left( \bigotimes_{j=1}^{m} X^{a_j} Z^{b_j} \right) \nonumber \\
\label{eq:hw_commute}
 & = (-1)^{a' b^T + b' a^T} D(a', b') D(\vecnot{a}, \vecnot{b}) .
\end{align}
Here we have used the property of the Kronecker product that $(A \otimes B)(C \otimes D) = AC \otimes BD$.
Similarly, it can be shown that elements of $HW_N$ also satisfy 
\begin{align}
D(a,b)^T = (-1)^{ab^T} D(a,b)\ \ \text{and} \ \ D(a,b) D(a',b') = (-1)^{a'b^T} D(a+a',b+b').
\end{align}

\begin{definition}
The standard \emph{symplectic inner product} in $\mathbb{F}_2^{2m}$ is defined as 
\begin{align}
\label{eq:symp_inner_pdt}
\syminn{[a,b]}{[a',b']} \triangleq a' b^T + b' a^T = [a,b]\ \Omega \ [a',b']^T \ (\bmod\ 2) ,
\end{align}
where the symplectic form in $\mathbb{F}_2^{2m}$ is $\Omega = 
\begin{bmatrix}
0 & I_m \\ 
I_m & 0
\end{bmatrix}$ (see~\cite{Calderbank-it98*2}).
\end{definition}

We observe that two operators $D(a,b)$ and $D(a',b')$ commute if and only if $\syminn{[a,b]}{[a',b']} = 0$.
Hence, the mapping $\gamma \colon HW_N/\langle \iota^{\kappa} I_N \rangle \rightarrow \mathbb{F}_2^{2m}$ defined by 
\begin{align}
\label{eq:gamma}
\gamma(D(a,b)) \triangleq [a,b]
\end{align}
is an isomorphism that enables the representation of elements of $HW_N$ (up to multiplication by scalars) as binary vectors.

\begin{remark}
Formally, a \emph{symplectic geometry} is a pair $(V,\beta)$ where $V$ is a finite-dimensional vector space over a field $K$ and $\beta \colon V \times V \rightarrow K$ is a non-degenerate alternating bilinear form (see~\cite[Chap. 1]{Gosson-2006}).
A vector space $V$ equipped with a non-degenerate alternating bilinear form is called a \emph{symplectic vector space}.
The function $\beta$ is \emph{bilinear} if for any $u,v,w \in V$ and any $k \in K$ it satisfies
\[ \beta(u + k v, w) = \beta(u, w) + k \beta(v, w) \ {\rm and}\ \beta(w, u + k v) = \beta(w, u) + k \beta(w, v) . \]
It is \emph{alternating} if for any $v \in V$ it satisfies $\beta(v, v) = 0$.
Notice that expanding $\beta(v + w, v + w) = 0 \Rightarrow \beta(v, w) = - \beta(w, v)$ for any $v, w \in V$.
Finally, $\beta$ is \emph{non-degenerate} if $\beta(v, w) = 0$ for all $w \in V$ implies that $v = 0$.
For this paper, we set $V = \mathbb{F}_2^{2m}, K = \mathbb{F}_2$ and $\beta = \syminn{\, \cdot \,}{\, \cdot \, }$ defined in~\eqref{eq:symp_inner_pdt}.
\end{remark}

\subsection{The Clifford Group and Symplectic Matrices}
\label{sec:clifford_gp}

Let $\mathbb{U}_N$ be the group of unitary operators on vectors in $\mathbb{C}^N$.
The Clifford group ${\rm Cliff}_N \subset \mathbb{U}_N$ consists of all unitary matrices $g \in \mathbb{U}_N$ that permute the elements of $HW_N$ under conjugation.
Formally,
\begin{align}
\text{Cliff}_N \triangleq \left\{ g \in \mathbb{U}_N \mid g D(a,b) g^{\dagger} \in HW_N \ \forall \ D(a,b) \in HW_N \right\} ,
\end{align}
where $g^{\dagger}$ is the adjoint (or Hermitian transpose) of $g$~\cite{Gottesman-arxiv09}.
Note that $HW_N \subset \text{Cliff}_N$.

\begin{definition}
Let $G$ and $H$ be subgroups of a group.
The set of elements $f \in G$ such that $f H f^{-1} = H$ is defined to be the \emph{normalizer} of $H$ in $G$, denoted as $\mathcal{N}_G(H)$.
The condition $f H f^{-1} = H$ can be restated as requiring that the left coset $fH$ be equal to the right coset $Hf$.
If $H$ is a subgroup of $G$, then $\mathcal{N}_G(H)$ is also a subgroup containing $H$.
In this case, $H$ is a normal subgroup of $\mathcal{N}_G(H)$.
\end{definition}

Hence, the Clifford group is the \emph{normalizer} of $HW_N$ in the unitary group $\mathbb{U}_N$, i.e., $\text{Cliff}_N = \mathcal{N}_{\mathbb{U}_N}(HW_N)$.
We regard operators in ${\rm Cliff}_N$ as physical operators acting on quantum states in $\mathbb{C}^N$, to be implemented by quantum circuits.

\begin{definition}
\label{def:automorphism}
Let $A$ be a collection of objects.
The automorphism group Aut($A$) of $A$ is the group of functions $f \colon A \rightarrow A$ (with the composition operation) that preserve the structure of $A$.
If $A$ is a group then every $f \in \text{Aut}(A)$ preserves the multiplication table of the group.
The inner automorphism group Inn($A$) is a subgroup of Aut($A$) defined as $\text{Inn}(A) \triangleq \{ f_a \mid a \in A \}$, where $f_a \in \text{Aut}(G)$ is defined by $f_a(x) = a x a^{-1}$, i.e., these are automorphisms of $A$ induced by conjugation with elements of $A$.
\end{definition}

As a corollary of Theorem~\ref{thm:symp_action}, it holds that $\text{Aut}(HW_N) = \text{Cliff}_N$, i.e. the automorphisms induced by conjugation form the full automorphism group of $HW_N$ in $\mathbb{U}_N$.
We sometimes refer to elements of $\text{Cliff}_N$ as \emph{unitary automorphisms} of $HW_N$.

\begin{fact}
\label{fact:XnZn}
The action of any unitary automorphism of $HW_N$ is defined by its action on the following two maximal commutative subgroups of $HW_N$ that generate $HW_N$:
\begin{align}
\label{eq:XnZn}
X_N \triangleq \left\{ D(a,0) \in HW_N \mid a \in \mathbb{F}_2^m \right\} , \ \ Z_N \triangleq \left\{ D(0,b) \in HW_N \mid b \in \mathbb{F}_2^m \right\} .
\end{align}
\end{fact}

\begin{definition}
Given a group $G$ and an element $h \in G$, a \emph{conjugate} of $h$ in $G$ is an element $ghg^{-1}$, where $g \in G$.
Conjugacy defines an equivalence relation on $G$ and the equivalence classes are called \emph{conjugacy classes} of $G$.
\end{definition}

\begin{lemma}
The set of all conjugacy classes of $HW_N$ is given by
\begin{align*}
\text{Class}(HW_N) = \left( \bigcup_{\substack{D(a,b) \in HW_N\\(a,b) \neq (0,0)}} \left\{ \{ D(a,b), -D(a,b) \}, \{ \iota D(a,b), -\iota D(a,b) \} \right\} \right) \cup \left( \bigcup_{\kappa=0}^{3} \left\{ \iota^\kappa I_N \right\} \right).
\end{align*}
\begin{IEEEproof}
For an element $D(a,b) \in HW_N$, its conjugacy class is given by
$ \{ D(c,d) D(a,b) D(c,d)^{-1} \mid D(c,d) \in HW_N \} $.
We know that $D(c,d) D(a,b) = (-1)^{cb^T + da^T} D(a,b) D(c,d)$ which implies
\[ D(c,d) D(a,b) D(c,d)^{-1} = (-1)^{cb^T + da^T} D(a,b) D(c,d) D(c,d)^{-1} = (-1)^{cb^T + da^T} D(a,b) . \]
Therefore $D(a,b)$ is mapped either to itself, if $cb^T + da^T = 0$, or to $- D(a,b)$, if $cb^T + da^T = 1$.
This does not change if $D(c,d)$ has a leading $\pm \iota$ because the inverse will cancel it.
So the conjugacy class of $\pm D(a,b)$ is $\{ D(a,b), -D(a,b) \}$ and that of $\pm \iota D(a,b)$ is $\{ \iota D(a,b), -\iota D(a,b) \}$.
For the special case of $D(a,b) = \iota^\kappa I_N$ with $\kappa \in \{ 0,1,2,3 \}$, the corresponding conjugacy class is a singleton $\{ \iota^\kappa I_N \}$ since $D(c,d) D(a,b) = \iota^\kappa D(c,d)$.
Hence the result follows.
\end{IEEEproof}
\end{lemma}

\begin{corollary}
\label{cor:conjugacy}
The elements of $HW_N$ have order either $1, 2$ or $4$.
Any automorphism of $HW_N$ must preserve the order of all elements since by Definition~\ref{def:automorphism} it must preserve the multiplication table of the group.
Also, the inner automorphisms defined by conjugation with matrices $\iota^\kappa D(a,b) \in HW_N$ preserve every conjugacy class of $HW_N$, because~\eqref{eq:hw_commute} implies that elements in $HW_N$ either commute or anti-commute.
\end{corollary}

\begin{definition}
Given row vectors $a,b \in \mathbb{F}_2^{m}$ we define the $m$-qubit Hermitian operator ${E(a,b) \triangleq \iota^{ab^T} D(a,b)}$.
Note that $E(a,b)^2 = I_N$ and hence $E(a,b)$ is also unitary.
\end{definition}

Using similar techniques as in Section~\ref{sec:heisenberg_weyl}, we can show the following result.

\begin{lemma}
\label{lem:Eab}
Given $[a,b], [a',b'] \in \mathbb{F}_2^{2m}$ we have
\begin{align*}
E(a,b) E(a',b') = (-1)^{a'b^T} \iota^{\syminn{[a,b]}{[a',b']}} E(a+a',b+b') .
\end{align*}
\end{lemma}
This can be rewritten as $E(a,b) E(a',b') = \iota^{a'b^T - b'a^T} E(a+a',b+b')$, where the exponent is computed modulo $4$.
Hence, if $E(a,b)$ and $E(a',b')$ commute then $E(a,b) E(a',b') = \pm E(a+a',b+b')$, and if not then $E(a,b) E(a',b') = \pm \iota E(a+a',b+b')$.

It can also be shown that the matrices $\frac{1}{\sqrt{N}} E(a,b)$ form an orthonormal basis for the real vector space of $N \times N$ Hermitian matrices, under the trace inner product.

%\begin{lemma}
%The matrices $\frac{1}{\sqrt{N}} E(a,b)$ form an orthonormal basis for the real vector space of $N \times N$ Hermitian matrices.
%\begin{IEEEproof}
%The set of $N \times N$ Hermitian matrices is a real vector space under the trace inner product defined as $\langle A,B \rangle = {\rm Tr}(AB)$.
%Define $\hat{E}(a,b) \triangleq \frac{1}{\sqrt{N}} E(a,b)$.
%Then we see that $\langle \hat{E}(a,b),\hat{E}(a,b) \rangle = {\rm Tr}\left( \frac{1}{N} E(a,b)^2 \right) = {\rm Tr}\left( \frac{1}{N} I_N \right) = 1$, so that the matrices $\hat{E}(a,b)$ are normalized.
%Next, observe that ${\rm Tr}(X) = {\rm Tr}(Z) = {\rm Tr}(XZ) = 0$ where $X,Z,XZ \in HW_2$.
%Also the Kronecker product has the property that ${\rm Tr}\left( A \otimes B \right) = {\rm Tr}(A) {\rm Tr}(B)$.
%Therefore, for $a,b \in \mathbb{F}_2^m$ ($N=2^m$) such that not both $a = 0$ and $b = 0$ simultaneously, ${\rm Tr}(\hat{E}(a,b)) = {\rm Tr}(E(a,b)) = {\rm Tr}(D(a,b)) = 0$.
%This implies that $\langle \hat{E}(a,b), \hat{E}(c,d) \rangle = 0$ whenever not both $a = c$ and $b = d$ simultaneously.
%Hence the matrices $\hat{E}(a,b)$ form an orthonormal basis for a space of dimension $N^2$.
%As the real vector space of $N \times N$ Hermitian matrices has dimension $N^2$ and the matrices $\hat{E}(a,b)$ are Hermitian, the set of all such matrices spans this vector space.
%\end{IEEEproof}
%\end{lemma}

\begin{definition}
\label{def:symp_matrix}
An invertible matrix $F \in \mathbb{F}_2^{2m \times 2m}$ is said to be a \emph{symplectic matrix} if it preserves the symplectic inner product between vectors in $\mathbb{F}_2^{2m}$ (see \cite{Can-2017a,Gottesman-arxiv09}).
In other words, a symplectic matrix $F$ satisfies $\syminn{[a,b]F}{[a',b']F} = \syminn{[a,b]}{[a',b']}$ for all $[a,b], [a',b'] \in \mathbb{F}_2^{2m}$.
This implies that $[a,b]\ F\Omega F^T \ [a',b']^T = [a,b]\ \Omega \ [a',b']^T$ and hence an equivalent condition for a symplectic matrix is that it must satisfy $F \Omega F^T = \Omega$.
\end{definition}

If we represent a symplectic matrix as $F = 
\begin{bmatrix}
A & B \\
C & D
\end{bmatrix} $ then the condition $F \Omega F^T = \Omega$ can be explicitly written as 
\begin{align}
A B^T = B A^T, \ C D^T = D C^T, \ A D^T + B C^T = I_m .
\end{align}

\begin{definition}
The group of symplectic $2m \times 2m$ binary matrices is called the \emph{symplectic group} over $\mathbb{F}_2^{2m}$ and is denoted by $\text{Sp}(2m,\mathbb{F}_2)$.
It can be interpreted as a generalization of the orthogonal group to the symplectic inner product.
The size of the symplectic group is well-known to be $2^{m^2} \prod_{j=1}^{m} (4^j - 1)$ (e.g., see~\cite{Calderbank-it98*2}).
\end{definition}

\begin{definition}
\label{def:symp_basis}
A \emph{symplectic basis} for $\mathbb{F}_2^{2m}$ is a set of pairs $\{ (v_1,w_1), (v_2,w_2), \ldots, (v_m,w_m) \}$ such that $\syminn{v_i}{w_j} = \delta_{ij}$ and $\syminn{v_i}{v_j} = \syminn{w_i}{w_j} = 0$, where $i,j \in \{1,\ldots,m\}$, and $\delta_{ij} = 1$ if $i=j$ and $0$ if $i \neq j$.
\end{definition}

Note that the rows of any matrix in $\text{Sp}(2m,\mathbb{F}_2)$ form a symplectic basis for $\mathbb{F}_2^{2m}$.
Also, the top and bottom halves of a symplectic matrix satisfy 
$\begin{bmatrix} A & B \end{bmatrix} \Omega \begin{bmatrix} A & B \end{bmatrix}^T = \begin{bmatrix} C & D \end{bmatrix} \Omega \begin{bmatrix} C & D \end{bmatrix}^T = 0$.
There exists a symplectic Gram-Schmidt orthogonalization procedure that can produce a symplectic basis starting from the standard basis for $\mathbb{F}_2^{2m}$ and an additional vector $v \in \mathbb{F}_2^{2m}$ (see~\cite{Koenig-jmp14}).

Next we state a classical result which forms the foundation for our algorithm in Section~\ref{sec:discuss} to synthesize logical Clifford operators of stabilizer codes.
For completeness we also provide a proof here (adapted from~\cite{Can-2017a}).

\begin{theorem}
\label{thm:symp_action}
The automorphism induced by $g \in \text{Cliff}_N$ satisfies
\begin{align}
\label{eq:symp_action}
g E(a,b) g^{\dagger} = \pm E\left( [a,b] F_g \right) \ , \ {\rm where} \ \ F_g = 
\begin{bmatrix}
A_g & B_g \\
C_g & D_g
\end{bmatrix} \in \text{Sp}(2m,\mathbb{F}_2).
\end{align} 
\begin{IEEEproof}
First we show that there exists a well-defined $2m \times 2m$ binary transformation $F$ such that for all $[a,b] \in \mathbb{F}_2^{2m}$ we have $g E(a,b) g^{\dagger} = \pm E\left( [a,b] F \right)$.
Let the standard basis vectors of $\mathbb{F}_2^{m}$ be denoted as $e_i$, which have an entry $1$ in the $i$-th position and entries $0$ elsewhere.
Then $\{E(e_1,0),\ldots,E(e_m,0)\}$ and $\{E(0,e_1),\ldots,E(0,e_m)\}$ form a basis for $X_N$ and $Z_N$ defined in~\eqref{eq:XnZn}, respectively.
Since $g \in \text{Cliff}_N$ is an automorphism of $HW_N$, by Corollary~\ref{cor:conjugacy} it preserves the order of every element of $HW_N$, and hence maps $E(e_i,0) \mapsto \pm E(a_i',b_i'), E(0,e_j) \mapsto \pm E(c_j',d_j')$ for some $[a_i',b_i'], [c_j',d_j'] \in \mathbb{F}_2^{2m}$, where $i,j \in \{1,\ldots,m\}$.
%As $E(e_i,0), E(0,e_j) \in \text{Cliff}_N$, $\pm E(a_i',b_i'), \pm E(c_j',d_j')$ belong to the conjugacy classes of $E(e_i,0), E(0,e_j)$ in $\text{Cliff}_N$, respectively.
Therefore we can express the action of $g$ as $g E(e_i,0) g^{\dagger} = \pm E(a_i',b_i'), g E(0,e_j) g^{\dagger} = \pm E(c_j',d_j')$.
Using the same $i,j$ define a matrix $F$ with the $i$-th row being $[a_i',b_i']$ and the $(m+j)$-th row being $[c_j',d_j']$.
This matrix satisfies 
%$[e_i,0] F = [a_i',b_i']$ and $[0,e_j] F = [c_j',d_j']$ so that
\begin{align*}
g E(e_i,0) g^{\dagger} = \pm E\left( [e_i,0] F \right), \ \ g E(0,e_j) g^{\dagger} = \pm E\left( [0,e_j] F \right) .
\end{align*}
Using the fact that any $[a,b] \in \mathbb{F}_2^{2m}$ can be written as a linear combination of $[e_i,0]$ and $[0,e_j]$, the result from Lemma~\ref{lem:Eab}, and Corollary~\ref{cor:conjugacy} we have $g E(a,b) g^{\dagger} = \pm E\left( [a,b] F \right)$.
As the rows of $F$ are a result of the action of $g$ on $X_N$ and $Z_N$, we explicitly denote this dependence by $F_g$.
We are just left to show that $F_g$ is symplectic.

Conjugating both sides of the equation in Lemma~\ref{lem:Eab} by $g$ we get
\begin{align*}
\left( g E(a,b) g^{\dagger} \right) \left( g E(a',b') g^{\dagger} \right) &= (-1)^{a'b^T} \iota^{\syminn{[a,b]}{[a',b']}} \left( g E(a+a',b+b') g^{\dagger} \right) \\
\Rightarrow E\left( [a,b] F_g \right) E\left( [a',b'] F_g \right) &= \pm (-1)^{a'b^T} \iota^{\syminn{[a,b]}{[a',b']}} E\left( [a+a',b+b'] F_g \right) .
\end{align*}
Also, applying Lemma~\ref{lem:Eab} to $E\left( [a,b] F_g \right), E\left( [a',b'] F_g \right)$, and defining $[c,d] \triangleq [a,b] F_g, [c',d'] \triangleq [a',b'] F_g$, we get
\begin{align*}
E\left( [a,b] F_g \right) E\left( [a',b'] F_g \right) = (-1)^{c'd^T} \iota^{\syminn{[a,b] F_g}{[a',b'] F_g}} E\left( [a+a',b+b'] F_g \right) .
\end{align*}
Equating the two expressions on the right hand side for all $[a,b],[a',b']$ we get $F_g \Omega F_g^T = \Omega$, or that $F_g$ is symplectic.
\end{IEEEproof}
\end{theorem}

%Since any $g \in \text{Aut}(HW_N)$ maps to a $F_g \in \text{Sp}(2m,\mathbb{F}_2)$, which corresponds to a $g \in \text{Cliff}_N$, we have $\text{Aut}(HW_N) = \text{Cliff}_N$.
Since the action of any $g \in \text{Aut}(HW_N)$ on $HW_N$ can be expressed as mappings $[e_i,0] \mapsto [a_i',b_i'], [0,e_j] \mapsto [c_j',d_j']$ in $\mathbb{F}_2^{2m}$, that can be induced by a $g \in \text{Cliff}_N$ (under conjugation), we have $\text{Aut}(HW_N) = \text{Cliff}_N$.
The fact that $F_g$ is symplectic expresses the property that the automorphism induced by $g$ must respect commutativity in $HW_N$.
By Corollary~\ref{cor:conjugacy}, any inner automorphism induced by conjugation with $g \in HW_N$ has $F_g = I_{2m}$.
So the map $\phi \colon {\rm Cliff}_N \rightarrow \text{Sp}(2m,\mathbb{F}_2)$ defined by 
\begin{align}
\label{eq:phi}
\phi(g) \triangleq F_g
\end{align}
is a homomorphism with kernel $HW_N$, and every Clifford operator maps down to a matrix $F_g$.
Hence, $HW_N$ is a normal subgroup of ${\rm Cliff}_N$ and ${\rm Cliff}_N/HW_N \cong \text{Sp}(2m,\mathbb{F}_2)$.

\vspace{0.1cm}
In summary, every unitary automorphism $g \in \text{Cliff}_N$ of $HW_N$ induces a symplectic transformation $F_g \in \text{Sp}(2m,\mathbb{F}_2)$ via the map $\phi$, and two automorphisms that induce the same symplectic transformation can differ only by an inner automorphism that is also an element of $HW_N$.
The symplectic group $\text{Sp}(2m,\mathbb{F}_2)$ is generated by the set of elementary symplectic transformations given in the first column of Table~\ref{tab:std_symp} (also see~\cite{Dehaene-physreva03}).
The second column lists their corresponding unitary automorphisms, under the relation proven in Theorem~\ref{thm:symp_action}, i.e., $\bar{g} E(a,b) \bar{g}^{\dagger} = \pm E\left( [a,b] F_{\bar{g}} \right)$.
The third column relates these elementary forms with the elementary quantum gates that they can represent.
A discussion of these transformations and their circuits is provided in Appendix~\ref{sec:elem_symp}.
The terminology of logical and physical operators, and the notation $\bar{g}$ in Table~\ref{tab:std_symp}, are introduced in Section~\ref{sec:log_ops_qecc}.

Since it is a well-known fact that $\text{Cliff}_N = \langle HW_N, H, P, \text{CNOT\ or\ CZ} \rangle$ (e.g., see~\cite{Gottesman-arxiv09}), this verifies that the matrices in the first column generate the symplectic group $\text{Sp}(2m,\mathbb{F}_2)$ under the map $\phi$, and hence form a universal set for the same.
Here, and throughout this paper, $\langle \, \cdot \, \rangle$ represents the span of the elements inside and
\begin{align}
H \triangleq {\renewcommand\arraycolsep{1.35pt}
\frac{1}{\sqrt{2}}
\left[
\begin{array}{rr}
1 & 1 \\
1 & -1
\end{array}\right]},\ 
P \triangleq \begin{bmatrix}
1 & 0 \\
0 & \iota
\end{bmatrix},\ 
\text{CNOT}_{1 \rightarrow 2} \triangleq \ketbra{0} \otimes I_2 + \ketbra{1} \otimes X\ \text{and}\ \text{CZ}_{12} \triangleq \ketbra{0} \otimes I_2 + \ketbra{1} \otimes Z
\end{align}
are the Hadamard, Phase, Controlled-NOT and Controlled-$Z$ operators, respectively. 
For CNOT and CZ, the first qubit is the control and the second qubit is the target.
Since swapping the control and target qubits does not change CZ, we use the subscript ``$12$'' rather than ``$1 \rightarrow 2$'' as in CNOT.
Some important circuit identities involving these operators are listed in Appendix~\ref{sec:circuit_ids}.
Note that the Clifford group does not enable universal quantum computation since it is a finite group.
For example, the ``T'' gate defined by $T \triangleq \begin{bmatrix}
1 & 0 \\
0 & e^{\iota \pi/4}
\end{bmatrix} = \sqrt{P}$ does not belong to the Clifford group.
In fact, the Clifford group along with the T gate forms a universal set of gates to perform universal quantum computation~\cite{Amy-tcad13}.

\begin{definition}
\label{def:depth}
In a circuit, any set of gates acting sequentially on disjoint subsets of qubits can be applied concurrently.
These gates constitute one stage of the circuit, and the number of such stages is defined to be the \emph{depth} of the circuit.
\end{definition}

For example, consider $m=4$ and the quantum circuit $H_1$ --- $\cz{2}{3}$ --- $P_4$, where the subscripts indicate the qubit(s) on which the gate is applied.
More precisely, we can explicitly write $H_1 = H \otimes I_2 \otimes I_2 \otimes I_2$, where the subscript $2$ indicates the $2 \times 2$ identity matrix.
This is a circuit of depth $1$.
However the circuit $H_2$ --- $\cz{2}{3}$ --- $P_4$ has depth $2$.
Note that the unitary operator is computed as the matrix product $U = P_4 \cz{2}{3} H_2 = \cz{2}{3} P_4 H_2 = \cz{2}{3} H_2 P_4$, since $U$ acts on a quantum state $\ket{\psi}$ as $U \ket{\psi}$, i.e., on the right.
However, in the circuit the state goes through from left to right and hence the order is reversed.

\subsection{Symplectic Transvections}
\label{sec:symp_transvec}

\begin{definition}
Given a vector $h \in \mathbb{F}_2^{2m}$, a \emph{symplectic transvection}~\cite{Koenig-jmp14} is a map $Z_h \colon \mathbb{F}_2^{2m} \rightarrow \mathbb{F}_2^{2m}$ defined by
\begin{align}
\label{eq:symp_transvec}
Z_h(x) \triangleq x + \syminn{x}{h} h \ \Leftrightarrow \ F_h \triangleq I_{2m} + \Omega h^T h ,
\end{align}
where $F_h$ is its associated symplectic matrix.
A transvection does not correspond to a single elementary Clifford operator.
\end{definition}

\begin{fact}[{\hspace{1sp}\cite[Theorem 2.10]{Salam-laa08}}]
The symplectic group $\text{Sp}(2m,\mathbb{F}_2)$ is generated by the family of symplectic transvections.
\end{fact}

An important result that is involved in the proof of this fact is the following theorem from~\cite{Salam-laa08,Koenig-jmp14}, which we restate here for $\mathbb{F}_2^{2m}$ since we will build on this result to state and prove Theorem~\ref{thm:symp_lineq}.

\begin{theorem}
\label{thm:symp_transvec}
Let $x,y \in \mathbb{F}_2^{2m}$ be two non-zero vectors. Then $x$ can be mapped to $y$ by a product of at most two symplectic transvections.
\begin{IEEEproof}
There are two possible cases: $\syminn{x}{y} = 1$ or $0$.
First assume $\syminn{x}{y} = 1$.
Define $h \triangleq x + y$ so that 
\begin{align*}
x F_h = Z_h(x) = x + \syminn{x}{x+y} (x+y) = x + \left( \syminn{x}{x} + \syminn{x}{y} \right) (x+y) = x + (0+1) (x+y) = y .
\end{align*}
Next assume $\syminn{x}{y} = 0$.
Define $h_1 \triangleq w + y, h_2 \triangleq x + w$, where $w \in \mathbb{F}_2^{2m}$ is chosen such that $\syminn{x}{w} = \syminn{y}{w} = 1$.
Then we have
\begin{IEEEeqnarray*}{rCl+x*}
x F_{h_1} F_{h_2} = Z_{h_2} \left( Z_{h_1} (x) \right) & = & Z_{h_2} \left(x + \syminn{x}{w+y} (w+y) \right) = (x + w+ y) + \syminn{(x+w)+y}{x+w} (x+w) = y. &  \IEEEQEDhere
\end{IEEEeqnarray*}
\end{IEEEproof}
\end{theorem}

In Section~\ref{sec:discuss} we will use the above result to propose an algorithm (Alg.~\ref{alg:transvec}) which determines a symplectic matrix $F$ that satisfies $x_i F = y_i,\ i=1,2,\ldots,t \leq 2m$, where $x_i$ are linearly independent and satisfy $\syminn{x_i}{x_j} = \syminn{y_i}{y_j} \ \forall \ i,j \in \{1,\ldots,t\}$.

\begin{table}
\centering
%\vspace{-0.05cm}
%\vspace{-0.1cm}
\begin{tabular}{c|c|c}
Logical Operator $F_{\bar{g}}$ & Physical Operator $\bar{g}$ & Circuit Element\\
~ & ~ & ~ \\
\hline
~ & ~ & ~ \\
$\Omega = \begin{bmatrix} 0 & I_m \\ I_m & 0 \end{bmatrix}$ & $H_N = H^{\otimes m} = 
{\renewcommand\arraycolsep{1.35pt}
\frac{1}{(\sqrt{2})^{m}}
\left[
\begin{array}{rr}
1 & 1 \\
1 & -1
\end{array}\right]^{\otimes m}}$ & Transversal Hadamard \\
~ & ~ & ~ \\
$A_Q = \begin{bmatrix} Q & 0 \\ 0 & Q^{-T} \end{bmatrix}$ & $a_Q: e_{v} \mapsto e_{v Q}$ & \makecell{Controlled-NOT (CNOT)\\ Qubit Permutation} \\
~ & ~ & ~ \\
\makecell{$T_R = \begin{bmatrix} I_m & R \\ 0 & I_m \end{bmatrix}$\\~\\($R$ symmetric)} & \makecell{$t_R\ =\ {\rm diag}\left( \iota^{v R v^T} \right)$
%\\~\\(exponents\ $vPv^T$ read modulo $4$)
} & \makecell{Controlled-$Z$ (CZ)\\ Phase ($P$)}\\
~ & ~ & ~ \\
$G_k = \begin{bmatrix} L_{m-k} & U_k \\ U_k & L_{m-k} \end{bmatrix}$ & $g_k = H_{2^k} \otimes I_{2^{m-k}}$ & Partial Hadamards \\
~ & ~ & ~ \\
%$G_k \Omega^{-1} = \begin{bmatrix} U_k & L_{m-k} \\ L_{m-k} & U_k \end{bmatrix}$ & $g_k H_N = I_{2^k} \otimes H_{2^{m-k}}$ \\
%~ & ~ \\
\hline
\end{tabular}
\caption{\label{tab:std_symp}A universal set of logical operators for $\text{Sp}(2m,\mathbb{F}_2)$ and their corresponding physical operators in $\text{Cliff}_N$ ({\normalfont see Appendix~\ref{sec:elem_symp} for a detailed discussion and circuits}).
{\normalfont The number of $1$s in $Q$ and $R$ directly relates to number of gates. The $N$ coordinates are indexed by binary vectors $v \in \mathbb{F}_2^m$, and $e_v$ denotes the standard basis vector in $\mathbb{C}^N$ with an entry $1$ in position $v$ and all other entries $0$. 
More precisely, if $v = [v_1,v_2,\ldots,v_m], e_0 \triangleq \ket{0}, e_1 \triangleq \ket{1}$ then $e_v = e_{v_1} \otimes e_{v_2} \otimes \cdots e_{v_m} = \ket{v_1} \otimes \cdots \otimes \ket{v_m} = \ket{v}$.
Here $H_{2^k}$ denotes the Walsh-Hadamard matrix of size $2^k$, $U_k = {\rm diag}\left( I_k, 0_{m-k} \right)$ and $L_{m-k} = {\rm diag}\left( 0_k, I_{m-k} \right)$. }}
%\vspace{-0.4cm}
\end{table}

\subsection{Stabilizer Codes}
\label{sec:stabilizer_codes}

\begin{definition}
A \emph{stabilizer} is an abelian subgroup $S$ of $HW_N$ generated by commuting Hermitian matrices~\cite{Gottesman-phd97,Nielsen-2010}.
We denote the normalizer of $S$ in $HW_N$ as $S^{\perp}$, i.e., $S^{\perp} \triangleq \mathcal{N}_{HW_N}(S)$.
\end{definition}

\begin{definition}
The \emph{stabilizer code} corresponding to $S$ is the subspace $V(S)$ fixed pointwise by $S$, i.e., 
\begin{align} 
V(S) = \{ \ket{\psi} \in \mathbb{C}^N \mid g \ket{\psi} = \ket{\psi}\  \forall \ g \in S \} .
\end{align}
\end{definition}

\begin{figure}
\centering
\scalebox{1.1}{%
\begin{tikzpicture}

\node (V) at (0,0) {$V(S)$};
\node (S) at (0,1.5) {$S$};
\node (HW) at (0,3) {$HW_N$};
\node (Cliff) at (0,4.5) {$\text{Cliff}_N = \mathcal{N}_{\mathbb{U}_N}(HW_N)$};
\node (U) at (0,6) {$\mathbb{U}_N$};

\draw[dashed] (V) -- (S);
\path[draw,->] (S) -- (HW);
\path[draw,->] (HW) -- (Cliff);
\path[draw,->] (Cliff) -- (U);

\node (Sperp) at (3,2.25) {$S^{\perp} \triangleq \mathcal{N}_{HW_N}(S)$};
\node (Ncliff) at (3,3.75) {$\mathcal{N}_{\text{Cliff}_N}(S)$};
\node (NU) at (3,5.25) {$\mathcal{N}_{\mathbb{U}_N}(S)$};

\path[draw,->] (S) -- (Sperp);
\path[draw,->] (Sperp) -- (Ncliff);
\path[draw,->] (Ncliff) -- (NU);
\path[draw,->] (NU) -- (U);
\path[draw,->] (Sperp) -- (HW);
\path[draw,->] (Ncliff) -- (Cliff);

\end{tikzpicture}
}
\caption{\label{fig:subgroups}Lattice of subgroups for the unitary group $\mathbb{U}_N$. A solid arrow ``$A \longrightarrow B$'' indicates that $A$ is a subgroup of $B$. The dashed line implies that the stabilizer group $S$ fixes the subspace $V(S)$.}
\end{figure}
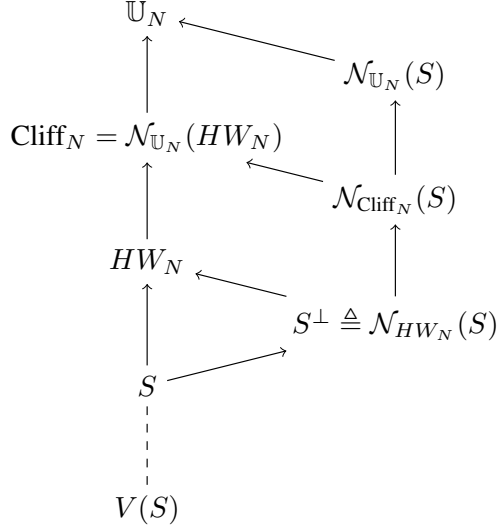

Therefore, for $V(S)$ to be non-trivial, a stabilizer $S$ has the additional property that if it contains an operator $g$, then it does not contain $-g$, i.e., $-I_N \notin S$.
Fig.~\ref{fig:subgroups} shows the lattice of subgroups for the unitary group $\mathbb{U}_N$.

For $a,b \in \mathbb{F}_2^m$, recollect that $\pm E(a,b) = \pm \iota^{ab^T} D(a,b)$ is Hermitian and $E(a,b)^2 = I_N$.
Then the operators $\frac{I_N \pm E(a,b)}{2}$ project onto the $\pm 1$ eigenspaces of $E(a,b)$, respectively.
Consider the stabilizer $S$ generated by Hermitian matrices $E(a_i,b_i)$, where $[a_i,b_i], i=1,2,\ldots,k$ are linearly independent vectors in $\mathbb{F}_2^{2m}$.
Observe that the operator
\begin{align}
\frac{I_N + E(a_1,b_1)}{2} \times \cdots \times \frac{I_N + E(a_k,b_k)}{2} 
\end{align}
projects onto $V(S)$, and that ${\rm dim}\ V(S) = 2^{m-k} \triangleq M$.
Such a code encodes $m-k$ \emph{logical} qubits into $m$ \emph{physical} qubits.
Hence an $\llbr m, m-k \rrbr$ QECC is an embedding of a $2^{m-k}$-dimensional Hilbert space into a $2^m$-dimensional Hilbert space. 
Note that all quantum codes are not necessarily stabilizer codes (e.g., see \cite{Calderbank-it98*2}).
Logical qubits are commonly referred to as \emph{encoded} qubits, or \emph{protected} qubits, and their operators are referred to as encoded operators.

\subsection{Logical Operators for QECCs}
\label{sec:log_ops_qecc}

%\begin{definition}
Unitary operators $g^L \in \mathbb{U}_{M}$, where $M = 2^{m-k}$, acting on the logical qubits are called \emph{logical} operators.
%\end{definition}
%
QECCs encode a \emph{logical} state in $\mathbb{C}^M$ into a \emph{physical} state in $\mathbb{C}^N$.
The process of \emph{synthesizing} a logical operator $g^L \in \mathbb{U}_M$ for a QECC refers to finding a \emph{physical} operator $\bar{g} \in \mathbb{U}_{N}$ that preserves the code space (i.e., normalizes $S$) and realizes the action of $g^L$ on the qubits it protects.

\begin{remark}
\label{rem:logical_op}
It is standard practice in the literature to refer to the physical implementation $\bar{g}$ itself as the logical operator.
We adopt this convention frequently but it should be clear from the context if we are referring to an operator in $\mathbb{U}_M$ or $\mathbb{U}_N$.
\end{remark}

Two well-known methods to synthesize logical Pauli operators were described in \cite{Gottesman-phd97} and \cite{Wilde-physreva09}.
For stabilizer codes, these imply that for all logical Paulis $h^L \in HW_M$ the associated physical operator $\bar{h} \in HW_N$ as well.
Hence for all $g^L \in \text{Cliff}_M$ we also have $\bar{g} \in \text{Cliff}_N$.
The physical operators $\bar{h}$ have a representation in $\mathbb{F}_2^{2m}$ via the map $\gamma$ defined in~\eqref{eq:gamma}.
Using the map $\phi$ defined in~\eqref{eq:phi}, we regard a logical Clifford operator $g^L \in \text{Cliff}_M$ as a symplectic matrix $F_g \in \text{Sp}(2(m-k), \mathbb{F}_2)$.
For stabilizer codes, in order to translate $g^L$ into a physical operator $\bar{g}$, there are multiple ways to embed $F_g$ into $F_{\bar{g}} \in \text{Sp}(2m, \mathbb{F}_2)$ such that the corresponding $\bar{g}$ operates on states in $\mathbb{C}^N$ and acts as desired on the states of the QECC.
For each $g^L \in \text{Cliff}_M$ our algorithm allows one to identify \emph{all} such embeddings.
The idea is as follows.
Applying Fact~\ref{fact:XnZn} for $HW_M$, we observe that the logical Clifford operators $g^L \in \text{Cliff}_M$ are uniquely defined by their conjugation relations with the logical Paulis $h^L$ (also see~\cite{Gottesman-arxiv97,Gottesman-arxiv09,Nielsen-2010}).
Therefore these relations can be directly translated to their physical equivalents $\bar{g}$ and $\bar{h}$ respectively, i.e., $g^L h^L (g^L)^{\dagger} = (h')^L \in HW_M \Rightarrow \bar{g} \bar{h} \bar{g}^{\dagger} = \bar{h}' \in HW_N$ as well.
Using the relation in~\eqref{eq:symp_action}, these conditions are translated into linear constraints on $F_{\bar{g}}$.
Then, linear constraints that require $F_{\bar{g}}$ to normalize $S$ are added.
The set of all $F_{\bar{g}} \in \text{Sp}(2m, \mathbb{F}_2)$ that satisfy these constraints identify all embeddings of $F_g$ into $\text{Sp}(2m, \mathbb{F}_2)$.
Since the space of symplectic operations is much smaller than the space of unitary operations, one can optimize over this space much more efficiently.

After we obtain $F_{\bar{g}}$, we can synthesize a corresponding physical operator $\bar{g}$ by factoring $F_{\bar{g}}$ into elementary symplectic matrices of the types listed in Table~\ref{tab:std_symp}.
Three algorithms for this purpose are given in~\cite{Dehaene-physreva03},\cite{Maslov-arxiv17} and~\cite{Can-2017a}.
We restate the relevant result given by Can in~\cite{Can-2017a} and sketch the idea of the proof.
The decompositions in~\cite{Dehaene-physreva03} and~\cite{Maslov-arxiv17} are similar.

\begin{theorem}
\label{thm:Trung}
Any binary symplectic transformation $F$ can be expressed as
\begin{align*}
F = A_{Q_1} \Omega\, T_{R_1} G_k T_{R_2} A_{Q_2} ,
\end{align*}
as per the notation used in Table~\ref{tab:std_symp}, where invertible matrices $Q_1, Q_2$ and symmetric matrices $R_1,R_2$ are chosen appropriately.
\begin{IEEEproof}
The idea is to perform row and column operations on the matrix $F$ via left and right multiplication by elementary symplectic transformations from Table~\ref{tab:std_symp}, and bring the matrix $F$ to the standard form $\Omega\, T_{R_1} \Omega$ (details in Appendix~\ref{sec:trung_proof}).
%Then we can multiply by appropriate inverse matrices on both sides to get the form given in the statement of the theorem.
\end{IEEEproof}
\end{theorem}

\section{Synthesis of Logical Clifford Operators for the $\llbr 6,4,2 \rrbr$ CSS Code}
\label{sec:css642}

As stated at the end of Section~\ref{sec:clifford_gp}, the logical Clifford group $\text{Cliff}_{2^4}$ is generated by $g^L \in \{ HW_{2^4}, P^L, H^L, \text{CZ}^L, \text{CNOT}^L \}$.
We will first discuss the construction of the stabilizer $S$ and the physical implementations $\bar{h} \in HW_{2^6}$ of the logical Pauli operators $h^L \in HW_{2^4}$ for this code. 
These are synthesized using the generator matrices of the classical codes from which the $\llbr 6,4,2 \rrbr$ code is constructed.
Then we will demonstrate our algorithm by determining the symplectic matrices corresponding to the physical equivalents  $\bar{g} \in \{ \lP_1, \lH_1, \lcz{1}{2}, \lcnot{2}{1} \}$ of the above generating set, where the subscripts indicate the logical qubit(s) involved in the logical operations realized by these physical operators.
The corresponding operators on other (combinations of) logical qubits can be synthesized via a similar procedure.

\subsection{Stabilizer of the Code}
\label{sec:css642_stabilizer}

The $[6,5,2]$ classical binary single parity-check code $\MCC$ is generated by
\begin{equation}
\label{eq:Cgen}
G_{\MCC} =
\begin{bmatrix}
H_{\MCC} \\
G_{\MCC/\MCCd}
\end{bmatrix}
\ ; \
G_{\MCC/\MCCd} \triangleq
\begin{bmatrix}
1 & 1 & 0 & 0 & 0 & 0 \\
1 & 0 & 1 & 0 & 0 & 0 \\
1 & 0 & 0 & 1 & 0 & 0 \\
1 & 0 & 0 & 0 & 1 & 0
\end{bmatrix} 
 ,
\end{equation}
where the parity-check matrix for $\MCC$ is $H_{\MCC} = [1\ 1\ 1\ 1\ 1\ 1]$.
Hence, the dual (repetition) code $\MCCd = \{ 000000, 111111 \}$ of $\MCC$ is generated by the matrix $H_{\MCC}$.
The rows $\vecnot{h}_i$ of $G_{\MCC/\MCCd}$, for $i=1,2,3,4$, generate all coset representatives for $\MCCd$ in $\MCC$, which determine the physical states of the code.
The CSS construction~\cite{Calderbank-physreva96,Steane-physreva96,Nielsen-2010} provides a $\llbr m,m-k \rrbr = \llbr 6,4 \rrbr$ stabilizer code $\MCQ$ spanned by the set of basis vectors $\{ \ket{\psi_x} \mid \vecnot{x} \in \mathbb{F}_2^4 \}$, where $\vecnot{x} \triangleq [x_1,\ x_2,\ x_3,\ x_4]$ and
\begin{equation}
\label{eq:css_state}
%\left\{ \ket{\psi_x} \triangleq
%\frac{1}{\sqrt{2}} \ket{(000000) + \sum_{j=1}^{4} x_j \vecnot{h}_j} + \frac{1}{\sqrt{2}} \ket{(111111) + \sum_{j=1}^{4} x_j \vecnot{h}_j} 
%,\  \vecnot{x} \in \mathbb{F}_2^4 \right\} . 
\ket{\psi_x} \triangleq
\frac{1}{\sqrt{2}} \ket{(000000) + \sum_{j=1}^{4} x_j \vecnot{h}_j} + \frac{1}{\sqrt{2}} \ket{(111111) + \sum_{j=1}^{4} x_j \vecnot{h}_j} . 
\end{equation}
%where $x_j \in \mathbb{F}_2,\ j=1,2,3,4$.
Let $X_t$ and $Z_t$ denote the $X$ and $Z$ operators, respectively, acting on the $t$-th physical qubit.
Then, the physical operators defined by the row of $H_{\MCC}$ are
\begin{align} 
\label{eq:642stabilizer}
\bg^X = D(111111,000000) = X_1 X_2 X_3 X_4 X_5 X_6 \ {\rm and}\  \bg^Z = D(000000,111111) = Z_1 Z_2 Z_3 Z_4 Z_5 Z_6 .
\end{align}
These generate the stabilizer group $S$ that determines $\MCQ$.
The notation $X_1 X_2 \cdots X_6$ is commonly used in the literature to represent $X \otimes X \otimes \cdots \otimes X$.
If subscript $i \in \{1,\ldots,m\}$ is omitted, then it implies that the operator $I_2$ acts on the $i$-th qubit.

\subsection{Logical Operators for Protected Qubits}
\label{sec:logical_ops}

For the generating set $g^L \in \{ X^L, Z^L, P^L, H^L, \text{CZ}^L, \text{CNOT}^L \}$ of logical Clifford operators $\text{Cliff}_{2^4}$, we now synthesize their corresponding physical operators $\bar{g}$ that realize the action of $g^L$ on the protected qubits.
Since the operator $\bar{g}$ must also preserve $\MCQ$, conjugation by $\bar{g}$ must preserve both the stabilizer $S$ and hence its normalizer $S^{\perp}$ in $HW_N$~\cite{Calderbank-it98*2}.
We note that $\bar{g}$ need not commute with every element of the stabilizer $S$, i.e., \emph{centralize} $S$, although this can be enforced if necessary (see Theorem~\ref{thm:normalize_centralize}).

\subsubsection{Logical Paulis}
\label{sec:css_pauli}

Let $\ket{\vecnot{x}}_L,\ \vecnot{x} = [x_1,x_2,x_3,x_4] \in \mathbb{F}_2^4,$ be the logical state protected by the physical state $\ket{\psi_x}$ defined in~\eqref{eq:css_state}. 
Then the generating set $\{ X_j^L, Z_j^L \in HW_{2^4} \mid j=1,2,3,4 \}$ for the logical Pauli operators are defined by the actions
\begin{align}
X_j^L \ket{\vecnot{x}}_L = \ket{\vecnot{x}'}_L ,\ & {\rm where}\ x_i' = \begin{cases} x_j \oplus 1 &,\  {\rm if}\ i=j \\ x_i &,\ {\rm if}\ i \neq j \end{cases} \ \ 
{\rm and}\ \  Z_j^L \ket{\vecnot{x}}_L = (-1)^{x_j} \ket{\vecnot{x}}_L .
\end{align}
As per notation in Section~\ref{sec:log_ops_qecc}, we denote their corresponding physical operators as $\lX_j$ and $\lZ_j$, respectively.
The rows of
\begin{align}
\label{eq:GCZ}
G_{\MCC/\MCCd}^X \triangleq G_{\MCC/\MCCd} =
\begin{bmatrix}
1 & 1 & 0 & 0 & 0 & 0 \\
1 & 0 & 1 & 0 & 0 & 0 \\
1 & 0 & 0 & 1 & 0 & 0 \\
1 & 0 & 0 & 0 & 1 & 0
\end{bmatrix} \ \text{and} \ \ 
G_{\MCC/\MCCd}^Z \triangleq
\begin{bmatrix}
0 & 1 & 0 & 0 & 0 & 1 \\
0 & 0 & 1 & 0 & 0 & 1 \\
0 & 0 & 0 & 1 & 0 & 1 \\
0 & 0 & 0 & 0 & 1 & 1
\end{bmatrix} 
\end{align}
are used to define these physical implementations $\lX_j, \lZ_j, j=1,2,3,4$ as follows.
\begin{equation}
\label{eq:css_pauli}
%\begin{array}{lc|cl}
%\lX_1 = X_1 X_2 & \ & \ & \lZ_1 = Z_2 Z_6 \\
%\lX_2 = X_1 X_3 & \ & \ & \lZ_2 = Z_3 Z_6 \\
%\lX_3 = X_1 X_4 & \ & \ & \lZ_3 = Z_4 Z_6 \\
%\lX_4 = X_1 X_5 & \ & \ & \lZ_4 = Z_5 Z_6  
%\end{array} . 
\begin{array}{cc|cc}
\lX_1 \triangleq D(110000,000000) = X_1 X_2 & \quad & \quad & \lZ_1 \triangleq D(000000,010001) = Z_2 Z_6 \\
\lX_2 \triangleq D(101000,000000) = X_1 X_3 & \quad & \quad & \lZ_2 \triangleq D(000000,001001) = Z_3 Z_6 \\
\lX_3 \triangleq D(100100,000000) = X_1 X_4 & \quad & \quad & \lZ_3 \triangleq D(000000,000101) = Z_4 Z_6 \\
\lX_4 \triangleq D(100010,000000) = X_1 X_5 & \quad & \quad & \lZ_4 \triangleq D(000000,000011) = Z_5 Z_6 
\end{array} . 
\end{equation}
Although these are the physical realizations of logical Pauli operators, it is standard practice in the literature to refer to $\lX_j, \lZ_j$ itself as the logical Pauli operators.
These operators commute with every element of the stabilizer $S$ and satisfy, as required,
\begin{equation} 
\lX_i \lZ_j = 
\begin{cases}
- \lZ_j \lX_i & {\rm if} \ i=j, \\
\lZ_j \lX_i & {\rm if} \ i \neq j 
\end{cases} . 
\end{equation}
Note that this is a translation of the commutation relations between $X_j^L$ and $Z_j^L$ as discussed in Section~\ref{sec:log_ops_qecc}.
In general, to define valid logical Pauli operators, it can be observed that the matrices $G_{\MCC/\MCCd}^X, G_{\MCC/\MCCd}^Z$ must satisfy 
\begin{equation}
G_{\MCC/\MCCd}^X \left( G_{\MCC/\MCCd}^Z \right)^T = I_{m-k} ,\ \ \text{and}\ \ 
G_{\MCC}^Z =
\begin{bmatrix}
H_{\MCC} \\
G_{\MCC/\MCCd}^Z
\end{bmatrix}
\end{equation} 
must form another generator matrix for the (classical) code $\MCC$
(see Lemma~\ref{lem:logical_Paulis_commute} below where we show that such a matrix $G_{\MCC/\MCCd}^Z$ always exists).
It can be verified that the above matrices satisfy these conditions and hence the set of operators in~\eqref{eq:css_pauli} indeed form a generating set for all logical Pauli operators.
Note that $S^{\perp}$ is generated by $S, \lX_i, \lZ_i$, i.e., $S^{\perp} = \langle S, \lX_i, \lZ_i \rangle$ (see~\cite{Gottesman-arxiv09}).
This completes the translation of operators in $HW_{2^4}$ to their physical realizations.
%Since these physical implementations of the Paulis, given in~\eqref{eq:css_pauli} above, do not couple two qubits in a single gate they are automatically fault-tolerant.

Now we discuss the synthesis of physical operators $\bar{g} \in \{ \lP_1, \lH_1, \lcz{1}{2}, \lcnot{2}{1} \}$ corresponding to the (remaining elements of the) generating set $\{ HW_{2^4}, P^L, H^L, \text{CZ}^L, \text{CNOT}^L \}$ for the logical Clifford operators $\text{Cliff}_{2^4}$.

\subsubsection{Logical Phase Gate}
\label{sec:css_phase}

The phase gate $\bar{g} = \lP_1$ on the first logical qubit is defined by the actions (see~\cite{Nielsen-2010})
\begin{equation}
\label{eq:phase}
\lP_1 \lX_j \lP_1^{\dagger} = 
\begin{cases}
\lY_j & {\rm if}\ j=1, \\
\lX_j & {\rm if}\ j\neq 1,
\end{cases} , \ \ 
\lP_1 \lZ_j \lP_1^{\dagger} = \lZ_j \ \forall \ j=1,2,3,4 .
\end{equation}
This is again a translation of the relations $P_1^L X_j^L (P_1^L)^{\dagger}$ to the physical space, as discussed in Section~\ref{sec:log_ops_qecc}.
One can express $\lP_1$ in terms of the physical Paulis $X_t, Z_t$ as follows.
The condition $\lP_1 \lX_1 \lP_1^{\dagger} = \lY_1$ implies $\lP_1$ must transform $\lX_1 = X_1 X_2$ into $\lY_1 \triangleq \iota \lX_1 \lZ_1 = \iota X_1 X_2 Z_2 Z_6 = X_1 (\iota X_2 Z_2) Z_6 = X_1 Y_2 Z_6$.
Similarly, the other conditions imply that all other $\lX_j$s and all $\lZ_j$s must remain unchanged.
Hence we can explicitly write the mappings as below.
\begin{equation}
\label{eq:phase_maps}
\hspace*{-0.25cm}
\begin{array}{lc|cl}
\lX_1 = X_1 X_2 \overset{\lP_1}{\longmapsto} \lX_1' = X_1 Y_2 Z_6 & & & \lZ_1 = Z_2 Z_6 \overset{\lP_1}{\longmapsto} \lZ_1' = Z_2 Z_6 \\
\lX_2 = X_1 X_3 \overset{\lP_1}{\longmapsto} \lX_2' = X_1 X_3 & & & \lZ_2 = Z_3 Z_6 \overset{\lP_1}{\longmapsto} \lZ_2' = Z_3 Z_6 \\
\lX_3 = X_1 X_4 \overset{\lP_1}{\longmapsto} \lX_3' = X_1 X_4 & & & \lZ_3 = Z_4 Z_6 \overset{\lP_1}{\longmapsto} \lZ_3' = Z_4 Z_6 \\
\lX_4 = X_1 X_5 \overset{\lP_1}{\longmapsto} \lX_4' = X_1 X_5 & & & \lZ_4 = Z_5 Z_6 \overset{\lP_1}{\longmapsto} \lZ_4' = Z_5 Z_6 
\end{array} . 
\end{equation}

Direct inspection of these conditions yields the circuit given below.
First we find an operator which transforms $X_2$ to $Y_2$ and leaves other Paulis unchanged; this is $P_2$, the phase gate on the second physical qubit.
Then we find an operator that transforms $Y_2$ into $Y_2 Z_6$, which is ${\rm CZ}_{26}$ as $X_2 {\rm CZ}_{26} X_2^{\dagger} = X_2 Z_6$ and $Z_i {\rm CZ}_{26} Z_i^{\dagger} = Z_i, i=1,2,\ldots,6$.
Here ${\rm CZ}_{26}$ is the controlled-$Z$ gate on physical qubits $2$ and $6$.
But this also transforms $X_6$ into $Z_2 X_6$ and hence the circuit ${\rm CZ}_{26} P_2$ does not fix the stabilizer $\bg^X$.
Therefore we include $P_6$ so that the full circuit $\lP_1 = P_6 {\rm CZ}_{26} P_2$ fixes $\bg^X$, fixes $\bg^Z$, and realizes $P_1^L$.
%The corresponding physical operator commutes with every element of the stabilizer $S$.
%
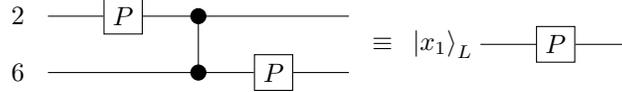
\begin{center}
\begin{tikzpicture}

\node[draw,rectangle] (S11) at (1,0.75) {$P$};
\node[draw,rectangle] (S30) at (3,0) {$P$};

\node at (-0.4,0) {$6$};
\node at (-0.4,0.75) {$2$};
\path[draw] (0,0) -- (S30) -- (4,0);
\path[draw] (0,0.75) -- (S11) -- (4,0.75);

\draw[fill=black] (2,0.75) circle (0.1);
\path[draw] (2,0.75) -- (2,0);
\draw[fill=black] (2,0) circle (0.1);

\node[align=center] at (4.45,0.375) {$\equiv$};

\node[draw,rectangle] (S10) at (6.75,0.375) {$P$};

\node at (5.25,0.375) {$\ket{x_1}_L$};
\path[draw] (5.75,0.375) -- (S10) -- (7.75,0.375);

\end{tikzpicture}
\end{center}
See Appendix~\ref{sec:circuit_ids} for the circuit identities used above. 
We now describe how this same circuit can be synthesized via symplectic geometry.
Let $F = \begin{bmatrix} A & B \\ C & D \end{bmatrix}$ be the symplectic matrix corresponding to $\lP_1$.
Using~\eqref{eq:symp_action}, the conditions imposed in~\eqref{eq:phase} on logical Pauli operators $\lX_j, j=1,2,3,4$ give
\begin{align*}
[110000,000000] F = [110000,010001] \ ({\rm i.e.,}\ X_1 X_2 \mapsto X_1 Y_2 Z_6) &
\Rightarrow [110000] A = [110000], [110000] B = [010001] , \\
[101000,000000] F = [101000,000000] \ ({\rm i.e.,}\ X_1 X_3 \mapsto X_1 X_3) &
 \Rightarrow [101000] A = [101000], [101000] B = [000000] , \\
[100100,000000] F = [100100,000000] \ ({\rm i.e.,}\ X_1 X_4 \mapsto X_1 X_4) &
 \Rightarrow [100100] A = [100100], [100100] B = [000000] , \\
[100010,000000] F = [100010,000000] \ ({\rm i.e.,}\ X_1 X_5 \mapsto X_1 X_5) &
 \Rightarrow [100010] A = [100010], [100010] B = [000000] .
\end{align*}
Let $\vecnot{e}_i \in \mathbb{F}_2^6$ be the standard basis vector with entry $1$ in the $i$-th location and zeros elsewhere, for $i=1,\ldots,6$.
Then the above conditions can be rewritten compactly as
\begin{align*}
(\vecnot{e}_1 + \vecnot{e}_2) A & = \vecnot{e}_1 + \vecnot{e}_2 ,\ (\vecnot{e}_1 + \vecnot{e}_2) B = \vecnot{e}_2 + \vecnot{e}_6 , \ \  
{\rm and}\ \ (\vecnot{e}_1 + \vecnot{e}_i) A = \vecnot{e}_1 + \vecnot{e}_i,\ (\vecnot{e}_1 + \vecnot{e}_i) B = \vecnot{0},\ i=3,4,5.
\end{align*}
Similarly, the conditions imposed on $\lZ_j, j=1,2,3,4$ give
\begin{align*}
[000000,010001] F = [000000,010001] &
 \Rightarrow [010001] C = [000000], [010001] D = [010001] , \\
[000000,001001] F = [000000,001001] &
 \Rightarrow [001001] C = [000000], [001001] D = [001001] , \\
[000000,000101] F = [000000,000101] &
 \Rightarrow [000101] C = [000000], [000101] D = [000101] , \\
[000000,000011] F = [000000,000011] &
 \Rightarrow [000011] C = [000000], [000011] D = [000011] . 
\end{align*}
Again these can be rewritten compactly as
\begin{align*}
(\vecnot{e}_i + \vecnot{e}_6) C = 0,\ (\vecnot{e}_i + \vecnot{e}_6) D = \vecnot{e}_i + \vecnot{e}_6,\ i=2,3,4,5.
\end{align*}
Although it is sufficient for $\lP_1$ to just normalize $S$, we can always require that the physical operator commute with every element of the stabilizer $S$ (see Theorem~\ref{thm:normalize_centralize}).
This gives the centralizing conditions
\begin{align*}
[111111,000000] F = [111111,000000] &
 \Rightarrow [111111] A = [111111], [111111] B = [000000] , \\
[000000,111111] F = [000000,111111] &
 \Rightarrow [111111] C = [000000], [111111] D = [111111] .
\end{align*}
Again these can be rewritten compactly as
\begin{align*}
(\vecnot{e}_1 + \ldots + \vecnot{e}_6) A & = \vecnot{e}_1 + \ldots + \vecnot{e}_6 = (\vecnot{e}_1 + \ldots + \vecnot{e}_6) D , \ \ 
(\vecnot{e}_1 + \ldots + \vecnot{e}_6) B = \vecnot{0} = (\vecnot{e}_1 + \ldots + \vecnot{e}_6) C .
\end{align*}
Note that, in addition to these linear constraints, $F$ also needs to satisfy the symplectic constraint $F \Omega F^T = \Omega$.
We obtain one solution using Algorithm~\ref{alg:transvec} as $F = T_{B}$ (see Table~\ref{tab:std_symp}), where
\begin{equation}
\label{eq:css_phase_B}
B \triangleq B_{P} = 
\begin{bmatrix}
0 & 0 & 0 & 0 & 0 & 0 \\
0 & 1 & 0 & 0 & 0 & 1 \\
0 & 0 & 0 & 0 & 0 & 0 \\
0 & 0 & 0 & 0 & 0 & 0 \\
0 & 0 & 0 & 0 & 0 & 0 \\
0 & 1 & 0 & 0 & 0 & 1 
\end{bmatrix} \Rightarrow F = \begin{bmatrix} I_6 & B_P \\ 0 & I_6 \end{bmatrix}  .
\end{equation} 
The resulting physical operator $\lP_1 = {\rm diag}\left( \iota^{vB_{P}v^T} \right)$ satisfies $\lP_1 = P_6 {\rm CZ}_{26} P_2$ and hence coincides with the above circuit (see the discussion in Appendix~\ref{sec:elem_symp} for this circuit decomposition).
Note that there can be multiple symplectic solutions to the set of linear constraints derived from~\eqref{eq:phase_maps} and each symplectic solution could correspond to multiple circuits depending on its decomposition into elementary symplectic forms from Table~\ref{tab:std_symp}.
The set of all symplectic solutions for $\lP_1$ were obtained using the result of Theorem~\ref{thm:symp_lineq_all} in Section~\ref{sec:discuss} below, and these are listed in Appendix~\ref{sec:css_phase_all}.
The above solution is the cheapest in this set in terms of the depth of the circuit (see Def.~\ref{def:depth}), and for all logical operators discussed below we also report their cheapest solutions.

\vspace{0.1cm}
Henceforth, for any logical operator in $\text{Cliff}_{2^4}$, we refer to its physical implementation $\bar{g}$ itself as the logical operator, since this is common terminology in the literature.
\vspace{0.1cm}

\subsubsection{Logical Controlled-Z (CZ)}
\label{sec:css_cz}

The logical operator $\bar{g} = \lcz{1}{2}$ is defined by its action on the logical Paulis as 
\begin{align}
\label{eq:cz}
\lcz{1}{2} \lX_j \lcz{1}{2}^{\dagger} & = 
\begin{cases}
\lX_1 \lZ_2 & {\rm if}\ j=1, \\
\lZ_1 \lX_2 & {\rm if}\ j=2,\\
\lX_j & {\rm if}\ j\neq 1,2
\end{cases} , \nonumber \\
\lcz{1}{2} \lZ_j \lcz{1}{2}^{\dagger} & = \lZ_j \ \forall \ j=1,2,3,4 .
\end{align}
We first express the logical operator $\lcz{1}{2}$, on the first two logical qubits, in terms of the physical Pauli operators $X_t, Z_t$.
\begin{equation}
\label{eq:cz_maps}
\begin{array}{lc|cl}
\lX_1 = X_1 X_2 \overset{\lcz{1}{2}}{\longmapsto} X_1 X_2 Z_3 Z_6 & & & \lZ_1 = Z_2 Z_6 \overset{\lcz{1}{2}}{\longmapsto} Z_2 Z_6 \\
\lX_2 = X_1 X_3 \overset{\lcz{1}{2}}{\longmapsto} X_1 X_3 Z_2 Z_6 & & &  \lZ_2 = Z_3 Z_6 \overset{\lcz{1}{2}}{\longmapsto} Z_3 Z_6 \\
\lX_3 = X_1 X_4 \overset{\lcz{1}{2}}{\longmapsto} X_1 X_4 & & &  \lZ_3 = Z_4 Z_6 \overset{\lcz{1}{2}}{\longmapsto} Z_4 Z_6 \\
\lX_4 = X_1 X_5 \overset{\lcz{1}{2}}{\longmapsto} X_1 X_5 & & &  \lZ_4 = Z_5 Z_6 \overset{\lcz{1}{2}}{\longmapsto} Z_5 Z_6 
\end{array} . 
\end{equation}
%$\lX_3, \lX_4, \lZ_3, \lZ_4$ are left unchanged by $\lcz{1}{2}$.
As with the phase gate, we translate these conditions into linear equations involving the constituents of the corresponding symplectic transformation $F$.
The conditions imposed by the $\lX_j$s are
\begin{align*}
(\vecnot{e}_1 + \vecnot{e}_i) A = \vecnot{e}_1 + \vecnot{e}_i, \ i=2,3,4,5, \ (\vecnot{e}_1 + \vecnot{e}_2) B = \vecnot{e}_3 + \vecnot{e}_6, \ (\vecnot{e}_1 + \vecnot{e}_3) B = \vecnot{e}_2 + \vecnot{e}_6, \ (\vecnot{e}_1 + \vecnot{e}_i) B = \vecnot{0}, \ i=4,5 .
\end{align*}
The conditions imposed by the $\lZ_j$s are
\begin{align*}
(\vecnot{e}_i + \vecnot{e}_6) C = \vecnot{0}, \ (\vecnot{e}_i + \vecnot{e}_6) D = \vecnot{e}_i + \vecnot{e}_6, \ i=2,3,4,5 .
\end{align*}
Although it is sufficient for $\lcz{1}{2}$ to just normalize $S$, we can always require that the physical operator commute with every element of the stabilizer $S$ (see Theorem~\ref{thm:normalize_centralize}).
This gives the centralizing conditions
\begin{align*}
(\vecnot{e}_1 + \ldots + \vecnot{e}_6) A & = \vecnot{e}_1 + \ldots + \vecnot{e}_6 = (\vecnot{e}_1 + \ldots + \vecnot{e}_6) D , \ \ 
(\vecnot{e}_1 + \ldots + \vecnot{e}_6) B = \vecnot{0} = (\vecnot{e}_1 + \ldots + \vecnot{e}_6) C .
\end{align*}
We again obtain one solution using Algorithm~\ref{alg:transvec} as $F = T_{B}$, where
\begin{equation}
B \triangleq B_{\rm CZ} = 
\begin{bmatrix}
0 & 0 & 0 & 0 & 0 & 0 \\
0 & 0 & 1 & 0 & 0 & 1 \\
0 & 1 & 0 & 0 & 0 & 1 \\
0 & 0 & 0 & 0 & 0 & 0 \\
0 & 0 & 0 & 0 & 0 & 0 \\
0 & 1 & 1 & 0 & 0 & 0 
\end{bmatrix} .
\end{equation} 
We find that the physical operator $\lcz{1}{2} = {\rm diag}\left( \iota^{vB_{\rm CZ}v^T} \right)$ commutes with the stabilizer $\bg^Z$ but not with $\bg^X$; it takes $X^{\otimes 6}$ to $-X^{\otimes 6}$.
This is remedied through post multiplication by $Z_6$ to obtain $\lcz{1}{2} = {\rm diag}\left( \iota^{vB_{\rm CZ}v^T} \right) Z_6$, which does not modify the symplectic matrix $F$ as $Z_6 \in HW_N$ and $HW_N$ is the kernel of the map $\phi$ defined in~\eqref{eq:phi}.
The resulting physical operator $\lcz{1}{2}$ corresponds to the same circuit obtained by Chao and Reichardt in~\cite{Chao-arxiv17b}, i.e., $\lcz{1}{2} = \cz{3}{6} \cz{2}{6} \cz{2}{3} Z_6$:
\begin{center}
\begin{tikzpicture}

\node[draw,rectangle] (Z10) at (1,0) {$Z$};
\node at (-0.4,0) {$6$};

\path[draw] (0,0) -- (Z10) -- (4,0);
\path[draw] (0,0.5) -- (4,0.5) node[at start,pos=-0.1] {$3$};
\path[draw] (0,1) -- (4,1) node[at start,pos=-0.1] {$2$};

\draw[fill=black] (1,1) circle (0.1);
\draw[fill=black] (2,1) circle (0.1);
\draw[fill=black] (3,0.5) circle (0.1);
\path[draw] (1,1) -- (1,0.5);
\path[draw] (2,1) -- (2,0);
\path[draw] (3,0.5) -- (3,0);
\draw[fill=black] (1,0.5) circle (0.1);
\draw[fill=black] (2,0) circle (0.1);
\draw[fill=black] (3,0) circle (0.1);

\node[align=center] at (4.625,0.5) {$\equiv$};

\path[draw] (6,0.1) -- (8,0.1) node[at start,pos=-0.2] {$\ket{x_2}_L$};
\path[draw] (6,0.9) -- (8,0.9) node[at start,pos=-0.2] {$\ket{x_1}_L$};

\draw[fill=black] (7,0.9) circle (0.1);
\path[draw] (7,0.9) -- (7,0.1);
\draw[fill=black] (7,0.1) circle (0.1);

\end{tikzpicture}
\end{center}
The set of all symplectic solutions for $\lcz{1}{2}$ were obtained using the result of Theorem~\ref{thm:symp_lineq_all} in Section~\ref{sec:discuss} below, and these are listed in Appendix~\ref{sec:css_cz_all}.
As for $\lP_1$, the above solution is the cheapest in this set in terms of the depth of the circuit.
\vspace{0.1cm}

\subsubsection{Logical Controlled-NOT (CNOT)}
\label{sec:css_cnot}

The logical operator $\bar{g} = \lcnot{2}{1}$, where logical qubit $2$ controls $1$, is defined by
\begin{align}
\label{eq:lcnot21}
\lcnot{2}{1} \lX_j \lcnot{2}{1}^{\dagger} & = 
\begin{cases}
\lX_1 \lX_2 & {\rm if}\ j=2,\\
\lX_j & {\rm if}\ j\neq 2
\end{cases} , \nonumber \\ 
\lcnot{2}{1} \lZ_j \lcnot{2}{1}^{\dagger} & = 
\begin{cases}
\lZ_1 \lZ_2 & {\rm if}\ j=1,\\
\lZ_j & {\rm if}\ j\neq 1 .
\end{cases}\end{align}
We approach synthesis via symplectic geometry, and express the operator $\lcnot{2}{1}$ in terms of the physical operators $X_t, Z_t$ as shown below.
\begin{equation}
\label{eq:lcnot21_maps}
\begin{array}{lc|cl}
\lX_1 = X_1 X_2 \overset{2 \rightarrow 1}{\longmapsto} X_1 X_2 & \ & \ & \lZ_1 = Z_2 Z_6 \overset{2 \rightarrow 1}{\longmapsto} Z_2 Z_3 \\
\lX_2 = X_1 X_3 \overset{2 \rightarrow 1}{\longmapsto} X_2 X_3 & \ & \ &  \lZ_2 = Z_3 Z_6 \overset{2 \rightarrow 1}{\longmapsto} Z_3 Z_6 \\
\lX_3 = X_1 X_4 \overset{2 \rightarrow 1}{\longmapsto} X_1 X_4 & \ & \ &  \lZ_3 = Z_4 Z_6 \overset{2 \rightarrow 1}{\longmapsto} Z_4 Z_6 \\
\lX_4 = X_1 X_5 \overset{2 \rightarrow 1}{\longmapsto} X_1 X_5 & \ & \ &  \lZ_4 = Z_5 Z_6 \overset{2 \rightarrow 1}{\longmapsto} Z_5 Z_6 
\end{array} . 
\end{equation}
Note that only $\lX_2$ and $\lZ_1$ are modified by $\lcnot{2}{1}$.
%$\lX_3, \lX_4, \lZ_3, \lZ_4$ are again left unchanged by $\lcnot{2}{1}$.
As before, we translate these conditions into linear equations involving the constituents of the corresponding symplectic transformation $F$.
The conditions imposed by $\lX_j$s are
\begin{align*}
(\vecnot{e}_1 + \vecnot{e}_3) A = \vecnot{e}_2 + \vecnot{e}_3, \ (\vecnot{e}_1 + \vecnot{e}_i) A = \vecnot{e}_1 + \vecnot{e}_i, \ i=2,4,5, \ (\vecnot{e}_1 + \vecnot{e}_i) B = \vecnot{0}, \ i=2,3,4,5 .
\end{align*}
The conditions imposed by $\lZ_j$s are
\begin{align*}
(\vecnot{e}_i + \vecnot{e}_6) C = \vecnot{0}, \ i=2,3,4,5, \ (\vecnot{e}_2 + \vecnot{e}_6) D = \vecnot{e}_2 + \vecnot{e}_3, \ (\vecnot{e}_i + \vecnot{e}_6) D = \vecnot{e}_i + \vecnot{e}_6, \ i=3,4,5 .
\end{align*}
Although it is sufficient for $\lcnot{2}{1}$ to just normalize $S$, we can always require that the physical operator commute with every element of the stabilizer $S$ (see Theorem~\ref{thm:normalize_centralize}).
This gives the centralizing conditions
\begin{align*}
(\vecnot{e}_1 + \ldots + \vecnot{e}_6) A & = \vecnot{e}_1 + \ldots + \vecnot{e}_6 = (\vecnot{e}_1 + \ldots + \vecnot{e}_6) D , \ \ 
(\vecnot{e}_1 + \ldots + \vecnot{e}_6) B = \vecnot{0} = (\vecnot{e}_1 + \ldots + \vecnot{e}_6) C .
\end{align*}
We again obtain one solution using Algorithm~\ref{alg:transvec} as $F = \begin{bmatrix} A & 0 \\ 0 & A^{-T} \end{bmatrix}$, where
\begin{equation}
A = 
\begin{bmatrix}
1 & 0 & 0 & 0 & 0 & 0 \\
0 & 1 & 0 & 0 & 0 & 0 \\
1 & 1 & 1 & 0 & 0 & 0 \\
0 & 0 & 0 & 1 & 0 & 0 \\
0 & 0 & 0 & 0 & 1 & 0 \\
1 & 1 & 0 & 0 & 0 & 1 
\end{bmatrix} ,\ A^{-T} = 
\begin{bmatrix}
1 & 0 & 1 & 0 & 0 & 1 \\
0 & 1 & 1 & 0 & 0 & 1 \\
0 & 0 & 1 & 0 & 0 & 0 \\
0 & 0 & 0 & 1 & 0 & 0 \\
0 & 0 & 0 & 0 & 1 & 0 \\
0 & 0 & 0 & 0 & 0 & 1 
\end{bmatrix} . 
\end{equation} 
The action of $\lcnot{2}{1}$ on logical qubits is related to the action on physical qubits through the generator matrix $G_{\MCC/\MCCd}$.
The map $\vecnot{v} \mapsto \vecnot{v} A$ fixes the code $\MCC$ (i.e., $e_v = \ket{\vecnot{v}} \mapsto e_{vA} = \ket{\vecnot{v} A}$ fixes $\MCQ$ and hence its stabilizers $\bg^X$ and $\bg^Z$) and induces a linear transformation on the coset space $\MCC/\MCCd$ (which defines the CSS state).
The action $K$ on logical qubits is related to the action $A$ on physical qubits by ${K \cdot G_{C/C^{\perp}}^X = G_{C/C^{\perp}}^X \cdot A}$ and we obtain
\begin{align}
K = 
\begin{bmatrix}
1 & 0 & 0 & 0 \\
1 & 1 & 0 & 0 \\
0 & 0 & 1 & 0 \\
0 & 0 & 0 & 1
\end{bmatrix}
\end{align}
as desired.
The circuit on the left below implements the operator $e_v \mapsto e_{vA}$, i.e., $\lcnot{2}{1}$, where $e_v$ is a standard basis vector in $\mathbb{C}^{N}$ as defined in Table~\ref{tab:std_symp}.
The circuit on the right implements $e_x \mapsto e_{xK}$, i.e., $\cnot{2}{1}^L$, where $x \in \mathbb{F}_2^4$.
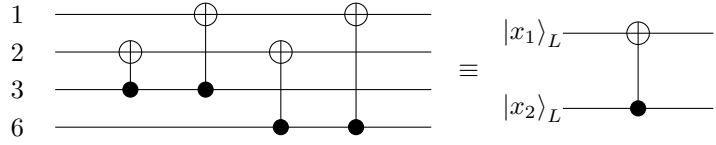
\begin{center}
\begin{tikzpicture}

\node at (-0.5,1.5) {$1$};
\node at (-0.5,1) {$2$};
\node at (-0.5,0.5) {$3$};
\node at (-0.5,0) {$6$};

\path[draw] (0,0) -- (5,0);
\path[draw] (0,0.5) -- (5,0.5);
\path[draw] (0,1) -- (5,1);
\path[draw] (0,1.5) -- (5,1.5);

\draw (1,1) circle (0.15);
\draw (2,1.5) circle (0.15);
\draw (3,1) circle (0.15);
\draw (4,1.5) circle (0.15);
\path[draw] (1,0.5) -- (1,1.15);
\path[draw] (2,0.5) -- (2,1.65);
\path[draw] (3,0) -- (3,1.15);
\path[draw] (4,0) -- (4,1.65);
\draw[fill=black] (1,0.5) circle (0.1);
\draw[fill=black] (2,0.5) circle (0.1);
\draw[fill=black] (3,0) circle (0.1);
\draw[fill=black] (4,0) circle (0.1);

\node[align=center] at (5.5,0.75) {$\equiv$};

\path[draw] (6.75,1.25) -- (8.75,1.25) node[at start,pos=-0.2] {$\ket{x_1}_L$};
\path[draw] (6.75,0.25) -- (8.75,0.25) node[at start,pos=-0.2] {$\ket{x_2}_L$};

\draw[fill=black] (7.75,0.25) circle (0.1);
\draw (7.75,1.25) circle (0.15);
\path[draw] (7.75,1.4) -- (7.75,0.25);

\end{tikzpicture}
\end{center}
We note that~\cite{Grassl-isit13} discusses codes and operators where $A$ is a permutation matrix corresponding to an automorphism of $\MCC$.
The set of all symplectic solutions for $\lcnot{2}{1}$ were obtained using the result of Theorem~\ref{thm:symp_lineq_all} in Section~\ref{sec:discuss} below, and these are listed in Appendix~\ref{sec:css_cnot_all}.
As for $\lP_1$ and $\lcz{1}{2}$, the above solution is the cheapest in this set in terms of the circuit depth.

\noindent\emph{Remark}: 
To implement $\cnot{2}{1}^L$ we can also use the circuit identity (see Appendix~\ref{sec:circuit_ids} for a useful set of circuit identities)
\begin{center}
\begin{tikzpicture}

\path[draw] (0,0) -- (2,0) node[at start,pos=-0.25] {$\ket{x_2}_L$};
\path[draw] (0,0.75) -- (2,0.75) node[at start,pos=-0.25] {$\ket{x_1}_L$};

\draw[fill=black] (1,0) circle (0.1);
\draw (1,0.75) circle (0.15);
\path[draw] (1,0) -- (1,0.9);

\node[align=center] at (2.5,0.375) {$=$};

\node[draw,rectangle] (H50) at (5,0.75) {$H$};
\node[draw,rectangle] (H70) at (7,0.75) {$H$};

\node at (3.6,0.75) {$\ket{x_1}_L$};
\path[draw] (4,0.75) -- (H50) -- (H70) -- (8,0.75);
\path[draw] (4,0) -- (8,0) node[at start,pos=-0.1] {$\ket{x_2}_L$};

\draw[fill=black] (6,0.75) circle (0.1);
\path[draw] (6,0.75) -- (6,0);
\draw[fill=black] (6,0) circle (0.1);

\end{tikzpicture}
\end{center}
where $H_1^L$ is the targeted Hadamard operator (synthesized below).
However, this construction might require more gates.
\vspace{0.1cm}

\subsubsection{Logical Targeted Hadamard}
\label{sec:css_had}

The Hadamard gate $\bar{g} = \lH_1$ on the first logical qubit is defined by the actions
\begin{equation}
\label{eq:hadamard}
\lH_1 \lX_j \lH_1^{\dagger} = 
\begin{cases}
\lZ_j & {\rm if}\ j=1, \\
\lX_j & {\rm if}\ j\neq 1,
\end{cases} , \ \ 
\lH_1 \lZ_j \lH_1^{\dagger} = 
\begin{cases}
\lX_j & {\rm if}\ j=1, \\
\lZ_j & {\rm if}\ j\neq 1,
\end{cases} . 
\end{equation}
As for the other gates, we express the targeted Hadamard $\lH_1$ in terms of the physical Pauli operators $X_t, Z_t$.
\begin{equation}
\label{eq:hadamard_maps}
\begin{array}{lc|cl}
\lX_1 = X_1 X_2 \overset{\lH_1}{\longmapsto} Z_2 Z_6 & \ & \ &  \lZ_1 = Z_2 Z_6 \overset{\lH_1}{\longmapsto} X_1 X_2 \\
\lX_2 = X_1 X_3 \overset{\lH_1}{\longmapsto} X_1 X_3 & \ & \ &  \lZ_2 = Z_3 Z_6 \overset{\lH_1}{\longmapsto} Z_3 Z_6 \\
\lX_3 = X_1 X_4 \overset{\lH_1}{\longmapsto} X_1 X_4 & \ & \ &  \lZ_3 = Z_4 Z_6 \overset{\lH_1}{\longmapsto} Z_4 Z_6 \\
\lX_4 = X_1 X_5 \overset{\lH_1}{\longmapsto} X_1 X_5 & \ & \ &  \lZ_4 = Z_5 Z_6 \overset{\lH_1}{\longmapsto} Z_5 Z_6 
\end{array} . 
\end{equation}
As before, we translate these conditions into linear equations involving the constituents of the corresponding symplectic transformation $F$.
The conditions imposed by $\lX_j$s are
\begin{align*}
(\vecnot{e}_1 + \vecnot{e}_2) A = \vecnot{0}, \ (\vecnot{e}_1 + \vecnot{e}_i) A = \vecnot{e}_1 + \vecnot{e}_i, \ i=3,4,5, \ (\vecnot{e}_1 + \vecnot{e}_2) B = \vecnot{e}_2 + \vecnot{e}_6, \ (\vecnot{e}_1 + \vecnot{e}_i) B = \vecnot{0}, \ i=3,4,5 .
\end{align*}
The conditions imposed by $\lZ_j$s are
\begin{align*}
(\vecnot{e}_2 + \vecnot{e}_6) C = \vecnot{e}_1 + \vecnot{e}_2, \ (\vecnot{e}_i + \vecnot{e}_6) C = \vecnot{0}, \ i=3,4,5, \ (\vecnot{e}_2 + \vecnot{e}_6) D = \vecnot{0}, \ (\vecnot{e}_i + \vecnot{e}_6) D = \vecnot{e}_i + \vecnot{e}_6, \ i=3,4,5 .
\end{align*}
Although it is sufficient for $\lH_1$ to just normalize $S$, we can always require that the physical operator commute with every element of the stabilizer $S$ (see Theorem~\ref{thm:normalize_centralize}).
This gives the centralizing conditions
\begin{align*}
(\vecnot{e}_1 + \ldots + \vecnot{e}_6) A & = \vecnot{e}_1 + \ldots + \vecnot{e}_6 = (\vecnot{e}_1 + \ldots + \vecnot{e}_6) D , \ \ 
(\vecnot{e}_1 + \ldots + \vecnot{e}_6) B = \vecnot{0} = (\vecnot{e}_1 + \ldots + \vecnot{e}_6) C .
\end{align*}
We again obtain one solution using Algorithm~\ref{alg:transvec} as 
\begin{align}
\label{eq:css_H1}
A = 
\begin{bmatrix}
1 & 0 & 0 & 0 & 0 & 0 \\
1 & 0 & 0 & 0 & 0 & 0 \\
0 & 0 & 1 & 0 & 0 & 0 \\
0 & 0 & 0 & 1 & 0 & 0 \\
0 & 0 & 0 & 0 & 1 & 0 \\
1 & 1 & 0 & 0 & 0 & 1 
\end{bmatrix} , \ 
B = 
\begin{bmatrix}
0 & 0 & 0 & 0 & 0 & 0 \\
0 & 1 & 0 & 0 & 0 & 1 \\
0 & 0 & 0 & 0 & 0 & 0 \\
0 & 0 & 0 & 0 & 0 & 0 \\
0 & 0 & 0 & 0 & 0 & 0 \\
0 & 1 & 0 & 0 & 0 & 1  
\end{bmatrix} , \ 
C = 
\begin{bmatrix}
1 & 1 & 0 & 0 & 0 & 0 \\
1 & 1 & 0 & 0 & 0 & 0 \\
0 & 0 & 0 & 0 & 0 & 0 \\
0 & 0 & 0 & 0 & 0 & 0 \\
0 & 0 & 0 & 0 & 0 & 0 \\
0 & 0 & 0 & 0 & 0 & 0  
\end{bmatrix} , \ 
D = 
\begin{bmatrix}
1 & 1 & 0 & 0 & 0 & 1 \\
0 & 0 & 0 & 0 & 0 & 1 \\
0 & 0 & 1 & 0 & 0 & 0 \\
0 & 0 & 0 & 1 & 0 & 0 \\
0 & 0 & 0 & 0 & 1 & 0 \\
0 & 0 & 0 & 0 & 0 & 1 
\end{bmatrix} .  
\end{align}
%It can be verified that these also satisfy $AB^T = BA^T,\ CD^T = DC^T,\ AD^T - BC^T = I$.
The unitary operation corresponding to this solution commutes with each stabilizer element.
Another solution for $\lH_1$ which fixes $Z^{\otimes 6}$ but takes $X^{\otimes 6} \leftrightarrow (111111,000000)$ to $Y^{\otimes 6} \leftrightarrow (111111,111111)$ is given by just changing $B$ above to
\begin{align}
B =
\begin{bmatrix}
0 & 0 & 0 & 0 & 0 & 1 \\
0 & 1 & 0 & 0 & 0 & 0 \\
0 & 0 & 0 & 0 & 0 & 1 \\
0 & 0 & 0 & 0 & 0 & 1 \\
0 & 0 & 0 & 0 & 0 & 1 \\
1 & 0 & 1 & 1 & 1 & 1  
\end{bmatrix} .  
\end{align}
However, for both these solutions the resulting symplectic transformation does not correspond to any of the elementary forms in Table~\ref{tab:std_symp}.
Hence the unitary needs to be determined by expressing $F$ as a sequence of elementary transformations and then multiplying the corresponding unitaries.
An algorithm for this is given by Can in~\cite{Can-2017a} and restated in Theorem~\ref{thm:Trung} above.
For the solution~\eqref{eq:css_H1}, we verified that the symplectic matrix corresponds to the following circuit on the left given by Chao and Reichardt in~\cite{Chao-arxiv17b}.
On the right we produce the circuit obtained by using Theorem~\ref{thm:Trung} to decompose the same matrix~\eqref{eq:css_H1}.
%
%\begin{center}
%\begin{tikzpicture}
%
%\node[draw,rectangle] (H11) at (1.5,1) {$H$};
%\node[draw,rectangle] (H35) at (3.5,0.5) {$H$};
%\node[draw,rectangle] (H51) at (5.5,1) {$H$};
%\node[draw,rectangle] (X71) at (7.5,1) {$X$};
%\node[draw,rectangle] (Z70) at (7.5,0) {$Z$};
%
%\node at (-0.25,1) {$1$};
%\node at (-0.25,0.5) {$2$};
%\node at (-0.25,0) {$6$};
%
%\path[draw] (0,0) -- (Z70) -- (8.5,0);
%\path[draw] (0,0.5) -- (H35) -- (8.5,0.5);
%\path[draw] (0,1) -- (H11) -- (H51) -- (X71) -- (8.5,1);
%
%\draw[fill=black] (0.5,0) circle (0.1);
%\draw[fill=black] (0.5,0.5) circle (0.1);
%\draw[fill=black] (6.5,0) circle (0.1);
%\draw[fill=black] (6.5,0.5) circle (0.1);
%\path[draw] (0.5,0.5) -- (0.5,0);
%\path[draw] (6.5,0.5) -- (6.5,0);
%
%\draw[fill=black] (2.5,1) circle (0.1);
%\draw (2.5,0.5) circle (0.15);
%\path[draw] (2.5,1) -- (2.5,0.35);
%\draw[fill=black] (4.5,1) circle (0.1);
%\draw (4.5,0.5) circle (0.15);
%\path[draw] (4.5,1) -- (4.5,0.35);
%
%%\foreach \x in {0.5,1.5,2.5,3.5,4.5,5.5,6.5,7.5}
%%	\path[draw,dashed] (\x,1.5) -- (\x,-0.5);
%
%\end{tikzpicture}
%\end{center}
%
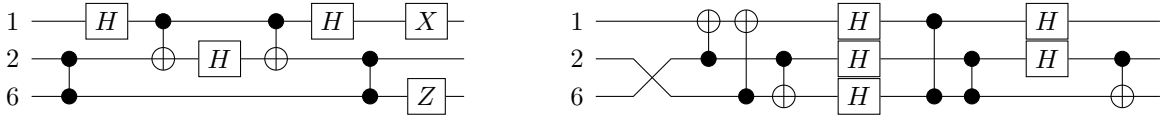
\begin{center}
\begin{tikzpicture}

\node[draw,rectangle] (H11) at (1,1) {$H$};
\node[draw,rectangle] (H25) at (2.5,0.5) {$H$};
\node[draw,rectangle] (H41) at (4,1) {$H$};
\node[draw,rectangle] (X51) at (5.25,1) {$X$};
\node[draw,rectangle] (Z50) at (5.25,0) {$Z$};

\node at (-0.25,1) {$1$};
\node at (-0.25,0.5) {$2$};
\node at (-0.25,0) {$6$};

\path[draw] (0,0) -- (Z50) -- (5.75,0);
\path[draw] (0,0.5) -- (H25) -- (5.75,0.5);
\path[draw] (0,1) -- (H11) -- (H41) -- (X51) -- (5.75,1);

\draw[fill=black] (0.5,0) circle (0.1);
\draw[fill=black] (0.5,0.5) circle (0.1);
\draw[fill=black] (4.5,0) circle (0.1);
\draw[fill=black] (4.5,0.5) circle (0.1);
\path[draw] (0.5,0.5) -- (0.5,0);
\path[draw] (4.5,0.5) -- (4.5,0);

\draw[fill=black] (1.75,1) circle (0.1);
\draw (1.75,0.5) circle (0.15);
\path[draw] (1.75,1) -- (1.75,0.35);
\draw[fill=black] (3.25,1) circle (0.1);
\draw (3.25,0.5) circle (0.15);
\path[draw] (3.25,1) -- (3.25,0.35);

%\foreach \x in {0.5,1.5,2.5,3.5,4.5,5.5,6.5,7.5}
%	\path[draw,dashed] (\x,1.5) -- (\x,-0.5);

\coordinate (ref) at (7.5,0);

\node[draw,rectangle] (H31) at ($(ref) + (3.5,1)$) {$H$};
\node[draw,rectangle] (H35) at ($(ref) + (3.5,0.5)$) {$H$};
\node[draw,rectangle] (H30) at ($(ref) + (3.5,0)$) {$H$};
\node[draw,rectangle] (H61) at ($(ref) + (6,1)$) {$H$};
\node[draw,rectangle] (H65) at ($(ref) + (6,0.5)$) {$H$};

\node at ($(ref) + (-0.25,1)$) {$1$};
\node at ($(ref) + (-0.25,0.5)$) {$2$};
\node at ($(ref) + (-0.25,0)$) {$6$};

\path[draw] ($(ref) + (0,0)$) -- ($(ref) + (0.5,0)$) -- ($(ref) + (1,0.5)$) -- (H35) -- (H65) -- ($(ref) + (7.5,0.5)$);
\path[draw] ($(ref) + (0,0.5)$) -- ($(ref) + (0.5,0.5)$) -- ($(ref) + (1,0)$) -- (H30) -- ($(ref) + (7.5,0)$);
\path[draw] ($(ref) + (0,1)$) -- (H31) -- (H61) -- ($(ref) + (7.5,1)$);

\draw[fill=black] ($(ref) + (4.5,0)$) circle (0.1);
\draw[fill=black] ($(ref) + (4.5,1)$) circle (0.1);
\draw[fill=black] ($(ref) + (5,0)$) circle (0.1);
\draw[fill=black] ($(ref) + (5,0.5)$) circle (0.1);
\path[draw] ($(ref) + (4.5,0)$) -- ($(ref) + (4.5,1)$);
\path[draw] ($(ref) + (5,0)$) -- ($(ref) + (5,0.5)$);

\draw[fill=black] ($(ref) + (1.5,0.5)$) circle (0.1);
\draw ($(ref) + (1.5,1)$) circle (0.15);
\path[draw] ($(ref) + (1.5,1.15)$) -- ($(ref) + (1.5,0.5)$);

\draw[fill=black] ($(ref) + (2,0)$) circle (0.1);
\draw ($(ref) + (2,1)$) circle (0.15);
\path[draw] ($(ref) + (2,1.15)$) -- ($(ref) + (2,0)$);

\draw[fill=black] ($(ref) + (2.5,0.5)$) circle (0.1);
\draw ($(ref) + (2.5,0)$) circle (0.15);
\path[draw] ($(ref) + (2.5,-0.15)$) -- ($(ref) + (2.5,0.5)$);

\draw[fill=black] ($(ref) + (7,0.5)$) circle (0.1);
\draw ($(ref) + (7,0)$) circle (0.15);
\path[draw] ($(ref) + (7,-0.15)$) -- ($(ref) + (7,0.5)$);

\end{tikzpicture}
\end{center}
Note that these are only two of all possible circuits, even given the specific symplectic matrix~\eqref{eq:css_H1}.
The set of all symplectic solutions for $\lH_1$ were obtained using the result of Theorem~\ref{thm:symp_lineq_all} in Section~\ref{sec:discuss} below, and these are listed in Appendix~\ref{sec:css_had_all}.
%As for $\lP_1, \lcz{1}{2}$ and $\lcnot{2}{1}$, the above solution is the cheapest in this set in terms of the circuit depth.

%\subsubsection{Logical Transversal Hadamard}

As noted in~\cite{Chao-arxiv17b}, for this code, the logical transversal Hadamard operator $\lH^{\otimes 4}$, applied to all logical qubits simultaneously, is easy to construct.
This operator must satisfy the conditions $\lH_j \lX_j \lH_j = \lZ_j, \lH_j \lZ_j \lH_j = \lX_j$ for $j=1,2,3,4$.
If we apply the physical Hadamard operator $H$ transversally, i.e. $H_1 H_2 \cdots H_6$, we get the mappings
\[ X_1 X_{i+1} \mapsto Z_1 Z_{i+1} \quad , \quad Z_{i+1} Z_6 \mapsto X_{i+1} X_6 . \]
To complete the logical transversal Hadamard we now have to just swap physical qubits $1$ and $6$.
%This will give the desired mappings
%\[ \begin{array}{cc|cc}
%\lX_1' = Z_2 Z_6 & \quad & \quad & \lZ_1' = X_1 X_2 \\
%\lX_2' = Z_3 Z_6 & \quad & \quad & \lZ_2' = X_1 X_3 \\
%\lX_3' = Z_4 Z_6 & \quad & \quad & \lZ_3' = X_1 X_4 \\
%\lX_4' = Z_5 Z_6 & \quad & \quad & \lZ_4' = X_1 X_5 
%\end{array} . \]
We note from Table~\ref{tab:std_symp} that the symplectic transformation associated with physical transversal Hadamard is $\Omega$ and the symplectic transformation associated with swapping qubits $1$ and $6$ is $\begin{bmatrix} A & 0 \\ 0 & A \end{bmatrix}$, where 
\begin{align}
A = 
\begin{bmatrix}
0 & 0 & 0 & 0 & 0 & 1 \\
0 & 1 & 0 & 0 & 0 & 0 \\
0 & 0 & 1 & 0 & 0 & 0 \\
0 & 0 & 0 & 1 & 0 & 0 \\
0 & 0 & 0 & 0 & 1 & 0 \\
1 & 0 & 0 & 0 & 0 & 0 
\end{bmatrix} .
\end{align}
Hence the symplectic transformation associated with the logical transversal Hadamard operator is 
\begin{align}
{F = \begin{bmatrix} 0 & I_6 \\ I_6 & 0 \end{bmatrix} \begin{bmatrix} A & 0 \\ 0 & A \end{bmatrix} = \begin{bmatrix} 0 & A \\ A & 0 \end{bmatrix}} .
\end{align}
%It can be verified that $F$ satisfies the constraints $[\vecnot{e}_1 + \vecnot{e}_i, \vecnot{0}] F = [\vecnot{0}, \vecnot{e}_i + \vecnot{e}_6], \ [\vecnot{0}, \vecnot{e}_i + \vecnot{e}_6] F = [\vecnot{e}_1 + \vecnot{e}_i, \vecnot{0}]$ for $i=2,3,4,5$.
Note that this solution swaps $X^{\otimes 6}$ and $Z^{\otimes 6}$ and hence only normalizes the stabilizer.
Therefore, in general, the simplest circuit to realize a logical operator might not always fix the stabilizer element-wise, i.e., it might not centralize the stabilizer.

\section{Generic Algorithm for Synthesis of Logical Clifford Operators}
\label{sec:discuss}

The synthesis of logical Paulis by Gottesman~\cite{Gottesman-phd97} and by Wilde~\cite{Wilde-physreva09} exploits symplectic geometry over the binary field. 
Building on their work we have demonstrated, using the $\llbr 6,4,2 \rrbr$ code as an example, that symplectic geometry provides a systematic framework for synthesizing physical implementations of any logical operator in the logical Clifford group ${\rm Cliff}_M$ for stabilizer codes. 
In other words, symplectic geometry provides a \emph{control plane} where effects of Clifford operators can be analyzed efficiently.
For each logical Clifford operator, one can obtain all symplectic solutions using the algorithm below.

\begin{enumerate}

\item Collect all the linear constraints on $F$, obtained from the conjugation relations of the desired Clifford operator with the stabilizer generators and logical Paulis, to obtain a system of equations $UF = V$.

\item Then vectorize both sides to get $\left( I_{2m} \otimes U \right) {\rm vec}(F) = {\rm vec}(V)$.

\item Perform Gaussian elimination on the augmented matrix $\left[ \left( I_{2m} \otimes U \right),\ {\rm vec}(V) \right]$.
If $\ell$ is the number of non-pivot variables in the row-reduced echelon form, then there are $2^{\ell}$ solutions to the linear system.

\item For each such solution, check if it satisfies $F \Omega F^T = \Omega$.
If it does, then it is a feasible symplectic solution for $\bar{g}$.

\end{enumerate}

Clearly, this algorithm is not very efficient since $\ell$ could be very large.
Specifically, for codes that do not encode many logical qubits this number will be very large as the system $UF = V$ will be very under-constrained.
We now state and prove two theorems that enable us to determine all symplectic solutions for each logical Clifford operator much more efficiently.
%The first theorem uses symplectic transvections discussed in Section~\ref{sec:symp_transvec} and builds on Theorem~\ref{thm:symp_transvec}.

\begin{theorem}
\label{thm:symp_lineq}
Let $x_i, y_i \in \mathbb{F}_2^{2m}, i=1,2,\ldots,t \leq 2m$ be a collection of (row) vectors such that $\syminn{x_i}{x_j} = \syminn{y_i}{y_j}$.
Assume that the $x_i$ are linearly independent.
Then a solution $F \in \text{Sp}(2m,\mathbb{F}_2)$ to the system of equations $x_i F = y_i$ can be obtained as the product of a sequence of at most $2t$ symplectic transvections $F_{h} \triangleq I_{2m} + \Omega h^T h$, where $h \in \mathbb{F}_2^{2m}$ is a row vector.
\begin{IEEEproof}
We will prove this result by induction.
For $i=1$ we can simply use Theorem~\ref{thm:symp_transvec} to find $F_1 \in \text{Sp}(2m,\mathbb{F}_2)$ as follows. 
If $\syminn{x_1}{y_1} = 1$ then $F_1 \triangleq F_{h_1}$ with $h_1 \triangleq x_1 + y_1$, or if $\syminn{x_1}{y_1} = 0$ then $F_1 \triangleq F_{h_{11}} F_{h_{12}}$ with $h_{11} \triangleq w_1 + y_1, h_{12} \triangleq x_1 + w_1$, where $w_1$ is chosen such that $\syminn{x_1}{w_1} = \syminn{y_1}{w_1} = 1$.
In any case $F_1$ satisfies $x_1 F_1 = y_1$.
Next consider $i = 2$.
Let $\tilde{x}_2 \triangleq x_2 F_1$ so that $\syminn{x_1}{x_2} = \syminn{y_1}{y_2} = \syminn{y_1}{\tilde{x}_2}$, since $F_1$ is symplectic and hence preserves symplectic inner products.
Similar to Theorem~\ref{thm:symp_transvec} we have two cases: $\syminn{\tilde{x}_2}{y_2} = 1$ or $0$.
For the former, we set $h_2 \triangleq \tilde{x}_2 + y_2$ so that we clearly have $\tilde{x}_2 F_{h_2} = Z_{h_2}(\tilde{x}_2) = y_2$ (see Section~\ref{sec:symp_transvec} for the definition of $Z_h(\cdot)$).
We also observe that
\begin{align*}
y_1 F_{h_2} = Z_{h_2}(y_1) = y_1 + \syminn{y_1}{\tilde{x}_2 + y_2} (\tilde{x}_2 + y_2) = y_1 + (\syminn{y_1}{y_2} + \syminn{y_1}{y_2}) (\tilde{x}_2 + y_2) = y_1 .
\end{align*}
Hence in this case $F_2 \triangleq F_1 F_{h_2}$ satisfies $x_1 F_2 = y_1, x_2 F_2 = y_2$.
For the case $\syminn{\tilde{x}_2}{y_2} = 0$ we again find a $w_2$ that satisfies $\syminn{\tilde{x}_2}{w_2} = \syminn{y_2}{w_2} = 1$ and set $h_{21} \triangleq w_2 + y_2, h_{22} \triangleq \tilde{x}_2 + w_2$.
Then by Theorem~\ref{thm:symp_transvec} we clearly have $\tilde{x}_2 F_{h_{21}} F_{h_{22}} = y_2$.
For $y_1$ we observe that
\begin{align*}
y_1 F_{h_{21}} F_{h_{22}} & = Z_{h_{22}} \left( Z_{h_{21}}(y_1) \right) \\
  & = Z_{h_{22}} \left( y_1 + \syminn{y_1}{w_2 + y_2} (w_2 + y_2) \right) \\
  & = y_1 + \syminn{y_1}{w_2 + y_2} (w_2 + y_2) + \left( \syminn{y_1}{\tilde{x}_2 + w_2} + \syminn{y_1}{w_2 + y_2} \syminn{w_2 + y_2}{\tilde{x}_2 + w_2} \right) (\tilde{x}_2 + w_2) \\
  & = y_1 + \syminn{y_1}{w_2 + y_2} (\tilde{x}_2 + y_2) \ (\because \syminn{y_1}{\tilde{x}_2} = \syminn{y_1}{y_2}, \syminn{w_2 + y_2}{\tilde{x}_2 + w_2} = 1+0+0+1 = 0) \\
  & = y_1 \ \text{if and only if}\ \syminn{y_1}{w_2} = \syminn{y_1}{y_2} .
\end{align*}
Hence, we pick a $w_2$ such that $\syminn{\tilde{x}_2}{w_2} = \syminn{y_2}{w_2} = 1$ and $\syminn{y_1}{w_2} = \syminn{y_1}{y_2}$, and then set $F_2 \triangleq F_1 F_{h_{21}} F_{h_{22}}$.
Again, for this case $F_2$ satisfies $x_1 F_2 = y_1, x_2 F_2 = y_2$ as well.

By induction, assume $F_{i-1}$ satisfies $x_j F_{i-1} = y_j$ for all $j=1,\ldots,i-1$, where $i \geq 3$.
Using the same idea as for $i=2$ above, let $x_i F_{i-1} = \tilde{x}_i$. 
If $\syminn{\tilde{x}_i}{y_i} = 1$, we simply set $F_i \triangleq F_{i-1} F_{h_i}$, where $h_i \triangleq \tilde{x}_i + y_i$. 
If $\syminn{\tilde{x}_i}{y_i} = 0$, we find a $w_i$ that satisfies $\syminn{\tilde{x}_i}{w_i} = \syminn{y_i}{w_i} = 1$ and $\syminn{y_j}{w_i} = \syminn{y_j}{y_i} \ \forall \ j < i$. 
Then we define $h_{i1} \triangleq w_i + y_i, h_{i2} \triangleq \tilde{x}_i + w_i$ and observe that for $j < i$ we have
\begin{align*}
y_j F_{h_{i1}} F_{h_{i2}} = Z_{h_{i2}} \left( Z_{h_{i1}}(y_j) \right) = y_j + \syminn{y_j}{w_i + y_i} (\tilde{x}_i + y_i) = y_j .
\end{align*}
Again, by Theorem~\ref{thm:symp_transvec}, we clearly have $\tilde{x}_i F_{h_{i1}} F_{h_{i2}} = y_i$.
Hence we set $F_i \triangleq F_{i-1} F_{h_{i1}} F_{h_{i2}}$ in this case. 
In both cases $F_i$ satisfies $x_j F_i = y_j \ \forall \ j=1,\ldots,i$. 
Setting $F \triangleq F_t$ completes the inductive proof and it is clear that $F$ is the product of at most $2t$ symplectic transvections.
\end{IEEEproof}
\end{theorem}

\renewcommand{\algorithmicrequire}{\textbf{Input:}}
\renewcommand{\algorithmicensure}{\textbf{Output:}}

The algorithm defined implicitly by the above proof is stated explicitly in Algorithm~\ref{alg:transvec}.
\begin{algorithm}
%\DontPrintSemicolon
%\vspace{0.3cm}
\caption{Algorithm to find $F \in \text{Sp}(2m,\mathbb{F}_2)$ satisfying a linear system of equations, using Theorem~\ref{thm:symp_lineq}}	
\label{alg:transvec}
  \begin{algorithmic}[1]	
\REQUIRE $x_i ,y_i \in \mathbb{F}_2^{2m}$ s.t. $\syminn{x_i}{x_j} = \syminn{y_i}{y_j} \ \forall \ i,j \in \{1,\ldots,t\}$. 
\ENSURE $F \in \text{Sp}(2m,\mathbb{F}_2)$ satisfying $x_i F = y_i \ \forall \ i \in \{1,\ldots,t\}$
\IF{$\syminn{x_1}{y_1} = 1$}
	\STATE set $h_1 \triangleq x_1 + y_1$ and $F_1 \triangleq F_{h_1}$.		
\ELSE
	\STATE $h_{11} \triangleq w_1 + y_1, h_{12} \triangleq x_1 + w_1$ and $F_1 \triangleq F_{h_{11}} F_{h_{12}}$.
\ENDIF
		
\FOR{$i = 2,\ldots,t$}
	\STATE Calculate $\tilde{x}_i \triangleq x_i F_{i-1}$ and $\syminn{\tilde{x}_i}{y_i}$.
	\IF{$\tilde{x}_i = y_i$}
		\STATE Set $F_i \triangleq F_{i-1}$. \textbf{Continue}.
	\ENDIF
	\IF{$\syminn{\tilde{x}_i}{y_i} = 1$}
		\STATE Set $h_i \triangleq \tilde{x}_i + y_i, F_i \triangleq F_{i-1} F_{h_i}$.
%	\ENDIF
%	\IF{$\syminn{\tilde{x}_i}{y_i} = 0$}
	\ELSE
		\STATE Find a $w_i$ s.t. $\syminn{\tilde{x}_i}{w_i} = \syminn{y_i}{w_i} = 1$ and $\syminn{y_j}{w_i} = \syminn{y_j}{y_i} \ \forall \ j < i$. 
		\STATE Set $h_{i1} \triangleq w_i + y_i, h_{i2} \triangleq \tilde{x}_i + w_i, F_i \triangleq F_{i-1} F_{h_{i1}} F_{h_{i2}}$.
	\ENDIF
\ENDFOR
\RETURN $F \triangleq F_t$.
\end{algorithmic}
\end{algorithm}

Now we state our main theorem, which enables one to determine all symplectic solutions for a system of linear equations.

\begin{theorem}
\label{thm:symp_lineq_all}
Let $\{(u_a, v_a),\ a \in \{1,\ldots,m\}\}$ be a collection of pairs of (row) vectors that form a symplectic basis for $\mathbb{F}_2^{2m}$, where $u_a, v_a \in \mathbb{F}_2^{2m}$. 
Consider the system of linear equations $u_i F = u_i', v_j F = v_j'$, where $i \in \mathcal{I} \subseteq \{1,\ldots,m\}, {j \in \mathcal{J} \subseteq \{1,\ldots,m\}}$ and $F \in \text{Sp}(2m,\mathbb{F}_2)$.
Assume that the given vectors satisfy $\syminn{u_{i_1}}{u_{i_2}} = \syminn{u_{i_1}'}{u_{i_2}'} = 0, \syminn{v_{j_1}}{v_{j_2}} = \syminn{v_{j_1}'}{v_{j_2}'} = 0, \syminn{u_{i}}{v_{j}} = \syminn{u_{i}'}{v_{j}'} = \delta_{ij}$, where $i_1,i_2 \in \mathcal{I}, \ j_1,j_2 \in \mathcal{J}$, since symplectic transformations $F$ must preserve symplectic inner products.
Let $\alpha \triangleq |\bar{\mathcal{I}}| + |\bar{\mathcal{J}}|$, where $\bar{\mathcal{I}}, \bar{\mathcal{J}}$ denote the set complements of $\mathcal{I},\mathcal{J}$ in $\{1,\ldots,m\}$, respectively.
Then there are $2^{\alpha(\alpha+1)/2}$ solutions $F$ to the given linear system.
\begin{IEEEproof}
By the definition of a symplectic basis (Definition~\ref{def:symp_basis}), we have $\syminn{u_a}{v_b} = \delta_{ab}$ and $\syminn{u_a}{u_b} = \syminn{v_a}{v_b} = 0$, where $a,b \in \{1,\ldots,m\}$.
The same definition extends to any (symplectic) subspace of $\mathbb{F}_2^{2m}$.
The linear system under consideration imposes constraints only on $u_i, i \in \mathcal{I}$ and $v_j, j \in \mathcal{J}$.
Let $W$ be the subspace of $\mathbb{F}_2^{2m}$ spanned by the symplectic pairs $(u_c,v_c)$ where $c \in \mathcal{I} \cap \mathcal{J}$ and $W^{\perp}$ be its orthogonal complement under the symplectic inner product, i.e., $W \triangleq \langle \{ (u_c,v_c),\ c \in \mathcal{I} \cap \mathcal{J} \} \rangle$ and $W^{\perp} \triangleq \langle \{ (u_d, v_d),\ d \in \bar{\mathcal{I}} \cup \bar{\mathcal{J}} \} \rangle$, where $\bar{\mathcal{I}}, \bar{\mathcal{J}}$ denote the set complements of $\mathcal{I},\mathcal{J}$ in $\{1,\ldots,m\}$, respectively.
%If $\bar{\mathcal{I}} = \bar{\mathcal{J}} = \phi$ then the system is fully constrained and we have exactly one solution $F$.
%Otherwise, we call $F$ to be a \emph{partial} symplectic matrix since it is under-constrained.

Using the result of Theorem~\ref{thm:symp_lineq} we first compute one solution $F_0$ for the given system of equations.
In the subspace $W$, $F_0$ maps $(u_c,v_c) \mapsto (u_c',v_c')$ for all $c \in \mathcal{I} \cap \mathcal{J}$ and hence we now have $W = \langle \{ (u_c',v_c'),\ c \in \mathcal{I} \cap \mathcal{J} \} \rangle$ spanned by its new basis pairs $(u_c',v_c')$.
However in $W^{\perp}$, $F_0$ maps $(u_d,v_d) \mapsto (u_d',\tilde{v}_d')$ or $(u_d,v_d) \mapsto (\tilde{u}_d',v_d')$ or $(u_d,v_d) \mapsto (\tilde{u}_d',\tilde{v}_d')$ depending on whether $d \in \mathcal{I} \cap \bar{\mathcal{J}}$ or $d \in \bar{\mathcal{I}} \cap \mathcal{J}$ or $d \in \bar{\mathcal{I}} \cap \bar{\mathcal{J}}$, respectively ($d \notin \mathcal{I} \cap \mathcal{J}$  by definition of $W^{\perp}$).
Note however that the subspace $W^{\perp}$ itself is fixed.
We observe that such $\tilde{u}_d'$ and $\tilde{v}_d'$ are not specified by the given linear system and hence form only a particular choice for the new symplectic basis of $W^{\perp}$. 
These can be mapped to arbitrary choices $\tilde{u}_d$ and $\tilde{v}_d$, while fixing other $u_d'$ and $v_d'$, as long as the new choices still complete a symplectic basis for $W^{\perp}$.
Hence, these form the degrees of freedom for the solution set of the given system of linear equations.
The number of such ``free'' vectors is exactly $|\bar{\mathcal{I}}| + |\bar{\mathcal{J}}| = \alpha$.
This can be verified by observing that the number of basis vectors for $W^{\perp}$ is $2 |\bar{\mathcal{I}} \cup \bar{\mathcal{J}}|$ and making the following calculation.
\begin{align*}
\text{Number\ of\ constrained\ vectors\ in\ the\ new\ basis\ for}\ W^{\perp} &= |\mathcal{I} \setminus \mathcal{J}| + |\mathcal{J} \setminus \mathcal{I}| \\
  &= |\mathcal{I}| - |\mathcal{I} \cap \mathcal{J}| + |J| - |\mathcal{I} \cap \mathcal{J}| \\
  &= (m - |\bar{\mathcal{I}}|) + (m - |\bar{\mathcal{J}}|) - 2 (m - |\bar{\mathcal{I}} \cup \bar{\mathcal{J}}|) \\
  &= 2 |\bar{\mathcal{I}} \cup \bar{\mathcal{J}}| - (|\bar{\mathcal{I}}| + |\bar{\mathcal{J}}|) \\
  &= 2 |\bar{\mathcal{I}} \cup \bar{\mathcal{J}}| - \alpha .
\end{align*}

For convenience, we relabel the subscripts of these basis vectors for $W^{\perp}$ with $d, d_1, d_2 \in \{1,\ldots,|\bar{\mathcal{I}} \cup \bar{\mathcal{J}}|\}$.
The constraints on free vectors $\tilde{u}_d$ and $\tilde{v}_d$ are that $\syminn{\tilde{u}_{d_1}}{v_{d_2}'} = \syminn{u_{d_1}'}{\tilde{v}_{d_2}} = \syminn{\tilde{u}_{d_1}}{\tilde{v}_{d_2}} = \delta_{d_1 d_2}$ and all other pairs of vectors in the new basis set for $W^{\perp}$ be orthogonal to each other.
In the $d$-th symplectic pair --- $(\tilde{u}_d,v_d')$ or $(u_d',\tilde{v}_d)$ or $(\tilde{u}_d,\tilde{v}_d)$ --- of its new symplectic basis there is at least one free vector --- $\tilde{u}_d$ or $\tilde{v}_d$ or both, respectively.
For the first of the $\alpha$ free vectors, there are $2 |\bar{\mathcal{I}} \cup \bar{\mathcal{J}}| - \alpha$ symplectic inner product constraints (which are linear constraints) imposed by the $2 |\bar{\mathcal{I}} \cup \bar{\mathcal{J}}| - \alpha$ constrained vectors $u_d',v_d'$. 
Since $W^{\perp}$ has (binary) vector space dimension $2 |\bar{\mathcal{I}} \cup \bar{\mathcal{J}}|$ and each linearly independent constraint decreases the dimension by $1$, this leads to $2^{\alpha}$ possible choices for the first free vector.
For the second free vector, there are $\alpha-1$ degrees of freedom as it has an additional inner product constraint from the first free vector.
This leads to $2^{\alpha-1}$ possible choices for the second free vector, and so on.
Therefore, the given linear system has $\prod_{\ell=1}^{\alpha} 2^{\ell} = 2^{\alpha(\alpha+1)/2}$ symplectic solutions.

Finally we show how to get each symplectic matrix $F$ for the given linear system. 
First form the matrix $A$ whose rows are the new symplectic basis vectors for $\mathbb{F}_2^{2m}$ obtained under the action of $F_0$, i.e., the first $m$ rows are $u_c',u_d',\tilde{u}_d'$ and the last $m$ rows are $v_c',v_d',\tilde{v}_d'$. 
Observe that this matrix is symplectic and invertible.
Then form a matrix $B = A$ and replace the rows corresponding to free vectors with a particular choice of free vectors, chosen to satisfy the conditions mentioned above.
Note that $B$ and $A$ differ in exactly $\alpha$ rows, and that $B$ is also symplectic and invertible.
Determine the symplectic matrix $F' = A^{-1} B$ which fixes all new basis vectors obtained for $W$ and $W^{\perp}$ under $F_0$ except the free vectors in the basis for $W^{\perp}$.
Then this yields a new solution $F = F_0 F'$ for the given system of linear equations.
Note that if $\tilde{u}_d = \tilde{u}_d'$ and $\tilde{v}_d = \tilde{v}_d'$ for all free vectors, where $\tilde{u}_d', \tilde{v}_d'$ were obtained under the action of $F_0$ on $W^{\perp}$, then $F' = I_{2m}$.
Repeating this process for all $2^{\alpha(\alpha+1)/2}$ choices of free vectors enumerates all the solutions for the linear system under consideration.
\end{IEEEproof}
\end{theorem}

\begin{remark}
For any system of symplectic linear equations $x_i F = y_i,\ i=1,\ldots,t$ where the $x_i$ do not form a symplectic basis for $\mathbb{F}_2^{2m}$, we first calculate a symplectic basis $(u_j,v_j), \ j=1,\ldots,m$ using the symplectic Gram-Schmidt orthogonalization procedure discussed in~\cite{Koenig-jmp14}.
Then we transform the given system into an equivalent system of constraints on these basis vectors $u_j,v_j$ and apply Theorem~\ref{thm:symp_lineq_all} to obtain all symplectic solutions.
\end{remark}

The algorithm defined implicitly by the above proof is stated explicitly in Algorithm~\ref{alg:symp_lineq_all} below.

\begin{algorithm}
\caption{Algorithm to determine all $F \in \text{Sp}(2m,\mathbb{F}_2)$ satisfying a linear system of equations, using Theorem~\ref{thm:symp_lineq_all}}
\label{alg:symp_lineq_all}
\begin{algorithmic}[1]

\REQUIRE $u_a,v_b \in \mathbb{F}_2^{2m}$ s.t. $\syminn{u_a}{v_b} = \delta_{ab}$ and $\syminn{u_a}{u_b} = \syminn{v_a}{v_b} = 0$, where $a,b \in \{1,\ldots,m\}$.

$u_i', v_j' \in \mathbb{F}_2^{2m}$ s.t. $\syminn{u_{i_1}'}{u_{i_2}'} = 0, \syminn{v_{j_1}'}{v_{j_2}'} = 0, \syminn{u_{i}'}{v_{j}'} = \delta_{ij}$, where $i,i_1,i_2 \in \mathcal{I}, \ j,j_1,j_2 \in \mathcal{J},\ \mathcal{I}, \mathcal{J} \subseteq \{1,\ldots,m\}$.

\ENSURE $\mathcal{F} \subset \text{Sp}(2m,\mathbb{F}_2)$ such that each $F \in \mathcal{F}$ satisfies $u_i F = u_i' \ \forall\ i \in \mathcal{I}$, and $v_j F = v_j' \ \forall \ j \in \mathcal{J}$.

\STATE Determine a particular symplectic solution $F_0$ for the linear system using Algorithm~\ref{alg:transvec}.

\STATE Form the matrix $A$ whose $a$-th row is $u_a F_0$ and $(m+b)$-th row is $v_b F_0$, where $a,b \in \{1,\ldots,m\}$. 

\STATE Compute the inverse of this matrix, $A^{-1}$, in $\mathbb{F}_2$.

\STATE Set $\mathcal{F} = \phi$ and $\alpha \triangleq |\bar{\mathcal{I}}| + |\bar{\mathcal{J}}|$, where $\bar{\mathcal{I}}, \bar{\mathcal{J}}$ denote the set complements of $\mathcal{I},\mathcal{J}$ in $\{1,\ldots,m\}$, respectively.

\FOR{$\ell = 1,\ldots,2^{\alpha(\alpha+1)/2}$}
	
	\STATE Form a matrix $B_{\ell} = A$.

	\STATE For $i \notin \mathcal{I}$ and $j \notin \mathcal{J}$ replace the $i$-th and $(m+j)$-th rows of $B_{\ell}$ with arbitrary vectors such that $B_{\ell} \Omega B_{\ell}^T = \Omega$ and $B_{\ell} \neq B_{\ell'}$ for $1 \leq \ell' < \ell$. \hfill {\rm $\boldsymbol{/\ast}$ See proof of Theorem~\ref{thm:symp_lineq_all} for details or Appendix~\ref{sec:alg2_matlab} for example \texttt{MATLAB\textsuperscript{\textregistered}} code $\boldsymbol{\ast /}$}
	
	\STATE Compute $F' = A^{-1} B$.
	
	\STATE Add $F_{\ell} \triangleq F_0 F'$ to $\mathcal{F}$.
	
\ENDFOR

\RETURN $\mathcal{F}$

\end{algorithmic}
\end{algorithm}

For a given system of linear (independent) equations, if $\alpha = 0$ then the symplectic matrix $F$ is fully constrained and there is a unique solution.
Otherwise, the system is partially constrained and we refer to a solution $F$ as a \emph{partial} symplectic matrix.

\emph{Example}:
As an application of this theorem, we discuss the procedure to determine all symplectic solutions for the logical Phase gate $\lP_1$ discussed in Section~\ref{sec:css_phase}.
First we define a symplectic basis for $\mathbb{F}_2^{2m}$ using the binary vector representation of the logical Pauli operators and stabilizer generators of the $\llbr 6,4,2 \rrbr$ code.
\begin{align}
u_1 \triangleq [110000,000000] \quad &, \quad v_1 \triangleq [000000,010001] , \nonumber \\ 
u_2 \triangleq [101000,000000] \quad &, \quad v_2 \triangleq [000000,001001] , \nonumber \\ 
u_3 \triangleq [100100,000000] \quad &, \quad v_3 \triangleq [000000,000101] , \nonumber \\ 
u_4 \triangleq [100010,000000] \quad &, \quad v_4 \triangleq [000000,000011] , \nonumber \\ 
u_5 \triangleq [111111,000000] \quad &, \quad v_5 \triangleq [000000,000001] , \nonumber \\ 
u_6 \triangleq [100000,000000] \quad &, \quad v_6 \triangleq [000000,111111] .
\end{align}
Note that $v_5$ and $u_6$ do not correspond to either a logical Pauli operator or a stabilizer element but were added to complete a symplectic basis.
Hence we have $\mathcal{I} = \{1,2,3,4,5\}, \mathcal{J} = \{1,2,3,4,6\}$ and $\alpha = 1 + 1 = 2$.
As discussed in Section~\ref{sec:css_phase}, we impose constraints on all $u_i,v_j$ except for $i=6$ and $j=5$.
Therefore, as per the notation in the above proof, we have $W \triangleq \langle \{(u_1,v_1), \ldots, (u_4,v_4)\} \rangle$ and $W^{\perp} \triangleq \langle \{(u_5,v_5), (u_6,v_6)\} \rangle$.
Using Algorithm~\ref{alg:transvec} we obtain a particular solution $F_0 = T_B$ where $B$ is given in~\eqref{eq:css_phase_B}.
Then we compute the action of $F_0$ on the bases for $W$ and $W^{\perp}$ to get
\begin{align}
u_i F_0 \triangleq u_i',\ v_j F_0 \triangleq v_j',\ i \in \mathcal{I},\ j \in \mathcal{J}, \ \ \text{and} \ \ u_6 F_0 = [100000,000000] \triangleq \tilde{u}_6', \ v_5 F_0 = [000000,000001] \triangleq \tilde{v}_5' ,
%u_1 F_0 = [110000,010001] \triangleq u_1' \quad &, \quad v_1 F_0 = [000000,010001] \triangleq v_1' \nonumber \\
%%
%u_2 F_0 = [101000,000000] \triangleq u_2' \quad &, \quad v_2 F_0 = [000000,001001] \triangleq v_2', \nonumber \\ 
%%
%u_3 F_0 = [100100,000000] \triangleq u_3' \quad &, \quad v_3 F_0 = [000000,000101] \triangleq v_3' , \nonumber \\ 
%%
%u_4 F_0 = [100010,000000] \triangleq u_4' \quad &, \quad v_4 F_0 = [000000,000011] \triangleq v_4' , \nonumber \\ 
%%
%u_5 F_0 = [111111,000000] \triangleq u_5' \quad &, \quad v_5 F_0 = [000000,000001] \triangleq \tilde{v}_5' , \nonumber \\ 
%%
%u_6 F_0 = [100000,000000] \triangleq \tilde{u}_6' \quad &, \quad v_6 F_0 = [000000,111111] \triangleq v_6'.
\end{align}
where $u_i', v_j'$ are the vectors obtained in Section~\ref{sec:css_phase}.
Then we identify $\tilde{v}_5$ and $\tilde{u}_6$ to be the free vectors and one particular solution is $\tilde{v}_5 = \tilde{v}_5', \tilde{u}_6 = \tilde{u}_6'$.
In this case we have $2^{\alpha} = 2^2 = 4$ choices to pick $\tilde{v}_5$ (since we need $\syminn{u_5}{\tilde{v}_5} = 1, \syminn{v_6}{\tilde{v}_5} = 0$) and for each such choice we have $2^{\alpha-1} = 2$ choices for $\tilde{u}_6$.
Next we form the matrix $A$ whose $i$-th row is $u_i'$ and $(6+j)$-th row is $v_j'$, where $i \in \mathcal{I},j \in \mathcal{J}$.
We set the $6$th row to be $\tilde{u}_6'$ and the $11$th row to be $\tilde{v}_5'$.
Then we form a matrix $B = A$ and replace rows $6$ and $11$ by one of the $8$ possible pair of choices for $\tilde{u}_6$ and $\tilde{v}_5$, respectively.
This yields the matrix $F' = A^{-1} B$ and the symplectic solution $F = F_0 F'$.
Looping through all the $8$ choices we obtain the solutions listed in Appendix~\ref{sec:css_phase_all}.

\begin{theorem}
\label{thm:stabilizer_solutions}
For an $\llbr m,m-k \rrbr$ stabilizer code, the number of solutions for each logical Clifford operator is $2^{k(k+1)/2}$.
\begin{IEEEproof}
Let $u_i, v_i \in \mathbb{F}_2^{2m}$ represent the logical Pauli operators $\lX_i, \lZ_i$, for $i=1,\ldots,m-k$, respectively, i.e., $\gamma(\lX_i) = u_i, \gamma(\lZ_i) = v_i$, where $\gamma$ is the map defined in~\eqref{eq:gamma}.
Since $\lX_i \lZ_i = -\lZ_i \lX_i$ and $\lX_i \lZ_j = \lZ_j \lX_i$ for all $j \neq i$, it is clear that $\syminn{u_i}{v_j} = \delta_{ij}$ for $i,j \in \{1,\ldots,m-k\}$ and hence they form a partial symplectic basis for $\mathbb{F}_2^{2m}$.
Let $u_{m-k+1},\ldots,u_m$ represent the stabilizer generators, i.e., $\gamma(S_j) = u_{m-k+j}$ where the stabilizer group is $S = \langle S_1,\ldots,S_k \rangle$.
Since by definition $\lX_i, \lZ_i$ commute with all stabilizer elements, it is clear that $\syminn{u_i}{u_j} = \syminn{v_i}{u_j} = 0$ for $i \in \{1,\ldots,m-k\}, j \in \{m-k+1,\ldots,m\}$.
To complete the symplectic basis we find vectors $v_{m-k+1},\ldots,v_m$ s.t. $\syminn{u_i}{v_j} = \delta_{ij} \ \forall \ i,j \in \{1,\ldots,m\}$.
Now we note that for any logical Clifford operator, the conjugation relations with logical Paulis yield $2(m-k)$ constraints, on $u_i,v_i$ for $i \in \{1,\ldots,m-k\}$, and the normalization condition on the stabilizer yields $k$ constraints, on $u_{m-k+1},\ldots,u_m$.
Hence we have $\bar{\mathcal{I}} = \phi, \bar{\mathcal{J}} = \{m-k+1,\ldots,m\}$, as per the notation in Theorem~\ref{thm:symp_lineq_all}, and thus $\alpha = |\bar{\mathcal{I}}| + |\bar{\mathcal{J}}| = k$.
\end{IEEEproof}
\end{theorem}

Note that for each symplectic solution there are multiple decompositions into elementary forms (from Table~\ref{tab:std_symp}) possible, and one possibility is given in Theorem~\ref{thm:Trung}.
Although each decomposition yields a different circuit, all of them will act identically on $X_N$ and $Z_N$, defined in~\eqref{eq:XnZn}, under conjugation.
Once a logical Clifford operator is defined by its conjugation with the logical Pauli operators, a physical realization of the operator could either normalize the stabilizer or centralize it, i.e., fix each element of the stabilizer group under conjugation.
We show that any obtained normalizing solution can be converted into a centralizing solution.

\begin{theorem}
\label{thm:normalize_centralize}
For an $\llbr m,m-k \rrbr$ stabilizer code with stabilizer $S$, each physical realization of a given logical Clifford operator that normalizes $S$ can be converted into a circuit that centralizes $S$ while realizing the same logical operation.
\begin{IEEEproof}
Let the symplectic solution for a specific logical Clifford operator $\bar{g} \in \text{Cliff}_N$ that normalizes the stabilizer $S$ be denoted by $F_n$. 
Define the logical Pauli groups $\lX \triangleq \langle \lX_1,\ldots,\lX_{m-k} \rangle$ and $\lZ \triangleq \langle \lZ_1,\ldots,\lZ_{m-k} \rangle$. 
Let $\gamma(\lX)$ and $\gamma(\lZ)$ denote the matrices whose rows are $\gamma(\lX_i)$ and $\gamma(\lZ_i)$, respectively, for $i=1,\ldots,m-k$, where $\gamma$ is the map defined in~\eqref{eq:gamma}.
Similarly, let $\gamma(S)$ denote the matrix whose rows are the images of the stabilizer generators under the map $\gamma$.
Then, by stacking these matrices as in the proof of Theorem~\ref{thm:stabilizer_solutions}, we observe that $F_n$ is a solution of the linear system
\begin{align*}
\begin{bmatrix}
\gamma(\lX) \\
\gamma(S) \\
\gamma(\lZ)
\end{bmatrix} F_n = 
\begin{bmatrix}
\gamma(\lX') \\
\gamma(S') \\
\gamma(\lZ')
\end{bmatrix} ,
\end{align*}
where $\lX', \lZ'$ are defined by the conjugation relations of $\bar{g}$ with the logical Paulis, i.e., $\bar{g} \lX_i \bar{g}^{\dagger} = \lX_i', \bar{g} \lZ_i \bar{g}^{\dagger} = \lZ_i'$, and $S'$ denotes the stabilizer group of the code generated by a different set of generators than that of $S$.
Note, however, that as a group $S' = S$.
The goal is to find a different solution $F_c$ that centralizes the stabilizer, i.e. we replace $\gamma(S')$ with $\gamma(S)$ above.

We first find a matrix $K \in \text{GL}(k, \mathbb{F}_2)$ such that $K \gamma(S') = \gamma(S)$, which always exists since generators of $S'$ span $S$ as well.
Then we determine a symplectic solution $H$ for the linear system
\begin{align*}
\begin{bmatrix}
\gamma(\lX) \\
\gamma(S) \\
\gamma(\lZ)
\end{bmatrix} H = 
\begin{bmatrix}
\gamma(\lX) \\
K \gamma(S) \\
\gamma(\lZ)
\end{bmatrix} ,
\end{align*}
so that $H$ satisfies $K \gamma(S) = \gamma(S) H$ while fixing $\gamma(\lX)$ and $\gamma(\lZ)$.
Then since $K$ is invertible we can write
\begin{align*}
\begin{bmatrix}
I_{m-k} &   &         \\
        & K &         \\
        &   & I_{m-k}
\end{bmatrix}
\begin{bmatrix}
\gamma(\lX) \\
\gamma(S) \\
\gamma(\lZ)
\end{bmatrix} F_n = 
\begin{bmatrix}
I_{m-k} &   &         \\
        & K &         \\
        &   & I_{m-k}
\end{bmatrix}
\begin{bmatrix}
\gamma(\lX') \\
\gamma(S') \\
\gamma(\lZ')
\end{bmatrix}
\Rightarrow 
\begin{bmatrix}
\gamma(\lX) \\
\gamma(S) \\
\gamma(\lZ)
\end{bmatrix} H F_n = 
\begin{bmatrix}
\gamma(\lX') \\
\gamma(S) \\
\gamma(\lZ')
\end{bmatrix} .
\end{align*}
Hence $F_c \triangleq H F_n$ is a centralizing solution for $\bar{g}$.
Note that there are $2^{k(k+1)/2}$ solutions for $H$, as per the result of Theorem~\ref{thm:stabilizer_solutions} with the operator being the identity operator on the logical qubits, and these produce all centralizing solutions for $\bar{g}$.
\end{IEEEproof}
\end{theorem}

The above result demonstrates the relationship between the two solutions for the targeted Hadamard operator discussed in Section~\ref{sec:css_had}.
As noted in that section, after the logical transversal Hadamard operator, although any normalizing solution can be converted into a centralizing solution, the optimal solution with respect to a suitable metric need not always centralize the stabilizer.
Anyhow, we can always setup the problem of identifying a symplectic matrix, representing the physical circuit, by constraining it to centralize the stabilizer.
The general procedure to determine all symplectic solutions, and their circuits, for a logical Clifford operator for a stabilizer code is summarized in Algorithm~\ref{alg:log_ops}.
For the $\llbr 6,4,2 \rrbr$ CSS code, we employed Algorithm~\ref{alg:log_ops} to determine the solutions listed in Appendix~\ref{sec:css642_all_phy_ops} for each of the operators discussed in Section~\ref{sec:logical_ops}.
%These solutions are listed in Appendix~\ref{sec:css642_all_phy_ops}.

\begin{algorithm}
\caption{Algorithm to determine all logical Clifford operators for a stabilizer code}
\label{alg:log_ops}
\begin{algorithmic}[1]

\STATE Determine the target logical operator $\bar{g}$ by specifying its action on logical Paulis $\lX_i, \lZ_i$~\cite{Gottesman-arxiv09}: $\bar{g} \lX_i \bar{g}^{\dagger} = \lX_i', \bar{g} \lZ_i \bar{g}^{\dagger} = \lZ_i'$ .

\STATE Transform the above relations into linear equations on $F \in \text{Sp}(2m,\mathbb{F}_2)$ using the map $\gamma$ in~\eqref{eq:gamma} and the result of Theorem~\ref{thm:symp_action}, i.e., $\gamma(\lX_i) F = \gamma(\lX_i'), \gamma(\lZ_i) F = \gamma(\lZ_i')$. 
Add the conditions for normalizing the stabilizer $S$, i.e., $\gamma(S) F = \gamma(S')$.

\STATE Calculate the feasible symplectic solution set $\mathcal{F}$ using Algorithm~\ref{alg:symp_lineq_all} by mapping $\lX_i, S, \lZ_i$ to $u_i, v_i$ as in Theorem~\ref{thm:stabilizer_solutions}.

\STATE Factor each $F \in \mathcal{F}$ into a product of elementary symplectic transformations listed in Table~\ref{tab:std_symp}, possibly using the algorithm given in~\cite{Can-2017a} (which is restated in Theorem~\ref{thm:Trung} here), and compute the physical Clifford operator $\bar{g}$ (also see Remark~\ref{rem:logical_op}).

\STATE Check for conjugation of $\bar{g}$ with the stabilizer generators and for the conditions derived in step 1.
If some signs are incorrect, post-multiply by an element from $HW_N$ as necessary to satisfy all these conditions (apply~\cite[Proposition 10.4]{Nielsen-2010} for $S^{\perp} = \langle S, \lX_i, \lZ_i \rangle$, using~\eqref{eq:gamma}). 
Since $HW_N$ is the kernel of the map $\phi$ in~\eqref{eq:phi}, post-multiplication does not change $F$.
%Note that every Pauli operator in $HW_N$ induces the symplectic transformation $I_{2m}$, so post-multiplication does not change the target symplectic matrix.

\STATE Express $\bar{g}$ as a sequence of physical Clifford gates corresponding to the elementary symplectic matrices obtained from the factorization in step 4 (see Appendix~\ref{sec:elem_symp} for the circuits for these matrices).

\end{algorithmic}
\end{algorithm}

The \texttt{MATLAB\textsuperscript{\textregistered}} programs for all algorithms in this paper are available at \url{https://github.com/nrenga/symplectic-arxiv18a}.
We executed our programs on a laptop running the Windows 10 operating system (64-bit) with an Intel\textsuperscript{\textregistered} Core\textsuperscript{\texttrademark} i7-5500U @ 2.40GHz processor and 8GB RAM.
For the $\llbr 6,4,2 \rrbr$ CSS code, it takes about 0.5 seconds to generate all 8 symplectic solutions and their circuits for one logical Clifford operator.
For the $\llbr 5,1,3 \rrbr$ perfect code, it takes about 20 seconds to generate all 1024 solutions and their circuits.
Note that for step 5 in Algorithm~\ref{alg:log_ops}, we use 1-qubit and 2-qubit unitary matrices (from $\text{Cliff}_{2^2}$) to calculate conjugations for the Pauli operator on each qubit, at each circuit element at each depth (see Def.~\ref{def:depth}), and then combine the results to compute the conjugation of $\bar{g}$ with a stabilizer generator or logical Pauli operator.
We observe that most of the time is consumed in computing Kronecker products and hence these conjugations, and not in calculating the symplectic solutions.

\section{Logical Pauli Operators for Calderbank-Shor-Steane (CSS) Codes}
\label{sec:css_operators}

In this section we propose a general method to construct logical Pauli operators for CSS codes.
The exposition here is closely related to Gottesman's algorithm in~\cite{Gottesman-phd97} and the Symplectic Gram-Schmidt Orthogonalization Procedure (SGSOP) discussed by Wilde in~\cite{Wilde-physreva09}. 
However, we provide a completely classical coding-theoretic description for constructing these operators which, to the best of our knowledge, has not appeared before in the literature.

The CSS construction of quantum codes was introduced by Calderbank and Shor~\cite{Calderbank-physreva96}, and Steane~\cite{Steane-physreva96}.
Given $[m,k_1]$ and $[m,k_2]$ classical codes $\MCC_1$ and $\MCC_2$, respectively, such that $\MCC_2 \subset \MCC_1$, this construction provides an $m$-qubit quantum code CSS($\MCC_1,\MCC_2$) of dimension $2^{k_1-k_2}$ (also see~\cite[Section 10.4.2]{Nielsen-2010}.
The code CSS($\MCC_1,\MCC_2$) is represented as an $\llbr m,k_1-k_2 \rrbr$ quantum code.
If $\MCC_1$ and $\MCC_2^{\perp}$ can correct $t$ (binary) errors, then the code CSS($\MCC_1,\MCC_2$) can correct an arbitrary Pauli error on up to $t$ qubits.
For simplicity, we consider CSS codes constructed from classical codes $\MCC_1 \triangleq \MCC$ and $\MCC_2 \triangleq \MCCd$ that satisfy $\MCCd \subset \MCC$, so that $\MCC_2$ is a self-orthogonal code.
The proposed construction easily extends to general CSS codes and we comment on this extension towards the end of this section.

\subsection{Binary Self-Orthogonal Codes}
\label{sec:self_ortho}

Let $\MCCd \subset \mathbb{F}_2^m$ be an $[m,k]$ classical binary self-orthogonal code with generator and parity-check matrices $G_{\MCCd}$ and $H_{\MCCd}$ respectively.
Then it is contained in its dual $\MCC$ which is an $[m,m-k]$ classical binary code with generator and parity-check matrices $G_{\MCC} = H_{\MCCd}$ and $H_{\MCC} = G_{\MCCd}$ respectively.
Since $\MCCd \subseteq \MCC$ we immediately have $k \leq \frac{m}{2}$, so that $\MCC$ has rate at least $1/2$.
As $\MCC$ is a subgroup of $\mathbb{F}_2^m$ and $\MCCd$ is a subgroup of $\MCC$, the quotient group $\MCC/\MCCd$ is the set of all \emph{distinct} cosets of $\MCCd$ in $\MCC$,
\begin{align}
\MCC/\MCCd = \left\{ \{ \vecnot{u} + \MCCd \} \, : \, \vecnot{u} \in \{ \vecnot{0} \} \cup (\MCC \setminus \MCCd) \right\} . 
\end{align}
From each coset $\{ \vecnot{u} + \MCCd \}$ select a vector $\vecnot{v}$ as the \emph{representative} of that coset.
Then the group $\MCC/\MCCd$ is isomorphic to the group of all such representatives $\vecnot{v}$ and we will denote this group by $\MCC/\MCCd$ as well.
Since this group is also a subspace of $\mathbb{F}_2^m$ over the field $\mathbb{F}_2$, we can find a basis for it.
Let $G_{\MCC/\MCCd}$ be the matrix whose rows form a basis for the subspace $\MCC/\MCCd$.
Then, since $\MCC$ is a self-orthogonal code we can split the rows of its generator matrix to obtain the form
\begin{equation}
\label{eq:gmatrix}
G_{\MCC} = 
\begin{bmatrix}
H_{\MCC} \\
G_{\MCC/\MCCd}
\end{bmatrix}_{(m-k) \times m}
=
\begin{bmatrix}
G_{\MCCd} \\
G_{\MCC/\MCCd}
\end{bmatrix}_{(m-k) \times m} ,
\end{equation}
where $H_{\MCC} = G_{\MCCd}$ is a $k \times m$ matrix and $G_{\MCC/\MCCd}$ is an $(m-2k) \times m$ matrix.
This representation was also used by Grassl and Roetteler~\cite{Grassl-isit13} in the context of leveraging automorphisms of a classical code to realize non-trivial logical operations.
Note that there is no unique choice for $G_{\MCC/\MCCd}$ as there are multiple bases for a vector space.
Denote the rows of $H_{\MCC}$ as $\vecnot{g}_i$ for $i=1,2,\ldots,k$ and the rows of $G_{\MCC/\MCCd}$ as $\vecnot{h}_j$ for $j=1,2,\ldots,m-2k$.
Then for some $\vecnot{x} \in \{0,1\}^{m-2k}$, a coset representative $\vecnot{v}$ can be expressed as $\vecnot{v} = \sum_{j=1}^{m-2k} x_j \vecnot{h}_j = \vecnot{x} \cdot G_{\MCC/\MCCd}$.

%\subsection*{Example}
%
%Let $\MCC$ be the $[6,5,2]$ single-parity check code with $m=6, k=1$ and minimum distance $d=2$.
%The dual code $\MCCd$ is the $[6,1,6]$ repetition code of length $m=6$ which contains only two codewords: $\MCCd = \{ 000000, 111111 \}$.
%So $G_{\MCCd} = H_{\MCC} = [1\ 1\ 1\ 1\ 1\ 1]$.
%This implies $\MCC$ contains all length $6$ binary vectors that contain an even number of $1$s.
%One possible generator matrix for $\MCC$ is
%\begin{equation*}
%G_{\MCC} =
%\begin{bmatrix}
%1 & 1 & 1 & 1 & 1 & 1 \\
%1 & 1 & 0 & 0 & 0 & 0 \\
%1 & 0 & 1 & 0 & 0 & 0 \\
%1 & 0 & 0 & 1 & 0 & 0 \\
%1 & 0 & 0 & 0 & 1 & 0
%\end{bmatrix}
%=
%\begin{bmatrix}
%\vecnot{g}_1 \\
%\vecnot{h}_1 \\
%\vecnot{h}_2 \\
%\vecnot{h}_3 \\
%\vecnot{h}_4
%\end{bmatrix}
%\quad ; \quad
%G_{\MCC/\MCCd} \triangleq
%\begin{bmatrix}
%1 & 1 & 0 & 0 & 0 & 0 \\
%1 & 0 & 1 & 0 & 0 & 0 \\
%1 & 0 & 0 & 1 & 0 & 0 \\
%1 & 0 & 0 & 0 & 1 & 0
%\end{bmatrix} 
% .
%\end{equation*}
%So the group $\MCC/\MCCd$ of $16$ coset representatives is generated by binary linear combinations of the rows of $G_{\MCC/\MCCd}$.

\subsection{Construction of the CSS Code}
\label{sec:css_construct}

Given a classical $[m,m-k]$ binary self-orthogonal code $\MCC$ (i.e., $\MCC$ contains its dual $\MCCd$), the CSS quantum code $\MCQ$ is constructed as follows.
Let $\vecnot{v} \in \mathbb{F}_2^m$ be a length-$m$ binary vector.
The quantum state corresponding to this vector is defined as
\begin{equation}
\ket{\psi_v} \equiv \ket{\vecnot{v} + \MCCd} \triangleq \frac{1}{\sqrt{|\MCCd|}} \sum_{\vecnot{c} \in \MCCd} \ket{\vecnot{c} + \vecnot{v}} ,
\end{equation}
where $\vecnot{c} + \vecnot{v} = \vecnot{c} \oplus \vecnot{v}$ is the component-wise modulo-$2$ addition of vectors.
Note that the vectors $\vecnot{c} + \vecnot{v}$ for all $\vecnot{c} \in \MCCd$ generate the coset $\vecnot{v} + \MCCd$ and hence the notation for the quantum state.

The CSS code $\MCQ$ is defined as the collection of all such \emph{distinct} quantum states generated by the coset representatives $\vecnot{v} \in \MCC/\MCCd$.
As $|\MCC| = 2^{m-k}$ and $|\MCCd| = 2^k$, by Lagrange's theorem we have $|\MCC/\MCCd| = 2^{m-2k}$ and so the (binary) dimension of $\MCC/\MCCd$ is $m-2k$.
Since each bit of $\vecnot{v}$ corresponds to a qubit of $\ket{\psi_v}$, which has dimension $2$, the dimension of the quantum code $\MCQ$ is $2^{m-2k}$.
Formally, we write $\MCQ$ as an $[[m,m-2k]]$ CSS quantum code.

Now recall that a coset representative can be expressed as $\vecnot{v} = \vecnot{x} \cdot G_{\MCC/\MCCd}$.
So if we have an $(m-2k)$-qubit state $\ket{\vecnot{x}}_L = \ket{x_1}_L \otimes \cdots \otimes \ket{x_{m-2k}}_L$, called the \emph{logical} state, then the CSS code will encode this into the quantum state $\ket{\psi_x}$, where
\begin{equation}
\label{eq:css_state2}
\ket{\vecnot{x}}_L \leftrightarrow \ket{\psi_x} \equiv \ket{\vecnot{x} \cdot G_{\MCC/\MCCd} + \MCCd} \triangleq \frac{1}{\sqrt{|\MCCd|}} \sum_{\vecnot{c} \in \MCCd} \ket{\vecnot{c} + \vecnot{x} \cdot G_{\MCC/\MCCd}} = \frac{1}{\sqrt{|\MCCd|}} \sum_{\vecnot{c} \in \MCCd} \ket{\vecnot{c} + \sum_{j=1}^{m-2k} x_j \vecnot{h}_j} ,
\end{equation}
where $\vecnot{h}_j$ is the $j$-th row of $G_{\MCC/\MCCd}$.
As mentioned before, the logical state $\ket{\vecnot{x}}_L$ is also called the \emph{encoded} state and its $(m-2k)$ component qubits are called \emph{encoded} qubits.

%\subsection*{Example}
%
%Consider again our example of the $[6,5]$ single-parity check code with $m=6,k=1$.
%The CSS construction gives us a $[[6,4]]$ quantum code $\MCQ$ with the states defined as
%\[ \ket{\psi_x} = \frac{1}{\sqrt{2}} \ket{(000000) + \sum_{i=1}^{4} x_i \vecnot{h}_i} + \frac{1}{\sqrt{2}} \ket{(111111) + \sum_{i=1}^{4} x_i \vecnot{h}_i} \ ; \ \vecnot{x} \in \{0,1\}^4 . \]

\subsection{Stabilizer for the CSS Code}
\label{sec:css_stabilizer}

Consider an $[[m,m-2k]]$ CSS code $\MCQ$ defined using an $[m,m-k]$ classical binary code $\MCC$ that contains its dual $\MCCd$.
We will now demonstrate that it is indeed a stabilizer code and give the set of generators for its stabilizer.
Particularly, if we can find commuting Hermitian operators $\bg_1,\bg_2,\ldots,\bg_{2k} \in HW_N$ such that they do not generate $-I_N$ and satisfy $\bg_i \ket{\psi_v} = \ket{\psi_v} \ \forall \ \ket{\psi_v} \in \MCQ, i=1,2,\ldots,2k$ then we have defined the stabilizer of $\MCQ$.

Consider the generator matrix representation for $\MCC$ given in~\eqref{eq:gmatrix}.
Again, denote the rows of $H_{\MCC}$ as $\vecnot{g}_1, \vecnot{g}_2, \ldots, \vecnot{g}_k$.
Then for $i \in \{1,\ldots,k\}$ we have $\vecnot{g}_i \cdot \vecnot{v} = 0$ for all $\vecnot{v} \in \MCC$ and particularly for all $\vecnot{v} \in \MCC/\MCCd$, which are the vectors that define the states in $\MCQ$.
Denote the elements of the vector  $\vecnot{g}_i$ as $g_{it}$ so that $\vecnot{g}_i = [g_{i1},\ g_{i2},\ \ldots,\ g_{im}]$.
Now define the $2k$ operators 
\begin{equation}
\label{eq:css_generators}
\bg_i^X \triangleq D(\vecnot{g}_i, \vecnot{0}) = \bigotimes_{t=1}^{m} X^{g_{it}} , \quad \bg_i^Z \triangleq D(\vecnot{0}, \vecnot{g}_i) = \bigotimes_{t=1}^{m} Z^{g_{it}} \ ; \ i=1,2,\ldots,k .
\end{equation}
%where $X_t$ and $Z_t$ denote the $X$ and $Z$ operators, respectively, on the $t$-th physical qubit.

\begin{theorem}
\label{thm:css_generators}
%\begin{shaded}
The set of $2k$ $m$-qubit operators $\{ \bg_i^X, \bg_i^Z \}$ defined in~\eqref{eq:css_generators} commute with each other and do not generate $-I_N$.
%\end{shaded}
\begin{IEEEproof}
See Appendix~\ref{proof:css_generators}.
\end{IEEEproof}
\end{theorem}
Therefore these operators generate a valid stabilizer $S$ for some subspace $V(S)$ of $m$ qubits.
We are left only to verify that $V(S) = \MCQ$.

\begin{theorem}
\label{thm:css_stabilizer}
%\begin{shaded}
The set of $2k$ $m$-qubit operators $\{ \bg_i^X, \bg_i^Z \}$ defined in~\eqref{eq:css_generators} generate the stabilizer for the CSS code $\MCQ$.
%\end{shaded}
\begin{IEEEproof}
See Appendix~\ref{proof:css_stabilizer}.
\end{IEEEproof}
\end{theorem}

%\subsection*{Example}
%
%For our running example of the $[[6,4]]$ CSS code, we have $k=1$ and $\vecnot{g}_1 = [1\ 1\ 1\ 1\ 1\ 1]$.
%This gives the stabilizers
%\[ \bg_1^X \triangleq X_1 X_2 X_3 X_4 X_5 X_6 = X^{\otimes 6} \quad , \quad \bg_1^Z \triangleq Z_1 Z_2 Z_3 Z_4 Z_5 Z_6 = Z^{\otimes 6} . \]
%Clearly we have 
%\[ \bg_1^X \bg_1^Z = X^{\otimes 6} Z^{\otimes 6} = (XZ)^{\otimes 6} = (-1)^6 (ZX)^{\otimes 6} = Z^{\otimes 6} X^{\otimes 6} = \bg_1^Z \bg_1^X . \]

\subsection{Logical Pauli Operators for the CSS Code}
\label{sec:css_ops}

We will now define the (physical realizations of) logical Pauli operators for each of the $(m-2k)$ logical qubits encoded by the CSS code $\MCQ$.
Let us now reiterate the representation of the generator matrix for the code $\MCC$ from~\eqref{eq:gmatrix}:
\begin{equation*}
G_{\MCC} = 
\begin{bmatrix}
H_{\MCC} \\
G_{\MCC/\MCCd}
\end{bmatrix}_{(m-k) \times m}
=
\begin{bmatrix}
G_{\MCCd} \\
G_{\MCC/\MCCd}
\end{bmatrix}_{(m-k) \times m} ,
\end{equation*}
where $H_{\MCC} = G_{\MCCd}$ is a $k \times m$ matrix and $G_{\MCC/\MCCd}$ is an $(m-2k) \times m$ matrix.
The $(m-2k)$ logical Pauli operators $\lX_j, \lZ_j, j \in \{1,\ldots,m-2k\}$ are defined from the rows of the generator matrix for $\MCC/\MCCd$, represented above as $G_{\MCC/\MCCd}$.
These logical operators need to satisfy the (anti-)commutation conditions 
\begin{equation} 
\label{eq:logical_commute}
\lX_i \lZ_j = 
\begin{cases}
- \lZ_j \lX_i & {\rm if} \ i=j, \\
\lZ_j \lX_i & {\rm if} \ i \neq j 
\end{cases} . 
\end{equation}
Hence for a general CSS code we will need two generator matrices for $\MCC/\MCCd$ which we represent as $G_{\MCC/\MCCd}^X, G_{\MCC/\MCCd}^Z$ because they will be used to define the logical $X$ and logical $Z$ operators respectively.
Denote the rows of $G_{\MCC/\MCCd}^X$ as $\vecnot{h}_1,\ldots,\vecnot{h}_{m-2k}$ and the rows of $G_{\MCC/\MCCd}^Z$ as $\vecnot{h}_1',\ldots,\vecnot{h}_{m-2k}'$.
The entries of $\vecnot{h}_j$ are denoted as $h_{j1},\ldots,h_{jm}$ and similarly the entries of $\vecnot{h}_j'$ are denoted as $h_{j1}',\ldots,h_{jm}'$.
Then define the $m$-qubit operators
\begin{equation}
\label{eq:css_pauli2}
\lX_j \triangleq D(\vecnot{h}_j, \vecnot{0}) = \bigotimes_{t=1}^{m} X^{h_{jt}}  , \quad \lZ_j \triangleq D(\vecnot{0}, \vecnot{h}_j') = \bigotimes_{t=1}^{m} Z^{h_{jt}'} , \quad \lY_j \triangleq \iota \lX_j \lZ_j .
\end{equation}
for $j=1,2,\ldots,m-2k$.

\begin{lemma}
\label{lem:logical_Paulis_commute}
%\begin{shaded}
The physical operators defined in~\eqref{eq:css_pauli2} satisfy the commutation relations given in~\eqref{eq:logical_commute} if and only if\\ ${G_{\MCC/\MCCd}^X \left( G_{\MCC/\MCCd}^Z \right)^T = I_{m-2k}}$, where $I_{m-2k}$ is the $(m-2k) \times (m-2k)$ identity matrix.
%\end{shaded}
\begin{IEEEproof}
See Appendix~\ref{proof:logical_Paulis_commute}.
\end{IEEEproof} 
\end{lemma}

We have the following theorem to verify that the operators $\lX_j$ and $\lZ_j$ execute logical bit-flip and phase-flip operations, respectively, by operating on the physical qubits.

\begin{theorem}
\label{thm:logical_Paulis}
%\begin{shaded}
Let $\ket{\vecnot{x}}_L$ be the logical state defined by $\vecnot{x} \in \{0,1\}^{m-2k}$ and let $\ket{\vecnot{x}'}_L$ be the logical state such that $x_i' = x_i \oplus 1$ for some $i \in \{1,\ldots,m-2k\}$ and $x_j' = x_j \ \forall \ j \in \{1,\ldots,m-2k\} \ {\rm s.t.} \ j \neq i$.
Then the operators defined in~\eqref{eq:css_pauli2} satisfy 
\[ \lX_i \ket{\psi_x} = \ket{\psi_{x'}} \quad, \quad \lX_i \ket{\psi_x} = (-1)^{x_i} \ket{\psi_{x}} , \]
where $\ket{\psi_x}$ is the CSS state defined in~\eqref{eq:css_state2}.
%\end{shaded}
\begin{IEEEproof}
See Appendix~\ref{proof:logical_Paulis}.
\end{IEEEproof}
\end{theorem}

The proof of Theorem~\ref{thm:logical_Paulis} requires that $G_{\MCC}^X =
\begin{bmatrix}
H_{\MCC} \\
G_{\MCC/\MCCd}^X
\end{bmatrix}$ and $G_{\MCC}^Z =
\begin{bmatrix}
H_{\MCC} \\
G_{\MCC/\MCCd}^Z
\end{bmatrix}$ form two generator matrices for the classical code $\MCC$.
For every row $\vecnot{h}_j$ of $G_{\MCC/\MCCd}^X$ there exists at least one vector $\vecnot{w} \in \MCC\setminus \MCCd$ such that $\vecnot{h}_j \cdot \vecnot{w} = 1$.
Otherwise, all vectors in $\MCC\setminus \MCCd$ are orthogonal to it and hence $\vecnot{h}_j$ must be in the dual code $\MCCd$ which is a contradiction.
Since the space $\MCC/\MCCd$ of coset representatives has dimension $m-2k$ and the condition in the above lemma imposes $m-2k$ linearly independent constraints on each row of $G_{\MCC/\MCCd}^Z$, there always exists a unique $G_{\MCC/\MCCd}^Z$ for a given $G_{\MCC/\MCCd}^X$.

\begin{theorem}
If $G_{\MCC/\MCCd}^Z$ forms another generator matrix for the space $\MCC/\MCCd$ of coset representatives and $G_{\MCC/\MCCd}^X \left( G_{\MCC/\MCCd}^Z \right)^T = I_{m-2k}$, then the physical operators defined in~\eqref{eq:css_pauli2} are valid logical Pauli operators.
\begin{IEEEproof}
By Lemma~\ref{lem:logical_Paulis_commute} the operators $\lX_j, \lZ_j$ satisfy the necessary commutation relations in~\eqref{eq:logical_commute} for logical Pauli operators.
By Theorem~\ref{thm:logical_Paulis} they execute the action of Pauli operators on the logical qubits.
Finally, we need to verify that these physical operators commute with the elements of the stabilizer of the code.
But this is directly true because $\vecnot{g}_i \cdot \vecnot{h}_j = 0$ and $\vecnot{g}_i \cdot \vecnot{h}_j' = 0 \ \forall \ i \in \{1,\ldots,k\}, j \in \{1,\ldots,m-2k\}$ since $\vecnot{g}_i \in \MCCd$ and $\vecnot{h}_j, \vecnot{h}_j' \in \MCC$.
Hence $\lX_j, \lZ_j$ are valid logical Pauli operators.
\end{IEEEproof}
\end{theorem}

\vspace{0.1cm}

Therefore, we have demonstrated a general construction for the logical Pauli operators of a CSS code constructed from classical binary self-orthogonal codes.
This construction can be suitably extended to more general CSS codes.
More specifically, consider $[m,k_1]$ and $[m,k_2]$ binary linear codes $\MCC_1$ and $\MCC_2$, respectively, that satisfy $\MCC_2 \subset \MCC_1$ and that $\MCC_1$ and $\MCC_2^{\perp}$ correct up to $t$ errors.
It is well known that the rows of the parity-check matrices $H(\MCC_1)$ and $H(\MCC_2^{\perp})$ give the $Z$ and $X$ stabilizers for the quantum code CSS($\MCC_1,\MCC_2$), respectively.
Then, in order to determine the logical Pauli operators, we decompose the generator matrices of $\MCC_1$ and $\MCC_2^{\perp}$ as
\begin{align*}
G_{\MCC_1} = 
\begin{bmatrix}
G_{\MCC_2} \\
G_{\MCC_1/\MCC_2}
\end{bmatrix}_{k_1 \times m}
\qquad \text{and} \qquad
G_{\MCC_2^{\perp}} = 
\begin{bmatrix}
G_{\MCC_1^{\perp}} \\
G_{\MCC_2^{\perp}/\MCC_1^{\perp}}
\end{bmatrix}_{(m-k_2) \times m} ,
\end{align*}
where $G_{\MCC_1/\MCC_2}$ and $G_{\MCC_2^{\perp}/\MCC_1^{\perp}}$ are $(k_1-k_2) \times m$ matrices.
We consider these $(k_1-k_2) \times m$ matrices as the equivalents of $G_{\MCC/\MCCd}^X$ and $G_{\MCC/\MCCd}^Z$ above and use their rows to define logical $X$ and logical $Z$ operators, respectively.

\subsection{CSS State Preparation}
\label{sec:css_state_prepare}

As a final note, we reiterate a representation of the CSS state given in~\cite{Chao-arxiv17b} that is potentially useful in calculating the effect of operators on the state.
We first observe that the $X$ stabilizers generated by $\bg_i^X$ satisfy $\bg_i^X \ket{\vecnot{u}} = \ket{\vecnot{u} + \vecnot{g}_i}$ for $i=1,2,\ldots,k$.
The set of all $X$ stabilizers is given by
$S^X = \{ \bg_c^X \ : \ \vecnot{c} \in \MCCd \} \ ; \ \bg_c^X \triangleq D(\vecnot{c}, \vecnot{0}) = \bigotimes_{j=1}^{m} X^{c_{j}} , \ \vecnot{c} = [c_1,\ c_2,\ \ldots, \ c_m]$.
Hence these vectors satisfy $\bg_c^X \ket{0}^{\otimes m} = \bg_c^X \ket{00\ldots 0} = \ket{\vecnot{c}}$.
Therefore, given logical qubits $\ket{\vecnot{x}}_L$ with $\vecnot{x} \in \{0,1\}^{m-2k}$ we can first prepare the physical state $\ket{0}^{\otimes m}$ and then arrive at the desired CSS state $\ket{\psi_x}$ as follows:
\begin{align}
\ket{\psi_x} & \triangleq \frac{1}{\sqrt{|\MCCd|}} \sum_{\vecnot{c} \in \MCCd} \ket{\vecnot{c} + \sum_{j=1}^{m-2k} x_j \vecnot{h}_j} \nonumber \\
  & = \frac{1}{\sqrt{|\MCCd|}} \sum_{\vecnot{c} \in \MCCd} \prod_{j=1}^{m-2k} \lX_j^{x_j} \ket{\vecnot{c}} \nonumber \\
%
%  & = \frac{1}{\sqrt{|\MCCd|}} \prod_{j=1}^{m-2k} \lX_j^{x_j} \sum_{\vecnot{c} \in \MCCd} \ket{\vecnot{c}} \nonumber \\
%
  & = \frac{1}{\sqrt{|\MCCd|}} \prod_{j=1}^{m-2k} \lX_j^{x_j} \sum_{\vecnot{c} \in \MCCd} \bg_c^X \ket{0}^{\otimes m} \nonumber \\
  & = \prod_{j=1}^{m-2k} \lX_j^{x_j} \frac{1}{\sqrt{|\MCCd|}} \sum_{\bg \in S^X} \bg \ket{0}^{\otimes m} .
\end{align}
%\begin{align}
%% \ket{\psi_x} & \triangleq \frac{1}{\sqrt{|\MCCd|}} \sum_{\vecnot{c} \in \MCCd} \ket{\vecnot{c} + \sum_{j=1}^{m-2k} x_j \vecnot{h}_j} %
%\ket{\psi_x} = \frac{1}{\sqrt{|\MCCd|}} \sum_{\vecnot{c} \in \MCCd} \prod_{j=1}^{m-2k} \lX_j^{x_j} \ket{\vecnot{c}} %  
%%  = \frac{1}{\sqrt{|\MCCd|}} \prod_{j=1}^{m-2k} \lX_j^{x_j} \sum_{\vecnot{c} \in \MCCd} \ket{\vecnot{c}} %
%  = \frac{1}{\sqrt{|\MCCd|}} \prod_{j=1}^{m-2k} \lX_j^{x_j} \sum_{\vecnot{c} \in \MCCd} \bg_c^X \ket{0}^{\otimes m} %
%  = \frac{1}{\sqrt{|\MCCd|}} \prod_{j=1}^{m-2k} \lX_j^{x_j} \sum_{\bg \in S^X} \bg \ket{0}^{\otimes m} .
%\end{align}
Note that this perspective requires us to apply \emph{all} stabilizer elements to the state $\ket{0}^{\otimes m}$, which can be impractical.
However, this representation of the CSS state could be potentially useful for arguing about the effects of operators applied externally to a CSS state.
One such use (based on the first three equalities above) can be observed in the argument for $\lZ_j$ in the proof of Theorem~\ref{thm:logical_Paulis}.
An application of the final expression can be found in an important claim proven in~\cite[Claim 2]{Chao-arxiv17b}.

\section{Conclusion}
\label{sec:conclusion}

In this work we have used symplectic geometry to propose a systematic algorithm for synthesizing physical implementations of logical Clifford operators for any stabilizer code.
This algorithm provides as a solution all symplectic matrices corresponding to the desired logical operator, each of which is subsequently transformed into a circuit by decomposing it into elementary forms.
%If the solution corresponds to one of the elementary forms in Table~\ref{tab:std_symp} then the physical operator (i.e., circuit) is readily obtained.
%Otherwise an algorithm, such as~\cite{Can-2017a}, is used to decompose the solution as a product of elementary symplectic forms.
%However, this decomposition is neither unique nor necessarily efficient.
%In our future work we will consider an optimization problem where we search for efficient physical operators given a symplectic matrix.
This decomposition is not unique, and in future work we will address optimization of the synthesis algorithm with respect to circuit complexity and fault-tolerance.

\section*{Acknowledgment}

The authors would like to thank Jungsang Kim and Jianfeng Lu for helpful discussions.
%We particularly thank Jungsang Kim for pointing us to the work of Chao and Reichardt.
S. Kadhe would like to thank Alex Sprintson for his continued support, and Robert Calderbank for his hospitality during S. Kadhe's visit to Duke University.

\bibliographystyle{ieeetr}
%\bibliography{WCLabrv,WCLnewbib}

\iffull

%\newpage

%\thispagestyle{empty}

\appendices

\section{Elementary Symplectic Transformations and their Circuits}
\label{sec:elem_symp}

In this section we verify that the physical operators listed in Table~\ref{tab:std_symp} are associated with the corresponding symplectic transformation~\cite{Can-2017a}.
Furthermore, we also provide circuits that realize these physical operators (also see~\cite{Dehaene-physreva03}).

Since each physical operator in Table~\ref{tab:std_symp} is a unitary Clifford operator, it is enough to consider their actions on elements of the Heisenberg-Weyl group $HW_N$, where $N=2^m$.
Let $e_v$ be a standard basis (column) vector in $\mathbb{C}^N$ indexed by the vector $v \in \mathbb{F}_2^m$ such that it has entry $1$ in position $v$ and $0$ elsewhere.
More precisely, if $v = [v_1,v_2,\ldots,v_m]$ then $e_v = e_{v_1} \otimes e_{v_2} \otimes \cdots e_{v_m}$, where $e_0 \triangleq \begin{bmatrix} 1 \\ 0 \end{bmatrix} = \ket{0}, e_1 \triangleq \begin{bmatrix} 0 \\ 1 \end{bmatrix} = \ket{1}$.
Hence we can simply write $e_v = \ket{v} = \ket{v_1} \otimes \cdots \otimes \ket{v_m}$.

\begin{enumerate}

\item $H_N = H^{\otimes m} \colon$ The single-qubit Hadamard operator $H \triangleq \frac{1}{\sqrt{2}} \begin{bmatrix}
1 & 1 \\
1 & -1
\end{bmatrix}$ satisfies $H X H^{\dagger} = Z, H Z H^{\dagger} = X$. 
Hence the action of $H_N$ on a $HW_N$ element $D(a,b)$ is given by
\begin{align*}
H_N D(a,b) H_N^{\dagger} = H_N D(a,0) D(0,b) H_N^{\dagger} &= (H_N D(a,0) H_N^{\dagger}) (H_N D(0,b) H_N^{\dagger}) = D(0,a) D(b,0) = (-1)^{ab^T} D(b,a) \\
\Rightarrow  H_N D(a,b) H_N^{\dagger} &= (-1)^{ab^T} D \left( [a,b] \Omega \right) \ , \ {\rm where}\ \Omega = \begin{bmatrix}
0 & I_m \\
I_m & 0
\end{bmatrix} .
\end{align*}
The circuit for $H_N$ is just $H$ applied to each of the $m$ qubits.

\item $GL(m,\mathbb{F}_2) \colon$ Each non-singular $m \times m$ binary matrix $Q$ is associated with a symplectic transformation $A_Q$ given by 
\begin{align*}
A_Q = 
\begin{bmatrix} 
Q & 0 \\ 
0 & Q^{-T} 
\end{bmatrix} ,
\end{align*}
where $Q^{-T} = (Q^T)^{-1} = (Q^{-1})^T$.
The matrix $Q$ is also associated with the unitary operator $a_Q$ which realizes the mapping $e_v \mapsto e_{vQ}$.
We verify this as follows.
Note that $D(c,0) e_v = e_{v+c}$ and $D(0,d) e_v = (-1)^{vd^T} e_v$.
\begin{align*}
(a_Q D(c,d) a_Q^{\dagger}) e_v & = a_Q D(c,0) D(0,d) e_{vQ^{-1}} \\
                               & = a_Q (-1)^{cd^T} D(0,d) D(c,0) e_{vQ^{-1}} \\
                               & = (-1)^{cd^T} a_Q (-1)^{(vQ^{-1}+c) d^T} e_{vQ^{-1}+c} \\
                               & = (-1)^{cd^T} (-1)^{(v+cQ) Q^{-1}d^T} e_{v+cQ} \\
                               & = (-1)^{cd^T} D(0,d(Q^{-1})^T) D(cQ,0) e_v \\
                               & = D(cQ,dQ^{-T}) e_v \\
                               & = D \left( [c,d] A_Q \right) e_v .
\end{align*}
Since the operator $a_Q$ realizes the map $\ket{v} \mapsto \ket{vQ}$, the circuit for the operator is equivalent to the binary circuit that realizes $v \mapsto vQ$.
This is the scenario encountered in Section~\ref{sec:css_cnot}.
Evidently, this elementary transformation encompasses CNOT operations and qubit permutations.
For the latter, $Q$ will be a permutation matrix.
Note that if $a_Q$ preserves the code space of a CSS code then the respective permutation must be in the automorphism group of the constituent classical code.
This is the special case that is discussed in detail by Grassl and Roetteler in~\cite{Grassl-isit13}.

\vspace{0.1cm}
For a general $Q$, one can use the LU decomposition over $\mathbb{F}_2$ to obtain $P_{\pi} Q = LU$, where $P_{\pi}$ is a permutation matrix, $L$ is lower triangular and $U$ is upper triangular.
Note that $L_{ii} = U_{ii} = 1 \ \forall \ i \in \{1,\ldots,m\}$.
Then the circuit for $Q$ first involves the permutation $P_{\pi}^T$ (or $\pi^{-1}$), then CNOTs for $L$ with control qubits in the order $1,2,\ldots,m$ and then CNOTs for $U$ with control qubits in reverse order $m,m-1,\ldots,1$.
The order is important because an entry $L_{ji} = 1$ implies a CNOT gate with qubit $j$ controlling qubit $i$ (with $j > i$), i.e, $\cnot{j}{i}$, and similarly $L_{kj} = 1$ implies the gate $\cnot{k}{j}$ (with $k > j$).
Since the gate $\cnot{j}{i}$ requires the value of qubit $j$ \emph{before} it is altered by $\cnot{k}{j}$, it needs to be implemented first.
A similar reasoning applies to the reverse order of control qubits for $U$.

\item $t_R = {\rm diag}\left( \iota^{vRv^T} \right) \colon$ Each symmetric matrix $R \in \mathbb{F}_2^{m \times m}$ is associated with a symplectic transformation $T_R$ given by
\begin{align*}
T_R = 
\begin{bmatrix} 
I_m & R \\ 
0 & I_m 
\end{bmatrix} ,
\end{align*}
and with a unitary operator $t_R$ that realizes the map $e_v \mapsto \iota^{vRv^T} e_v$.
%Here the exponent $vRv^T$ is calculated using modulo $4$ arithmetic.
%For example, if 
%$R = 
%\begin{bmatrix}
%0 & 1 & 0 \\
%1 & 0 & 1 \\
%0 & 1 & 1
%\end{bmatrix}
%$ then $\iota^{vRv^T} = -\iota$ for $v = [0,1,1]$.
We now verify that conjugation by $t_R$ induces the symplectic transformation $T_R$.
\begin{align*}
(t_R D(a,b) t_R^{\dagger}) e_v & = \iota^{-vRv^T} t_R (-1)^{ab^T} D(0,b) D(a,0) e_v \\
                               & = \iota^{-vRv^T} (-1)^{ab^T} t_R (-1)^{(v+a)b^T} e_{v+a} \\
                               & = (-1)^{ab^T} \iota^{-vRv^T} (-1)^{(v+a)b^T} \iota^{(v+a)R(v+a)^T} e_{v+a} \\
                               & = (-1)^{ab^T} \iota^{aRa^T} (-1)^{vRa^T + (v+a)b^T} e_{v+a} \\
                               & = (-1)^{ab^T} \iota^{-aRa^T} (-1)^{(v+a)(b+aR)^T} e_{v+a} \\
                               & = (-1)^{ab^T} \iota^{-aRa^T}  D(0,b+aR) D(a,0) e_v \\
                               & = (-1)^{ab^T} \iota^{-aRa^T} (-1)^{a(b+aR)^T} D(a,b+aR) e_v \\
                               & = \iota^{aRa^T} D \left( [a,b] T_R \right) e_v .
\end{align*}
Hence, for $E(a,b) \triangleq \iota^{ab^T} D(a,b)$, we have $t_R E(a,b) t_R^{\dagger} = \iota^{ab^T} \iota^{aRa^T} D(a,b+aR) = E \left( [a,b] T_R \right)$ as required.
%Note that the operator on the right hand side also squares to $I_N$ (since $D(a,b+aR)^2 = (-1)^{a(b+aR)^T} I_N$) as implied by~\eqref{eq:symp_action}.
We derive the circuit for this unitary operator by observing the action of $T_R$ on the standard basis vectors $[\vecnot{e}_1,\vecnot{0}],\ldots,[\vecnot{e}_m,\vecnot{0}]$, $[\vecnot{0},\vecnot{e}_1],\ldots,[\vecnot{0},\vecnot{e}_m]$ of $\mathbb{F}_2^{2m}$, where $i \in \{1,\ldots,m\}$, which captures the effect of $t_R$ on the (basis) elements $X_1,\ldots,X_m$, $Z_1,\ldots,Z_m$ of $HW_N$, respectively, under conjugation.

\vspace{0.1cm}
Assume as the first special case that $R$ has non-zero entries only in its (main) diagonal.
If $R_{ii} = 1$ then we have $[\vecnot{e}_i,\vecnot{0}] T_R = [\vecnot{e}_i, \vecnot{e}_i]$.
This indicates that $t_R$ maps $X_i \mapsto X_i Z_i \approx Y_i$.
Since we know that the phase gate $P_i$ on the $i$-th qubit performs exactly this map under conjugation, we conclude that the circuit for $t_R$ involves $P_i$.
We proceed similarly for every $i \in \{1,\ldots,m\}$ such that $R_{ii} = 1$.

\vspace{0.1cm}
Now consider the case where $R_{ij} = R_{ji} = 1$ (since $R$ is symmetric).
Then we have ${[\vecnot{e}_i,\vecnot{0}] T_R = [\vecnot{e}_i,\vecnot{e}_j], \ [\vecnot{e}_j,\vecnot{0}] T_R = [\vecnot{e}_j,\vecnot{e}_i]}$.
This indicates that $t_R$ maps $X_i \mapsto X_i Z_j$ and $X_j \mapsto Z_i X_j$.
Since we know that the controlled-$Z$ gate ${\rm CZ}_{ij}$ on qubits $(i,j)$ performs exactly this map under conjugation, we conclude that the circuit for $t_R$ involves ${\rm CZ}_{ij}$.
We proceed similarly for every pair $(i,j)$ such that $R_{ij} = R_{ji} = 1$.

\vspace{0.1cm}
Finally, we note that the symplectic transformation associated with the operator $H_N t_R H_N$ is 
$\Omega\, T_R\, \Omega = \begin{bmatrix} 
I_m & 0 \\ 
R & I_m 
\end{bmatrix}$.

\item $g_k = H_{2^k} \otimes I_{2^{m-k}} \colon$ Since $H_{2^k}$ is the $k$-fold Kronecker product of $H$ and since $D(a,b) = X^{a_1} Z^{b_1} \otimes \cdots \otimes X^{a_m} Z^{b_m}$ we have
\begin{align*}
g_k D(a,b) g_k^{\dagger} & = \left( Z^{a_1} X^{b_1} \otimes \cdots \otimes Z^{a_k} X^{b_k} \right) \otimes \left( X^{a_{k+1}} Z^{b_{k+1}} \otimes \cdots \otimes X^{a_m} Z^{b_m} \right) \\
  & = \left( (-1)^{a_1 b_1} X^{b_1} Z^{a_1} \otimes \cdots \otimes (-1)^{a_k b_k} X^{b_k} Z^{a_k} \right) \otimes \left( X^{a_{k+1}} Z^{b_{k+1}} \otimes \cdots \otimes X^{a_m} Z^{b_m} \right) .
\end{align*}
We write $(a,b) = (\hat{a} \bar{a}, \hat{b} \bar{b})$ where $\hat{a} \triangleq a_1 \cdots a_k, \bar{a} \triangleq a_{k+1} \cdots a_m, \ \hat{b} \triangleq b_1 \cdots b_k, \bar{b} \triangleq b_{k+1} \cdots b_m$.
Then we have 
\begin{align*}
g_k D(\hat{a} \bar{a}, \hat{b} \bar{b}) g_k^{\dagger} = (-1)^{\hat{a} \hat{b}^T} D(\hat{b} \bar{a}, \hat{a} \bar{b}) = (-1)^{\hat{a} \hat{b}^T} D \left( [\hat{a} \bar{a}, \hat{b} \bar{b}] G_k \right) , \ {\rm where} \ 
G_k = 
%\left[
\begin{bmatrix}
0 & 0 & I_k & 0 \\
0 & I_{m-k} & 0 & 0 \\
%\hline
I_k & 0 & 0 & 0 \\
0 & 0 & 0 & I_{m-k}
\end{bmatrix} .
%\right] 
\end{align*}
Defining $U_k \triangleq \begin{bmatrix} I_k & 0 \\ 0 & 0 \end{bmatrix}, L_{m-k} \triangleq \begin{bmatrix} 0 & 0 \\ 0 & I_{m-k} \end{bmatrix}$, we then write $G_k = 
\begin{bmatrix} 
L_{m-k} & U_k \\ 
U_k & L_{m-k} 
\end{bmatrix}$.
Similar to part 1 above, the circuit for $g_k$ is simply $H$ applied to each of the first $k$ qubits.
Although this is a special case where the Hadamard operator was applied to consecutive qubits, we note that the symplectic transformation for Hadamards applied to arbitrary non-consecutive qubits can be derived in a similar fashion.

\end{enumerate}
Hence we have demonstrated the elementary symplectic transformations in $\text{Sp}(2m,\mathbb{F}_2)$ that are associated with arbitrary Hadamard, Phase, Controlled-$Z$ and Controlled-NOT gates.
Since we know that these gates, along with $HW_N$, generate the full Clifford group~\cite{Gottesman-arxiv09}, these elementary symplectic transformations form a universal set corresponding to physical operators in the Clifford group.

\subsection{Proof of Theorem~\ref{thm:Trung}}
\label{sec:trung_proof}

Let $F = \begin{bmatrix}
A & B \\
C & D
\end{bmatrix}$ so that $\begin{bmatrix} A & B \end{bmatrix} \Omega \begin{bmatrix} A & B \end{bmatrix}^T = 0$ and $\begin{bmatrix} C & D \end{bmatrix} \Omega \begin{bmatrix} C & D \end{bmatrix}^T = 0$ since $F \Omega F^T = \Omega$.
We will perform a sequence of row and column operations to transform $F$ into the form $\Omega \, T_{R_1} \Omega$ for some symmetric $R_1$.
If rank$(A) = k$ then there exists a row transformation $Q_{11}^{-1}$ and a column transformation $Q_2^{-1}$ such that
\begin{align*}
Q_{11}^{-1} A Q_2^{-1} = \begin{bmatrix}
I_k & 0 \\
0 & 0
\end{bmatrix} .
\end{align*}
Using the notation for elementary symplectic transformations discussed above, we apply $Q_{11}^{-1}$ and $A_{Q_2^{-1}}$ to $\begin{bmatrix} A & B \end{bmatrix}$ and obtain
\begin{align*}
\begin{bmatrix} Q_{11}^{-1} A & Q_{11}^{-1} B \end{bmatrix} \begin{bmatrix} Q_2^{-1} & 0 \\ 0 & Q_2^T \end{bmatrix} =
\left[ \begin{array}{cc|cc}
I_k & 0 & R_k & E' \\
0 & 0 & E & B_{m-k}
\end{array} \right] \triangleq \begin{bmatrix} A' & B' \end{bmatrix} ,
\end{align*}
where $B_{m-k}$ is an $(m-k) \times (m-k)$ matrix.
Since the above result is again the top half of a symplectic matrix, we have $\begin{bmatrix} A' & B' \end{bmatrix} \Omega \begin{bmatrix} A' & B' \end{bmatrix}^T = 0$ which implies $R_k$ is symmetric, $E = 0$ and hence rank$(B_{m-k}) = m-k$.
Therefore we determine an invertible matrix $Q_{m-k}$ which transforms $B_{m-k}$ to $I_{m-k}$ under row operations.
Then we apply $Q_{12}^{-1} \triangleq \begin{bmatrix}
I_k & 0 \\
0 & Q_{m-k}
\end{bmatrix}$ on the left of the matrix $\begin{bmatrix} A' & B' \end{bmatrix}$ to obtain
\begin{align*}
\begin{bmatrix} Q_{12}^{-1} Q_{11}^{-1} A & Q_{12}^{-1} Q_{11}^{-1} B \end{bmatrix} \begin{bmatrix} Q_2^{-1} & 0 \\ 0 & Q_2^T \end{bmatrix} =
\left[ \begin{array}{cc|cc}
I_k & 0 & R_k & E' \\
0 & 0 & 0 & I_{m-k}
\end{array} \right] .
\end{align*}
Now we observe that we can apply row operations to this matrix and transform $E'$ to $0$.
We left multiply by $Q_{13}^{-1} \triangleq \begin{bmatrix}
I_k & E' \\
0 & I_{m-k}
\end{bmatrix}$ to obtain
\begin{align*}
\begin{bmatrix} Q_{13}^{-1} Q_{12}^{-1} Q_{11}^{-1} A & Q_{13}^{-1} Q_{12}^{-1} Q_{11}^{-1} B \end{bmatrix} \begin{bmatrix} Q_2^{-1} & 0 \\ 0 & Q_2^T \end{bmatrix} =
\left[ \begin{array}{cc|cc}
I_k & 0 & R_k & 0 \\
0 & 0 & 0 & I_{m-k}
\end{array} \right] .
\end{align*}
Since the matrix $R_2 \triangleq \begin{bmatrix}
R_k & 0 \\
0 & 0
\end{bmatrix}$ is symmetric, we apply the elementary transformation $T_{R_2}$ from the right to obtain
\begin{align*}
\left[ \begin{array}{cc|cc}
I_k & 0 & R_k & 0 \\
0 & 0 & 0 & I_{m-k}
\end{array} \right] 
\left[ \begin{array}{cc|cc}
I_k & 0 & R_k & 0 \\
0 & I_{m-k} & 0 & 0 \\
\hline
0 & 0 & I_k & 0 \\
0 & 0 & 0 & I_{m-k}
\end{array} \right] 
=
\left[ \begin{array}{cc|cc}
I_k & 0 & 0 & 0 \\
0 & 0 & 0 & I_{m-k}
\end{array} \right] .
\end{align*}
Finally we apply the elementary transformation $G_k \Omega = \begin{bmatrix} 
U_k & L_{m-k} \\
L_{m-k} & U_k  
\end{bmatrix}$ to obtain
\begin{align*}
\left[ \begin{array}{cc|cc}
I_k & 0 & 0 & 0 \\
0 & 0 & 0 & I_{m-k}
\end{array} \right]
\left[ \begin{array}{cc|cc}
I_k & 0 & 0 & 0 \\
0 & 0 & 0 & I_{m-k} \\
\hline
0 & 0 & I_k & 0 \\
0 & I_{m-k} & 0 & 0
\end{array} \right] 
=
\left[ \begin{array}{cc|cc}
I_k & 0 & 0 & 0 \\
0 & I_{m-k} & 0 & 0
\end{array} \right]
=
\begin{bmatrix}
I_m & 0
\end{bmatrix} .
\end{align*}
Hence we have transformed the matrix $F$ to the form $\Omega \, T_{R_1} \Omega = \begin{bmatrix}
I_m & 0 \\
R_1 & I_m
\end{bmatrix}$, i.e. if we define $Q_1^{-1} \triangleq Q_{13}^{-1} Q_{12}^{-1} Q_{11}^{-1}$ then we have
\begin{align*}
A_{Q_{1}^{-1}} F A_{Q_2^{-1}} T_{R_2} G_k \Omega = \Omega \, T_{R_1} \Omega .
\end{align*} 
Rearranging terms and noting that $A_{Q}^{-1} = A_{Q^{-1}},  \Omega^{-1} = \Omega, G_k^{-1} = G_k, T_{R_2}^{-1} = T_{R_2}$ we obtain
\begin{IEEEeqnarray*}{rCl+x*}
F & = & A_{Q_1} \Omega \, T_{R_1} \Omega^2 G_k T_{R_2} A_{Q_2} = A_{Q_1} \Omega \, T_{R_1} G_k T_{R_2} A_{Q_2} . & \IEEEQEDhere
\end{IEEEeqnarray*}

%\vfill
%\newpage

\section{{MATLAB\textsuperscript{\textregistered}} Code for Algorithm~\ref{alg:symp_lineq_all}}
\label{sec:alg2_matlab}

\begin{lstlisting}
function F_all = find_all_symp_mat(U, V, I, J)

I = I(:)';
J = J(:)';
Ibar = setdiff(1:m,I);
Jbar = setdiff(1:m,J);
alpha = length(Ibar) + length(Jbar); 
tot = 2^(alpha*(alpha+1)/2);
F_all = cell(tot,1);

% Find one solution using symplectic transvections (Algorithm 1)
F0 = find_symp_mat(U([I, m+J], :), V);

A = mod(U * F0, 2);
Ainv = gf2matinv(A);  
IbJb = union(Ibar,Jbar);
Basis = A([IbJb, m+IbJb],:);  % these rows span the subspace W^{\perp} in Theorem 23
Subspace = mod(de2bi((0:2^(2*length(IbJb))-1)',2*length(IbJb)) * Basis, 2);

% Collect indices of free vectors in the top and bottom halves of Basis
% Note: these are now row indices of Basis, not row indices of A!!
[~, Basis_fixed_I, ~] = intersect(IbJb,I);  % intersect(IbJb,I) = intersect(I,Jbar)
[~, Basis_fixed_J, ~] = intersect(IbJb,J);  % intersect(IbJb,J) = intersect(Ibar,J)
Basis_fixed = [Basis_fixed_I, length(IbJb) + Basis_fixed_J];
Basis_free = setdiff(1:2*length(IbJb), Basis_fixed);

Choices = cell(alpha,1);

% Calculate all choices for each free vector using just conditions imposed
% by the fixed vectors in Basis (or equivalently in A)
for i = 1:alpha
    ind = Basis_free(i);
    h = zeros(1,length(Basis_fixed));    
    % Impose symplectic inner product of 1 with the "fixed" symplectic pair
    if (i <= length(Ibar))
        h(Basis_fixed == length(IbJb) + ind) = 1;
    else
        h(Basis_fixed == ind - length(IbJb)) = 1;
    end    
    % Check the necessary conditions on the symplectic inner products
    Innpdts = mod(Subspace * fftshift(Basis(Basis_fixed,:), 2)', 2);
    Choices{i,1} = Subspace(bi2de(Innpdts) == bi2de(h), :);
end

% First free vector has 2^(alpha) choices, second has 2^(alpha-1) choices and so on
for l = 0:(tot - 1)
    Bl = A;
    W = zeros(alpha,2*m);   % Rows are choices made for free vectors
                                 % W(i,:) corresponds to Basis(Basis_free(i),:)
    lbin = de2bi(l,alpha*(alpha+1)/2,'left-msb');
    v1_ind = bi2de(lbin(1,1:alpha),'left-msb') + 1;
    W(1,:) = Choices{1,1}(v1_ind,:);
    for i = 2:alpha
        % vi_ind loops through the 2^(alpha-(i-1)) valid choices for the i-th free vector
        vi_ind = bi2de(lbin(1,sum(alpha:-1:alpha-(i-2)) + (1:(alpha-(i-1)))),'left-msb') + 1;
        Innprods = mod(Choices{i,1} * fftshift(W,2)', 2);        
        % Impose symplectic inner product of 0 with chosen free vectors
        h = zeros(1,alpha);
        % Handle case when Basis contains a symplectic pair of free vectors
        if (i > length(Ibar))
            h(Basis_free == Basis_free(i) - length(IbJb)) = 1;
        end
        % Check the necessary and sufficient conditions on the symplectic inner products
        Ch_i = Choices{i,1}(bi2de(Innprods) == bi2de(h), :);
        W(i,:) = Ch_i(vi_ind,:);  % use the vi_ind-th valid choice for the i-th free vector
    end
    Bl([Ibar, m+Jbar], :) = W;  % replace rows of free vectors with current choices
    F = mod(Ainv * Bl, 2);          % this is the matrix F' in Theorem 23
    F_all{l+1,1} = mod(F0 * F, 2);
end    

end
\end{lstlisting}

\section{Enumeration of All Physical Operators for the $\llbr 6,4,2 \rrbr$ Code}
\label{sec:css642_all_phy_ops}

Using the algorithms described in Section~\ref{sec:discuss} we enumerate all symplectic solutions for each logical operator described in Section~\ref{sec:css642}.
The physical circuits corresponding to these matrices can be obtained by decomposing them into products of elementary symplectic transformations in Table~\ref{tab:std_symp} and using their circuits described in Appendix~\ref{sec:elem_symp} above.
An algorithm for performing this decomposition is given in the proof of Theorem~\ref{thm:Trung} (from~\cite{Can-2017a}).
Note that this decomposition is not unique.
The \texttt{MATLAB\textsuperscript{\textregistered}} programs for reproducing the following results, along with their circuits obtained from the above decomposition, are available at \url{https://github.com/nrenga/symplectic-arxiv18a}.
These programs can perform this task for any stabilizer code.

\vfill
%\newpage

\subsection{Logical Phase Gate $(\lP_1)$}
\label{sec:css_phase_all}

There are $8$ possible symplectic solutions that satisfy the linear constraints imposed by~\eqref{eq:phase_maps} and they are listed below.

\begin{align*}
F_1 = 
\left[
\begin{array}{cccccc|cccccc}
1 & 0 & 0 & 0 & 0 & 0 & 0 & 0 & 0 & 0 & 0 & 0 \\
0 & 1 & 0 & 0 & 0 & 0 & 0 & 1 & 0 & 0 & 0 & 1 \\
0 & 0 & 1 & 0 & 0 & 0 & 0 & 0 & 0 & 0 & 0 & 0 \\
0 & 0 & 0 & 1 & 0 & 0 & 0 & 0 & 0 & 0 & 0 & 0 \\
0 & 0 & 0 & 0 & 1 & 0 & 0 & 0 & 0 & 0 & 0 & 0 \\
0 & 0 & 0 & 0 & 0 & 1 & 0 & 1 & 0 & 0 & 0 & 1 \\
\hline
0 & 0 & 0 & 0 & 0 & 0 & 1 & 0 & 0 & 0 & 0 & 0 \\
0 & 0 & 0 & 0 & 0 & 0 & 0 & 1 & 0 & 0 & 0 & 0 \\
0 & 0 & 0 & 0 & 0 & 0 & 0 & 0 & 1 & 0 & 0 & 0 \\
0 & 0 & 0 & 0 & 0 & 0 & 0 & 0 & 0 & 1 & 0 & 0 \\
0 & 0 & 0 & 0 & 0 & 0 & 0 & 0 & 0 & 0 & 1 & 0 \\
0 & 0 & 0 & 0 & 0 & 0 & 0 & 0 & 0 & 0 & 0 & 1 \\
\end{array}
\right] & ,
F_2 = 
\left[
\begin{array}{cccccc|cccccc}
1 & 0 & 0 & 0 & 0 & 0 & 1 & 1 & 1 & 1 & 1 & 1 \\
0 & 1 & 0 & 0 & 0 & 0 & 1 & 0 & 1 & 1 & 1 & 0 \\
0 & 0 & 1 & 0 & 0 & 0 & 1 & 1 & 1 & 1 & 1 & 1 \\
0 & 0 & 0 & 1 & 0 & 0 & 1 & 1 & 1 & 1 & 1 & 1 \\
0 & 0 & 0 & 0 & 1 & 0 & 1 & 1 & 1 & 1 & 1 & 1 \\
0 & 0 & 0 & 0 & 0 & 1 & 1 & 0 & 1 & 1 & 1 & 0 \\
\hline
0 & 0 & 0 & 0 & 0 & 0 & 1 & 0 & 0 & 0 & 0 & 0 \\
0 & 0 & 0 & 0 & 0 & 0 & 0 & 1 & 0 & 0 & 0 & 0 \\
0 & 0 & 0 & 0 & 0 & 0 & 0 & 0 & 1 & 0 & 0 & 0 \\
0 & 0 & 0 & 0 & 0 & 0 & 0 & 0 & 0 & 1 & 0 & 0 \\
0 & 0 & 0 & 0 & 0 & 0 & 0 & 0 & 0 & 0 & 1 & 0 \\
0 & 0 & 0 & 0 & 0 & 0 & 0 & 0 & 0 & 0 & 0 & 1 \\
\end{array}
\right] , \\~\\
F_3 = 
\left[
\begin{array}{cccccc|cccccc}
1 & 0 & 0 & 0 & 0 & 0 & 0 & 0 & 0 & 0 & 0 & 0 \\
0 & 1 & 0 & 0 & 0 & 0 & 0 & 1 & 0 & 0 & 0 & 1 \\
0 & 0 & 1 & 0 & 0 & 0 & 0 & 0 & 0 & 0 & 0 & 0 \\
0 & 0 & 0 & 1 & 0 & 0 & 0 & 0 & 0 & 0 & 0 & 0 \\
0 & 0 & 0 & 0 & 1 & 0 & 0 & 0 & 0 & 0 & 0 & 0 \\
0 & 0 & 0 & 0 & 0 & 1 & 0 & 1 & 0 & 0 & 0 & 1 \\
\hline
1 & 1 & 1 & 1 & 1 & 1 & 1 & 0 & 0 & 0 & 0 & 0 \\
1 & 1 & 1 & 1 & 1 & 1 & 0 & 1 & 0 & 0 & 0 & 0 \\
1 & 1 & 1 & 1 & 1 & 1 & 0 & 0 & 1 & 0 & 0 & 0 \\
1 & 1 & 1 & 1 & 1 & 1 & 0 & 0 & 0 & 1 & 0 & 0 \\
1 & 1 & 1 & 1 & 1 & 1 & 0 & 0 & 0 & 0 & 1 & 0 \\
1 & 1 & 1 & 1 & 1 & 1 & 0 & 0 & 0 & 0 & 0 & 1 \\
\end{array}
\right] & ,
F_4 = 
\left[
\begin{array}{cccccc|cccccc}
1 & 0 & 0 & 0 & 0 & 0 & 1 & 1 & 1 & 1 & 1 & 1 \\
0 & 1 & 0 & 0 & 0 & 0 & 1 & 0 & 1 & 1 & 1 & 0 \\
0 & 0 & 1 & 0 & 0 & 0 & 1 & 1 & 1 & 1 & 1 & 1 \\
0 & 0 & 0 & 1 & 0 & 0 & 1 & 1 & 1 & 1 & 1 & 1 \\
0 & 0 & 0 & 0 & 1 & 0 & 1 & 1 & 1 & 1 & 1 & 1 \\
0 & 0 & 0 & 0 & 0 & 1 & 1 & 0 & 1 & 1 & 1 & 0 \\
\hline
1 & 1 & 1 & 1 & 1 & 1 & 1 & 0 & 0 & 0 & 0 & 0 \\
1 & 1 & 1 & 1 & 1 & 1 & 0 & 1 & 0 & 0 & 0 & 0 \\
1 & 1 & 1 & 1 & 1 & 1 & 0 & 0 & 1 & 0 & 0 & 0 \\
1 & 1 & 1 & 1 & 1 & 1 & 0 & 0 & 0 & 1 & 0 & 0 \\
1 & 1 & 1 & 1 & 1 & 1 & 0 & 0 & 0 & 0 & 1 & 0 \\
1 & 1 & 1 & 1 & 1 & 1 & 0 & 0 & 0 & 0 & 0 & 1 \\
\end{array}
\right] , \\~\\
F_5 = 
\left[
\begin{array}{cccccc|cccccc}
0 & 1 & 1 & 1 & 1 & 1 & 0 & 0 & 0 & 0 & 0 & 0 \\
1 & 0 & 1 & 1 & 1 & 1 & 0 & 1 & 0 & 0 & 0 & 1 \\
1 & 1 & 0 & 1 & 1 & 1 & 0 & 0 & 0 & 0 & 0 & 0 \\
1 & 1 & 1 & 0 & 1 & 1 & 0 & 0 & 0 & 0 & 0 & 0 \\
1 & 1 & 1 & 1 & 0 & 1 & 0 & 0 & 0 & 0 & 0 & 0 \\
1 & 1 & 1 & 1 & 1 & 0 & 0 & 1 & 0 & 0 & 0 & 1 \\
\hline
0 & 0 & 0 & 0 & 0 & 0 & 0 & 1 & 1 & 1 & 1 & 1 \\
0 & 0 & 0 & 0 & 0 & 0 & 1 & 0 & 1 & 1 & 1 & 1 \\
0 & 0 & 0 & 0 & 0 & 0 & 1 & 1 & 0 & 1 & 1 & 1 \\
0 & 0 & 0 & 0 & 0 & 0 & 1 & 1 & 1 & 0 & 1 & 1 \\
0 & 0 & 0 & 0 & 0 & 0 & 1 & 1 & 1 & 1 & 0 & 1 \\
0 & 0 & 0 & 0 & 0 & 0 & 1 & 1 & 1 & 1 & 1 & 0 \\
\end{array}
\right] & , 
F_6 = 
\left[
\begin{array}{cccccc|cccccc}
0 & 1 & 1 & 1 & 1 & 1 & 1 & 1 & 1 & 1 & 1 & 1 \\
1 & 0 & 1 & 1 & 1 & 1 & 1 & 0 & 1 & 1 & 1 & 0 \\
1 & 1 & 0 & 1 & 1 & 1 & 1 & 1 & 1 & 1 & 1 & 1 \\
1 & 1 & 1 & 0 & 1 & 1 & 1 & 1 & 1 & 1 & 1 & 1 \\
1 & 1 & 1 & 1 & 0 & 1 & 1 & 1 & 1 & 1 & 1 & 1 \\
1 & 1 & 1 & 1 & 1 & 0 & 1 & 0 & 1 & 1 & 1 & 0 \\
\hline
0 & 0 & 0 & 0 & 0 & 0 & 0 & 1 & 1 & 1 & 1 & 1 \\
0 & 0 & 0 & 0 & 0 & 0 & 1 & 0 & 1 & 1 & 1 & 1 \\
0 & 0 & 0 & 0 & 0 & 0 & 1 & 1 & 0 & 1 & 1 & 1 \\
0 & 0 & 0 & 0 & 0 & 0 & 1 & 1 & 1 & 0 & 1 & 1 \\
0 & 0 & 0 & 0 & 0 & 0 & 1 & 1 & 1 & 1 & 0 & 1 \\
0 & 0 & 0 & 0 & 0 & 0 & 1 & 1 & 1 & 1 & 1 & 0 \\
\end{array}
\right] , \\~\\
F_7 = 
\left[
\begin{array}{cccccc|cccccc}
0 & 1 & 1 & 1 & 1 & 1 & 0 & 0 & 0 & 0 & 0 & 0 \\
1 & 0 & 1 & 1 & 1 & 1 & 0 & 1 & 0 & 0 & 0 & 1 \\
1 & 1 & 0 & 1 & 1 & 1 & 0 & 0 & 0 & 0 & 0 & 0 \\
1 & 1 & 1 & 0 & 1 & 1 & 0 & 0 & 0 & 0 & 0 & 0 \\
1 & 1 & 1 & 1 & 0 & 1 & 0 & 0 & 0 & 0 & 0 & 0 \\
1 & 1 & 1 & 1 & 1 & 0 & 0 & 1 & 0 & 0 & 0 & 1 \\
\hline
1 & 1 & 1 & 1 & 1 & 1 & 0 & 1 & 1 & 1 & 1 & 1 \\
1 & 1 & 1 & 1 & 1 & 1 & 1 & 0 & 1 & 1 & 1 & 1 \\
1 & 1 & 1 & 1 & 1 & 1 & 1 & 1 & 0 & 1 & 1 & 1 \\
1 & 1 & 1 & 1 & 1 & 1 & 1 & 1 & 1 & 0 & 1 & 1 \\
1 & 1 & 1 & 1 & 1 & 1 & 1 & 1 & 1 & 1 & 0 & 1 \\
1 & 1 & 1 & 1 & 1 & 1 & 1 & 1 & 1 & 1 & 1 & 0 \\
\end{array}
\right] & , 
F_8 = 
\left[
\begin{array}{cccccc|cccccc}
0 & 1 & 1 & 1 & 1 & 1 & 1 & 1 & 1 & 1 & 1 & 1 \\
1 & 0 & 1 & 1 & 1 & 1 & 1 & 0 & 1 & 1 & 1 & 0 \\
1 & 1 & 0 & 1 & 1 & 1 & 1 & 1 & 1 & 1 & 1 & 1 \\
1 & 1 & 1 & 0 & 1 & 1 & 1 & 1 & 1 & 1 & 1 & 1 \\
1 & 1 & 1 & 1 & 0 & 1 & 1 & 1 & 1 & 1 & 1 & 1 \\
1 & 1 & 1 & 1 & 1 & 0 & 1 & 0 & 1 & 1 & 1 & 0 \\
\hline
1 & 1 & 1 & 1 & 1 & 1 & 0 & 1 & 1 & 1 & 1 & 1 \\
1 & 1 & 1 & 1 & 1 & 1 & 1 & 0 & 1 & 1 & 1 & 1 \\
1 & 1 & 1 & 1 & 1 & 1 & 1 & 1 & 0 & 1 & 1 & 1 \\
1 & 1 & 1 & 1 & 1 & 1 & 1 & 1 & 1 & 0 & 1 & 1 \\
1 & 1 & 1 & 1 & 1 & 1 & 1 & 1 & 1 & 1 & 0 & 1 \\
1 & 1 & 1 & 1 & 1 & 1 & 1 & 1 & 1 & 1 & 1 & 0 \\
\end{array}
\right] .
\end{align*}
Note that $F_1$ is the solution discussed in Section~\ref{sec:css_phase}.

\subsection{Logical Controlled-Z Gate $(\lcz{1}{2})$}
\label{sec:css_cz_all}

There are  $8$ possible symplectic solutions that satisfy the linear constraints imposed by~\eqref{eq:cz_maps} and they are listed below.

\begin{align*}
F_1 = 
\left[
\begin{array}{cccccc|cccccc}
1 & 0 & 0 & 0 & 0 & 0 & 0 & 0 & 0 & 0 & 0 & 0 \\
0 & 1 & 0 & 0 & 0 & 0 & 0 & 0 & 1 & 0 & 0 & 1 \\
0 & 0 & 1 & 0 & 0 & 0 & 0 & 1 & 0 & 0 & 0 & 1 \\
0 & 0 & 0 & 1 & 0 & 0 & 0 & 0 & 0 & 0 & 0 & 0 \\
0 & 0 & 0 & 0 & 1 & 0 & 0 & 0 & 0 & 0 & 0 & 0 \\
0 & 0 & 0 & 0 & 0 & 1 & 0 & 1 & 1 & 0 & 0 & 0 \\
\hline
0 & 0 & 0 & 0 & 0 & 0 & 1 & 0 & 0 & 0 & 0 & 0 \\
0 & 0 & 0 & 0 & 0 & 0 & 0 & 1 & 0 & 0 & 0 & 0 \\
0 & 0 & 0 & 0 & 0 & 0 & 0 & 0 & 1 & 0 & 0 & 0 \\
0 & 0 & 0 & 0 & 0 & 0 & 0 & 0 & 0 & 1 & 0 & 0 \\
0 & 0 & 0 & 0 & 0 & 0 & 0 & 0 & 0 & 0 & 1 & 0 \\
0 & 0 & 0 & 0 & 0 & 0 & 0 & 0 & 0 & 0 & 0 & 1 \\
\end{array}
\right] & , 
F_2 = 
\left[
\begin{array}{cccccc|cccccc}
1 & 0 & 0 & 0 & 0 & 0 & 1 & 1 & 1 & 1 & 1 & 1 \\
0 & 1 & 0 & 0 & 0 & 0 & 1 & 1 & 0 & 1 & 1 & 0 \\
0 & 0 & 1 & 0 & 0 & 0 & 1 & 0 & 1 & 1 & 1 & 0 \\
0 & 0 & 0 & 1 & 0 & 0 & 1 & 1 & 1 & 1 & 1 & 1 \\
0 & 0 & 0 & 0 & 1 & 0 & 1 & 1 & 1 & 1 & 1 & 1 \\
0 & 0 & 0 & 0 & 0 & 1 & 1 & 0 & 0 & 1 & 1 & 1 \\
\hline
0 & 0 & 0 & 0 & 0 & 0 & 1 & 0 & 0 & 0 & 0 & 0 \\
0 & 0 & 0 & 0 & 0 & 0 & 0 & 1 & 0 & 0 & 0 & 0 \\
0 & 0 & 0 & 0 & 0 & 0 & 0 & 0 & 1 & 0 & 0 & 0 \\
0 & 0 & 0 & 0 & 0 & 0 & 0 & 0 & 0 & 1 & 0 & 0 \\
0 & 0 & 0 & 0 & 0 & 0 & 0 & 0 & 0 & 0 & 1 & 0 \\
0 & 0 & 0 & 0 & 0 & 0 & 0 & 0 & 0 & 0 & 0 & 1 \\
\end{array}
\right] , \\~\\
F_3 = 
\left[
\begin{array}{cccccc|cccccc}
1 & 0 & 0 & 0 & 0 & 0 & 0 & 0 & 0 & 0 & 0 & 0 \\
0 & 1 & 0 & 0 & 0 & 0 & 0 & 0 & 1 & 0 & 0 & 1 \\
0 & 0 & 1 & 0 & 0 & 0 & 0 & 1 & 0 & 0 & 0 & 1 \\
0 & 0 & 0 & 1 & 0 & 0 & 0 & 0 & 0 & 0 & 0 & 0 \\
0 & 0 & 0 & 0 & 1 & 0 & 0 & 0 & 0 & 0 & 0 & 0 \\
0 & 0 & 0 & 0 & 0 & 1 & 0 & 1 & 1 & 0 & 0 & 0 \\
\hline
1 & 1 & 1 & 1 & 1 & 1 & 1 & 0 & 0 & 0 & 0 & 0 \\
1 & 1 & 1 & 1 & 1 & 1 & 0 & 1 & 0 & 0 & 0 & 0 \\
1 & 1 & 1 & 1 & 1 & 1 & 0 & 0 & 1 & 0 & 0 & 0 \\
1 & 1 & 1 & 1 & 1 & 1 & 0 & 0 & 0 & 1 & 0 & 0 \\
1 & 1 & 1 & 1 & 1 & 1 & 0 & 0 & 0 & 0 & 1 & 0 \\
1 & 1 & 1 & 1 & 1 & 1 & 0 & 0 & 0 & 0 & 0 & 1 \\
\end{array}
\right] & , 
F_4 = 
\left[
\begin{array}{cccccc|cccccc}
1 & 0 & 0 & 0 & 0 & 0 & 1 & 1 & 1 & 1 & 1 & 1 \\
0 & 1 & 0 & 0 & 0 & 0 & 1 & 1 & 0 & 1 & 1 & 0 \\
0 & 0 & 1 & 0 & 0 & 0 & 1 & 0 & 1 & 1 & 1 & 0 \\
0 & 0 & 0 & 1 & 0 & 0 & 1 & 1 & 1 & 1 & 1 & 1 \\
0 & 0 & 0 & 0 & 1 & 0 & 1 & 1 & 1 & 1 & 1 & 1 \\
0 & 0 & 0 & 0 & 0 & 1 & 1 & 0 & 0 & 1 & 1 & 1 \\
\hline
1 & 1 & 1 & 1 & 1 & 1 & 1 & 0 & 0 & 0 & 0 & 0 \\
1 & 1 & 1 & 1 & 1 & 1 & 0 & 1 & 0 & 0 & 0 & 0 \\
1 & 1 & 1 & 1 & 1 & 1 & 0 & 0 & 1 & 0 & 0 & 0 \\
1 & 1 & 1 & 1 & 1 & 1 & 0 & 0 & 0 & 1 & 0 & 0 \\
1 & 1 & 1 & 1 & 1 & 1 & 0 & 0 & 0 & 0 & 1 & 0 \\
1 & 1 & 1 & 1 & 1 & 1 & 0 & 0 & 0 & 0 & 0 & 1 \\
\end{array}
\right] , \\~\\
F_5 = 
\left[
\begin{array}{cccccc|cccccc}
0 & 1 & 1 & 1 & 1 & 1 & 0 & 0 & 0 & 0 & 0 & 0 \\
1 & 0 & 1 & 1 & 1 & 1 & 0 & 0 & 1 & 0 & 0 & 1 \\
1 & 1 & 0 & 1 & 1 & 1 & 0 & 1 & 0 & 0 & 0 & 1 \\
1 & 1 & 1 & 0 & 1 & 1 & 0 & 0 & 0 & 0 & 0 & 0 \\
1 & 1 & 1 & 1 & 0 & 1 & 0 & 0 & 0 & 0 & 0 & 0 \\
1 & 1 & 1 & 1 & 1 & 0 & 0 & 1 & 1 & 0 & 0 & 0 \\
\hline
0 & 0 & 0 & 0 & 0 & 0 & 0 & 1 & 1 & 1 & 1 & 1 \\
0 & 0 & 0 & 0 & 0 & 0 & 1 & 0 & 1 & 1 & 1 & 1 \\
0 & 0 & 0 & 0 & 0 & 0 & 1 & 1 & 0 & 1 & 1 & 1 \\
0 & 0 & 0 & 0 & 0 & 0 & 1 & 1 & 1 & 0 & 1 & 1 \\
0 & 0 & 0 & 0 & 0 & 0 & 1 & 1 & 1 & 1 & 0 & 1 \\
0 & 0 & 0 & 0 & 0 & 0 & 1 & 1 & 1 & 1 & 1 & 0 \\
\end{array}
\right] & , 
F_6 = 
\left[
\begin{array}{cccccc|cccccc}
0 & 1 & 1 & 1 & 1 & 1 & 1 & 1 & 1 & 1 & 1 & 1 \\
1 & 0 & 1 & 1 & 1 & 1 & 1 & 1 & 0 & 1 & 1 & 0 \\
1 & 1 & 0 & 1 & 1 & 1 & 1 & 0 & 1 & 1 & 1 & 0 \\
1 & 1 & 1 & 0 & 1 & 1 & 1 & 1 & 1 & 1 & 1 & 1 \\
1 & 1 & 1 & 1 & 0 & 1 & 1 & 1 & 1 & 1 & 1 & 1 \\
1 & 1 & 1 & 1 & 1 & 0 & 1 & 0 & 0 & 1 & 1 & 1 \\
\hline
0 & 0 & 0 & 0 & 0 & 0 & 0 & 1 & 1 & 1 & 1 & 1 \\
0 & 0 & 0 & 0 & 0 & 0 & 1 & 0 & 1 & 1 & 1 & 1 \\
0 & 0 & 0 & 0 & 0 & 0 & 1 & 1 & 0 & 1 & 1 & 1 \\
0 & 0 & 0 & 0 & 0 & 0 & 1 & 1 & 1 & 0 & 1 & 1 \\
0 & 0 & 0 & 0 & 0 & 0 & 1 & 1 & 1 & 1 & 0 & 1 \\
0 & 0 & 0 & 0 & 0 & 0 & 1 & 1 & 1 & 1 & 1 & 0 \\
\end{array}
\right] , \\~\\
F_7 = 
\left[
\begin{array}{cccccc|cccccc}
0 & 1 & 1 & 1 & 1 & 1 & 0 & 0 & 0 & 0 & 0 & 0 \\
1 & 0 & 1 & 1 & 1 & 1 & 0 & 0 & 1 & 0 & 0 & 1 \\
1 & 1 & 0 & 1 & 1 & 1 & 0 & 1 & 0 & 0 & 0 & 1 \\
1 & 1 & 1 & 0 & 1 & 1 & 0 & 0 & 0 & 0 & 0 & 0 \\
1 & 1 & 1 & 1 & 0 & 1 & 0 & 0 & 0 & 0 & 0 & 0 \\
1 & 1 & 1 & 1 & 1 & 0 & 0 & 1 & 1 & 0 & 0 & 0 \\
\hline
1 & 1 & 1 & 1 & 1 & 1 & 0 & 1 & 1 & 1 & 1 & 1 \\
1 & 1 & 1 & 1 & 1 & 1 & 1 & 0 & 1 & 1 & 1 & 1 \\
1 & 1 & 1 & 1 & 1 & 1 & 1 & 1 & 0 & 1 & 1 & 1 \\
1 & 1 & 1 & 1 & 1 & 1 & 1 & 1 & 1 & 0 & 1 & 1 \\
1 & 1 & 1 & 1 & 1 & 1 & 1 & 1 & 1 & 1 & 0 & 1 \\
1 & 1 & 1 & 1 & 1 & 1 & 1 & 1 & 1 & 1 & 1 & 0 \\
\end{array}
\right] & , 
F_8 = 
\left[
\begin{array}{cccccc|cccccc}
0 & 1 & 1 & 1 & 1 & 1 & 1 & 1 & 1 & 1 & 1 & 1 \\
1 & 0 & 1 & 1 & 1 & 1 & 1 & 1 & 0 & 1 & 1 & 0 \\
1 & 1 & 0 & 1 & 1 & 1 & 1 & 0 & 1 & 1 & 1 & 0 \\
1 & 1 & 1 & 0 & 1 & 1 & 1 & 1 & 1 & 1 & 1 & 1 \\
1 & 1 & 1 & 1 & 0 & 1 & 1 & 1 & 1 & 1 & 1 & 1 \\
1 & 1 & 1 & 1 & 1 & 0 & 1 & 0 & 0 & 1 & 1 & 1 \\
\hline
1 & 1 & 1 & 1 & 1 & 1 & 0 & 1 & 1 & 1 & 1 & 1 \\
1 & 1 & 1 & 1 & 1 & 1 & 1 & 0 & 1 & 1 & 1 & 1 \\
1 & 1 & 1 & 1 & 1 & 1 & 1 & 1 & 0 & 1 & 1 & 1 \\
1 & 1 & 1 & 1 & 1 & 1 & 1 & 1 & 1 & 0 & 1 & 1 \\
1 & 1 & 1 & 1 & 1 & 1 & 1 & 1 & 1 & 1 & 0 & 1 \\
1 & 1 & 1 & 1 & 1 & 1 & 1 & 1 & 1 & 1 & 1 & 0 \\
\end{array}
\right] .
\end{align*}
Note that $F_1$ is the solution discussed in Section~\ref{sec:css_cz}.

\subsection{Logical Controlled-NOT Gate $(\lcnot{2}{1})$}
\label{sec:css_cnot_all}

There are $8$ possible symplectic solutions that satisfy the linear constraints imposed by~\eqref{eq:lcnot21_maps} and they are listed below.

\begin{align*}
F_1 = 
\left[
\begin{array}{cccccc|cccccc}
1 & 0 & 0 & 0 & 0 & 0 & 0 & 0 & 0 & 0 & 0 & 0 \\
0 & 1 & 0 & 0 & 0 & 0 & 0 & 0 & 0 & 0 & 0 & 0 \\
1 & 1 & 1 & 0 & 0 & 0 & 0 & 0 & 0 & 0 & 0 & 0 \\
0 & 0 & 0 & 1 & 0 & 0 & 0 & 0 & 0 & 0 & 0 & 0 \\
0 & 0 & 0 & 0 & 1 & 0 & 0 & 0 & 0 & 0 & 0 & 0 \\
1 & 1 & 0 & 0 & 0 & 1 & 0 & 0 & 0 & 0 & 0 & 0 \\
\hline
0 & 0 & 0 & 0 & 0 & 0 & 1 & 0 & 1 & 0 & 0 & 1 \\
0 & 0 & 0 & 0 & 0 & 0 & 0 & 1 & 1 & 0 & 0 & 1 \\
0 & 0 & 0 & 0 & 0 & 0 & 0 & 0 & 1 & 0 & 0 & 0 \\
0 & 0 & 0 & 0 & 0 & 0 & 0 & 0 & 0 & 1 & 0 & 0 \\
0 & 0 & 0 & 0 & 0 & 0 & 0 & 0 & 0 & 0 & 1 & 0 \\
0 & 0 & 0 & 0 & 0 & 0 & 0 & 0 & 0 & 0 & 0 & 1 \\
\end{array}
\right] & , 
F_2 = 
\left[
\begin{array}{cccccc|cccccc}
1 & 0 & 0 & 0 & 0 & 0 & 1 & 1 & 1 & 1 & 1 & 1 \\
0 & 1 & 0 & 0 & 0 & 0 & 1 & 1 & 1 & 1 & 1 & 1 \\
1 & 1 & 1 & 0 & 0 & 0 & 1 & 1 & 1 & 1 & 1 & 1 \\
0 & 0 & 0 & 1 & 0 & 0 & 1 & 1 & 1 & 1 & 1 & 1 \\
0 & 0 & 0 & 0 & 1 & 0 & 1 & 1 & 1 & 1 & 1 & 1 \\
1 & 1 & 0 & 0 & 0 & 1 & 1 & 1 & 1 & 1 & 1 & 1 \\
\hline
0 & 0 & 0 & 0 & 0 & 0 & 1 & 0 & 1 & 0 & 0 & 1 \\
0 & 0 & 0 & 0 & 0 & 0 & 0 & 1 & 1 & 0 & 0 & 1 \\
0 & 0 & 0 & 0 & 0 & 0 & 0 & 0 & 1 & 0 & 0 & 0 \\
0 & 0 & 0 & 0 & 0 & 0 & 0 & 0 & 0 & 1 & 0 & 0 \\
0 & 0 & 0 & 0 & 0 & 0 & 0 & 0 & 0 & 0 & 1 & 0 \\
0 & 0 & 0 & 0 & 0 & 0 & 0 & 0 & 0 & 0 & 0 & 1 \\
\end{array}
\right] , \\~\\
F_3 = 
\left[
\begin{array}{cccccc|cccccc}
1 & 0 & 0 & 0 & 0 & 0 & 0 & 0 & 0 & 0 & 0 & 0 \\
0 & 1 & 0 & 0 & 0 & 0 & 0 & 0 & 0 & 0 & 0 & 0 \\
1 & 1 & 1 & 0 & 0 & 0 & 0 & 0 & 0 & 0 & 0 & 0 \\
0 & 0 & 0 & 1 & 0 & 0 & 0 & 0 & 0 & 0 & 0 & 0 \\
0 & 0 & 0 & 0 & 1 & 0 & 0 & 0 & 0 & 0 & 0 & 0 \\
1 & 1 & 0 & 0 & 0 & 1 & 0 & 0 & 0 & 0 & 0 & 0 \\
\hline
1 & 1 & 1 & 1 & 1 & 1 & 1 & 0 & 1 & 0 & 0 & 1 \\
1 & 1 & 1 & 1 & 1 & 1 & 0 & 1 & 1 & 0 & 0 & 1 \\
1 & 1 & 1 & 1 & 1 & 1 & 0 & 0 & 1 & 0 & 0 & 0 \\
1 & 1 & 1 & 1 & 1 & 1 & 0 & 0 & 0 & 1 & 0 & 0 \\
1 & 1 & 1 & 1 & 1 & 1 & 0 & 0 & 0 & 0 & 1 & 0 \\
1 & 1 & 1 & 1 & 1 & 1 & 0 & 0 & 0 & 0 & 0 & 1 \\
\end{array}
\right] & , 
F_4 = 
\left[
\begin{array}{cccccc|cccccc}
1 & 0 & 0 & 0 & 0 & 0 & 1 & 1 & 1 & 1 & 1 & 1 \\
0 & 1 & 0 & 0 & 0 & 0 & 1 & 1 & 1 & 1 & 1 & 1 \\
1 & 1 & 1 & 0 & 0 & 0 & 1 & 1 & 1 & 1 & 1 & 1 \\
0 & 0 & 0 & 1 & 0 & 0 & 1 & 1 & 1 & 1 & 1 & 1 \\
0 & 0 & 0 & 0 & 1 & 0 & 1 & 1 & 1 & 1 & 1 & 1 \\
1 & 1 & 0 & 0 & 0 & 1 & 1 & 1 & 1 & 1 & 1 & 1 \\
\hline
1 & 1 & 1 & 1 & 1 & 1 & 1 & 0 & 1 & 0 & 0 & 1 \\
1 & 1 & 1 & 1 & 1 & 1 & 0 & 1 & 1 & 0 & 0 & 1 \\
1 & 1 & 1 & 1 & 1 & 1 & 0 & 0 & 1 & 0 & 0 & 0 \\
1 & 1 & 1 & 1 & 1 & 1 & 0 & 0 & 0 & 1 & 0 & 0 \\
1 & 1 & 1 & 1 & 1 & 1 & 0 & 0 & 0 & 0 & 1 & 0 \\
1 & 1 & 1 & 1 & 1 & 1 & 0 & 0 & 0 & 0 & 0 & 1 \\
\end{array}
\right]  , \\~\\
F_5 = 
\left[
\begin{array}{cccccc|cccccc}
0 & 1 & 1 & 1 & 1 & 1 & 0 & 0 & 0 & 0 & 0 & 0 \\
1 & 0 & 1 & 1 & 1 & 1 & 0 & 0 & 0 & 0 & 0 & 0 \\
0 & 0 & 0 & 1 & 1 & 1 & 0 & 0 & 0 & 0 & 0 & 0 \\
1 & 1 & 1 & 0 & 1 & 1 & 0 & 0 & 0 & 0 & 0 & 0 \\
1 & 1 & 1 & 1 & 0 & 1 & 0 & 0 & 0 & 0 & 0 & 0 \\
0 & 0 & 1 & 1 & 1 & 0 & 0 & 0 & 0 & 0 & 0 & 0 \\
\hline
0 & 0 & 0 & 0 & 0 & 0 & 0 & 1 & 0 & 1 & 1 & 0 \\
0 & 0 & 0 & 0 & 0 & 0 & 1 & 0 & 0 & 1 & 1 & 0 \\
0 & 0 & 0 & 0 & 0 & 0 & 1 & 1 & 0 & 1 & 1 & 1 \\
0 & 0 & 0 & 0 & 0 & 0 & 1 & 1 & 1 & 0 & 1 & 1 \\
0 & 0 & 0 & 0 & 0 & 0 & 1 & 1 & 1 & 1 & 0 & 1 \\
0 & 0 & 0 & 0 & 0 & 0 & 1 & 1 & 1 & 1 & 1 & 0 \\
\end{array}
\right] & , 
F_6 = 
\left[
\begin{array}{cccccc|cccccc}
0 & 1 & 1 & 1 & 1 & 1 & 1 & 1 & 1 & 1 & 1 & 1 \\
1 & 0 & 1 & 1 & 1 & 1 & 1 & 1 & 1 & 1 & 1 & 1 \\
0 & 0 & 0 & 1 & 1 & 1 & 1 & 1 & 1 & 1 & 1 & 1 \\
1 & 1 & 1 & 0 & 1 & 1 & 1 & 1 & 1 & 1 & 1 & 1 \\
1 & 1 & 1 & 1 & 0 & 1 & 1 & 1 & 1 & 1 & 1 & 1 \\
0 & 0 & 1 & 1 & 1 & 0 & 1 & 1 & 1 & 1 & 1 & 1 \\
\hline
0 & 0 & 0 & 0 & 0 & 0 & 0 & 1 & 0 & 1 & 1 & 0 \\
0 & 0 & 0 & 0 & 0 & 0 & 1 & 0 & 0 & 1 & 1 & 0 \\
0 & 0 & 0 & 0 & 0 & 0 & 1 & 1 & 0 & 1 & 1 & 1 \\
0 & 0 & 0 & 0 & 0 & 0 & 1 & 1 & 1 & 0 & 1 & 1 \\
0 & 0 & 0 & 0 & 0 & 0 & 1 & 1 & 1 & 1 & 0 & 1 \\
0 & 0 & 0 & 0 & 0 & 0 & 1 & 1 & 1 & 1 & 1 & 0 \\
\end{array}
\right] , \\~\\
F_7 = 
\left[
\begin{array}{cccccc|cccccc}
0 & 1 & 1 & 1 & 1 & 1 & 0 & 0 & 0 & 0 & 0 & 0 \\
1 & 0 & 1 & 1 & 1 & 1 & 0 & 0 & 0 & 0 & 0 & 0 \\
0 & 0 & 0 & 1 & 1 & 1 & 0 & 0 & 0 & 0 & 0 & 0 \\
1 & 1 & 1 & 0 & 1 & 1 & 0 & 0 & 0 & 0 & 0 & 0 \\
1 & 1 & 1 & 1 & 0 & 1 & 0 & 0 & 0 & 0 & 0 & 0 \\
0 & 0 & 1 & 1 & 1 & 0 & 0 & 0 & 0 & 0 & 0 & 0 \\
\hline
1 & 1 & 1 & 1 & 1 & 1 & 0 & 1 & 0 & 1 & 1 & 0 \\
1 & 1 & 1 & 1 & 1 & 1 & 1 & 0 & 0 & 1 & 1 & 0 \\
1 & 1 & 1 & 1 & 1 & 1 & 1 & 1 & 0 & 1 & 1 & 1 \\
1 & 1 & 1 & 1 & 1 & 1 & 1 & 1 & 1 & 0 & 1 & 1 \\
1 & 1 & 1 & 1 & 1 & 1 & 1 & 1 & 1 & 1 & 0 & 1 \\
1 & 1 & 1 & 1 & 1 & 1 & 1 & 1 & 1 & 1 & 1 & 0 \\
\end{array}
\right] & , 
F_8 = 
\left[
\begin{array}{cccccc|cccccc}
0 & 1 & 1 & 1 & 1 & 1 & 1 & 1 & 1 & 1 & 1 & 1 \\
1 & 0 & 1 & 1 & 1 & 1 & 1 & 1 & 1 & 1 & 1 & 1 \\
0 & 0 & 0 & 1 & 1 & 1 & 1 & 1 & 1 & 1 & 1 & 1 \\
1 & 1 & 1 & 0 & 1 & 1 & 1 & 1 & 1 & 1 & 1 & 1 \\
1 & 1 & 1 & 1 & 0 & 1 & 1 & 1 & 1 & 1 & 1 & 1 \\
0 & 0 & 1 & 1 & 1 & 0 & 1 & 1 & 1 & 1 & 1 & 1 \\
\hline
1 & 1 & 1 & 1 & 1 & 1 & 0 & 1 & 0 & 1 & 1 & 0 \\
1 & 1 & 1 & 1 & 1 & 1 & 1 & 0 & 0 & 1 & 1 & 0 \\
1 & 1 & 1 & 1 & 1 & 1 & 1 & 1 & 0 & 1 & 1 & 1 \\
1 & 1 & 1 & 1 & 1 & 1 & 1 & 1 & 1 & 0 & 1 & 1 \\
1 & 1 & 1 & 1 & 1 & 1 & 1 & 1 & 1 & 1 & 0 & 1 \\
1 & 1 & 1 & 1 & 1 & 1 & 1 & 1 & 1 & 1 & 1 & 0 \\
\end{array}
\right] .
\end{align*}
Note that $F_1$ is the solution discussed in Section~\ref{sec:css_cnot}.

\subsection{Logical Targeted Hadamard Gate $(\lH_1)$}
\label{sec:css_had_all}

There are $8$ possible symplectic solutions that satisfy the linear constraints imposed by~\eqref{eq:hadamard_maps} and they are listed below.

\begin{align*}
F_1 = 
\left[
\begin{array}{cccccc|cccccc}
1 & 0 & 0 & 0 & 0 & 0 & 0 & 0 & 0 & 0 & 0 & 0 \\
1 & 0 & 0 & 0 & 0 & 0 & 0 & 1 & 0 & 0 & 0 & 1 \\
0 & 0 & 1 & 0 & 0 & 0 & 0 & 0 & 0 & 0 & 0 & 0 \\
0 & 0 & 0 & 1 & 0 & 0 & 0 & 0 & 0 & 0 & 0 & 0 \\
0 & 0 & 0 & 0 & 1 & 0 & 0 & 0 & 0 & 0 & 0 & 0 \\
1 & 1 & 0 & 0 & 0 & 1 & 0 & 1 & 0 & 0 & 0 & 1 \\
\hline
1 & 1 & 0 & 0 & 0 & 0 & 1 & 1 & 0 & 0 & 0 & 1 \\
1 & 1 & 0 & 0 & 0 & 0 & 0 & 0 & 0 & 0 & 0 & 1 \\
0 & 0 & 0 & 0 & 0 & 0 & 0 & 0 & 1 & 0 & 0 & 0 \\
0 & 0 & 0 & 0 & 0 & 0 & 0 & 0 & 0 & 1 & 0 & 0 \\
0 & 0 & 0 & 0 & 0 & 0 & 0 & 0 & 0 & 0 & 1 & 0 \\
0 & 0 & 0 & 0 & 0 & 0 & 0 & 0 & 0 & 0 & 0 & 1 \\
\end{array}
\right] & , 
F_2 = 
\left[
\begin{array}{cccccc|cccccc}
1 & 0 & 0 & 0 & 0 & 0 & 1 & 1 & 1 & 1 & 1 & 1 \\
1 & 0 & 0 & 0 & 0 & 0 & 1 & 0 & 1 & 1 & 1 & 0 \\
0 & 0 & 1 & 0 & 0 & 0 & 1 & 1 & 1 & 1 & 1 & 1 \\
0 & 0 & 0 & 1 & 0 & 0 & 1 & 1 & 1 & 1 & 1 & 1 \\
0 & 0 & 0 & 0 & 1 & 0 & 1 & 1 & 1 & 1 & 1 & 1 \\
1 & 1 & 0 & 0 & 0 & 1 & 1 & 0 & 1 & 1 & 1 & 0 \\
\hline
1 & 1 & 0 & 0 & 0 & 0 & 1 & 1 & 0 & 0 & 0 & 1 \\
1 & 1 & 0 & 0 & 0 & 0 & 0 & 0 & 0 & 0 & 0 & 1 \\
0 & 0 & 0 & 0 & 0 & 0 & 0 & 0 & 1 & 0 & 0 & 0 \\
0 & 0 & 0 & 0 & 0 & 0 & 0 & 0 & 0 & 1 & 0 & 0 \\
0 & 0 & 0 & 0 & 0 & 0 & 0 & 0 & 0 & 0 & 1 & 0 \\
0 & 0 & 0 & 0 & 0 & 0 & 0 & 0 & 0 & 0 & 0 & 1 \\
\end{array}
\right] , \\~\\
F_3 = 
\left[
\begin{array}{cccccc|cccccc}
1 & 0 & 0 & 0 & 0 & 0 & 0 & 0 & 0 & 0 & 0 & 0 \\
1 & 0 & 0 & 0 & 0 & 0 & 0 & 1 & 0 & 0 & 0 & 1 \\
0 & 0 & 1 & 0 & 0 & 0 & 0 & 0 & 0 & 0 & 0 & 0 \\
0 & 0 & 0 & 1 & 0 & 0 & 0 & 0 & 0 & 0 & 0 & 0 \\
0 & 0 & 0 & 0 & 1 & 0 & 0 & 0 & 0 & 0 & 0 & 0 \\
1 & 1 & 0 & 0 & 0 & 1 & 0 & 1 & 0 & 0 & 0 & 1 \\
\hline
0 & 0 & 1 & 1 & 1 & 1 & 1 & 1 & 0 & 0 & 0 & 1 \\
0 & 0 & 1 & 1 & 1 & 1 & 0 & 0 & 0 & 0 & 0 & 1 \\
1 & 1 & 1 & 1 & 1 & 1 & 0 & 0 & 1 & 0 & 0 & 0 \\
1 & 1 & 1 & 1 & 1 & 1 & 0 & 0 & 0 & 1 & 0 & 0 \\
1 & 1 & 1 & 1 & 1 & 1 & 0 & 0 & 0 & 0 & 1 & 0 \\
1 & 1 & 1 & 1 & 1 & 1 & 0 & 0 & 0 & 0 & 0 & 1 \\
\end{array}
\right] & , 
F_4 = 
\left[
\begin{array}{cccccc|cccccc}
1 & 0 & 0 & 0 & 0 & 0 & 1 & 1 & 1 & 1 & 1 & 1 \\
1 & 0 & 0 & 0 & 0 & 0 & 1 & 0 & 1 & 1 & 1 & 0 \\
0 & 0 & 1 & 0 & 0 & 0 & 1 & 1 & 1 & 1 & 1 & 1 \\
0 & 0 & 0 & 1 & 0 & 0 & 1 & 1 & 1 & 1 & 1 & 1 \\
0 & 0 & 0 & 0 & 1 & 0 & 1 & 1 & 1 & 1 & 1 & 1 \\
1 & 1 & 0 & 0 & 0 & 1 & 1 & 0 & 1 & 1 & 1 & 0 \\
\hline
0 & 0 & 1 & 1 & 1 & 1 & 1 & 1 & 0 & 0 & 0 & 1 \\
0 & 0 & 1 & 1 & 1 & 1 & 0 & 0 & 0 & 0 & 0 & 1 \\
1 & 1 & 1 & 1 & 1 & 1 & 0 & 0 & 1 & 0 & 0 & 0 \\
1 & 1 & 1 & 1 & 1 & 1 & 0 & 0 & 0 & 1 & 0 & 0 \\
1 & 1 & 1 & 1 & 1 & 1 & 0 & 0 & 0 & 0 & 1 & 0 \\
1 & 1 & 1 & 1 & 1 & 1 & 0 & 0 & 0 & 0 & 0 & 1 \\
\end{array}
\right]  , \\~\\
F_5 = 
\left[
\begin{array}{cccccc|cccccc}
0 & 1 & 1 & 1 & 1 & 1 & 0 & 0 & 0 & 0 & 0 & 0 \\
0 & 1 & 1 & 1 & 1 & 1 & 0 & 1 & 0 & 0 & 0 & 1 \\
1 & 1 & 0 & 1 & 1 & 1 & 0 & 0 & 0 & 0 & 0 & 0 \\
1 & 1 & 1 & 0 & 1 & 1 & 0 & 0 & 0 & 0 & 0 & 0 \\
1 & 1 & 1 & 1 & 0 & 1 & 0 & 0 & 0 & 0 & 0 & 0 \\
0 & 0 & 1 & 1 & 1 & 0 & 0 & 1 & 0 & 0 & 0 & 1 \\
\hline
1 & 1 & 0 & 0 & 0 & 0 & 0 & 0 & 1 & 1 & 1 & 0 \\
1 & 1 & 0 & 0 & 0 & 0 & 1 & 1 & 1 & 1 & 1 & 0 \\
0 & 0 & 0 & 0 & 0 & 0 & 1 & 1 & 0 & 1 & 1 & 1 \\
0 & 0 & 0 & 0 & 0 & 0 & 1 & 1 & 1 & 0 & 1 & 1 \\
0 & 0 & 0 & 0 & 0 & 0 & 1 & 1 & 1 & 1 & 0 & 1 \\
0 & 0 & 0 & 0 & 0 & 0 & 1 & 1 & 1 & 1 & 1 & 0 \\
\end{array}
\right] & , 
F_6 = 
\left[
\begin{array}{cccccc|cccccc}
0 & 1 & 1 & 1 & 1 & 1 & 1 & 1 & 1 & 1 & 1 & 1 \\
0 & 1 & 1 & 1 & 1 & 1 & 1 & 0 & 1 & 1 & 1 & 0 \\
1 & 1 & 0 & 1 & 1 & 1 & 1 & 1 & 1 & 1 & 1 & 1 \\
1 & 1 & 1 & 0 & 1 & 1 & 1 & 1 & 1 & 1 & 1 & 1 \\
1 & 1 & 1 & 1 & 0 & 1 & 1 & 1 & 1 & 1 & 1 & 1 \\
0 & 0 & 1 & 1 & 1 & 0 & 1 & 0 & 1 & 1 & 1 & 0 \\
\hline
1 & 1 & 0 & 0 & 0 & 0 & 0 & 0 & 1 & 1 & 1 & 0 \\
1 & 1 & 0 & 0 & 0 & 0 & 1 & 1 & 1 & 1 & 1 & 0 \\
0 & 0 & 0 & 0 & 0 & 0 & 1 & 1 & 0 & 1 & 1 & 1 \\
0 & 0 & 0 & 0 & 0 & 0 & 1 & 1 & 1 & 0 & 1 & 1 \\
0 & 0 & 0 & 0 & 0 & 0 & 1 & 1 & 1 & 1 & 0 & 1 \\
0 & 0 & 0 & 0 & 0 & 0 & 1 & 1 & 1 & 1 & 1 & 0 \\
\end{array}
\right] , \\~\\
F_7 = 
\left[
\begin{array}{cccccc|cccccc}
0 & 1 & 1 & 1 & 1 & 1 & 0 & 0 & 0 & 0 & 0 & 0 \\
0 & 1 & 1 & 1 & 1 & 1 & 0 & 1 & 0 & 0 & 0 & 1 \\
1 & 1 & 0 & 1 & 1 & 1 & 0 & 0 & 0 & 0 & 0 & 0 \\
1 & 1 & 1 & 0 & 1 & 1 & 0 & 0 & 0 & 0 & 0 & 0 \\
1 & 1 & 1 & 1 & 0 & 1 & 0 & 0 & 0 & 0 & 0 & 0 \\
0 & 0 & 1 & 1 & 1 & 0 & 0 & 1 & 0 & 0 & 0 & 1 \\
\hline
0 & 0 & 1 & 1 & 1 & 1 & 0 & 0 & 1 & 1 & 1 & 0 \\
0 & 0 & 1 & 1 & 1 & 1 & 1 & 1 & 1 & 1 & 1 & 0 \\
1 & 1 & 1 & 1 & 1 & 1 & 1 & 1 & 0 & 1 & 1 & 1 \\
1 & 1 & 1 & 1 & 1 & 1 & 1 & 1 & 1 & 0 & 1 & 1 \\
1 & 1 & 1 & 1 & 1 & 1 & 1 & 1 & 1 & 1 & 0 & 1 \\
1 & 1 & 1 & 1 & 1 & 1 & 1 & 1 & 1 & 1 & 1 & 0 \\
\end{array}
\right] & , 
F_8 = 
\left[
\begin{array}{cccccc|cccccc}
0 & 1 & 1 & 1 & 1 & 1 & 1 & 1 & 1 & 1 & 1 & 1 \\
0 & 1 & 1 & 1 & 1 & 1 & 1 & 0 & 1 & 1 & 1 & 0 \\
1 & 1 & 0 & 1 & 1 & 1 & 1 & 1 & 1 & 1 & 1 & 1 \\
1 & 1 & 1 & 0 & 1 & 1 & 1 & 1 & 1 & 1 & 1 & 1 \\
1 & 1 & 1 & 1 & 0 & 1 & 1 & 1 & 1 & 1 & 1 & 1 \\
0 & 0 & 1 & 1 & 1 & 0 & 1 & 0 & 1 & 1 & 1 & 0 \\
\hline
0 & 0 & 1 & 1 & 1 & 1 & 0 & 0 & 1 & 1 & 1 & 0 \\
0 & 0 & 1 & 1 & 1 & 1 & 1 & 1 & 1 & 1 & 1 & 0 \\
1 & 1 & 1 & 1 & 1 & 1 & 1 & 1 & 0 & 1 & 1 & 1 \\
1 & 1 & 1 & 1 & 1 & 1 & 1 & 1 & 1 & 0 & 1 & 1 \\
1 & 1 & 1 & 1 & 1 & 1 & 1 & 1 & 1 & 1 & 0 & 1 \\
1 & 1 & 1 & 1 & 1 & 1 & 1 & 1 & 1 & 1 & 1 & 0 \\
\end{array}
\right] .
\end{align*}
Note that $F_1$ is the solution discussed in eqn.~\eqref{eq:css_H1} in Section~\ref{sec:css_had}.

\section{Proofs for Results in Section~\ref{sec:css_operators}}
\label{sec:css_proofs}

\subsection{Proof of Theorem~\ref{thm:css_generators}}
\label{proof:css_generators}

%\begin{IEEEproof}[]
Since the $X$ and $Z$ operators trivially commute with themselves, it is clear that $\bg_i^X \bg_j^X = \bg_j^X \bg_i^X$ and $\bg_i^Z \bg_j^Z = \bg_j^Z \bg_i^Z$ for all $i,j \in \{1,\ldots,k\}$.
This is written in commutation notation as
\[ \left[ \bg_i^X, \bg_j^X \right] \triangleq \bg_i^X \bg_j^X - \bg_j^X \bg_i^X = \boldsymbol{0} \quad , \quad \left[ \bg_i^Z, \bg_j^Z \right] \triangleq \bg_i^Z \bg_j^Z - \bg_j^Z \bg_i^Z = \boldsymbol{0} , \]
where $\boldsymbol{0}$ is the zero operator, i.e. a matrix with all entries set to $0$.

However, the $X$ operator anti-commutes with the $Z$ operator so that $XZ = -ZX$.
So to check if $\bg_i^X$ and $\bg_j^Z$ commute or anti-commute we only have to count the number of indices $t$ with $g_{it} = 1$ and $g_{jt} = 1$, where $\vecnot{g}_i = [g_{i1},\ldots,g_{im}]$ is the $i$-th row of $H_{\MCC}$.
Now observe that since $\MCCd$ is a self-orthogonal code and $\vecnot{g}_i \in \MCCd$, we have $\vecnot{g}_i \cdot \vecnot{g}_j = 0 \ \forall \ i,j \in \{1,\ldots,k\}$.
This implies $b_{ij} \triangleq |\{ t \in \{1,\ldots,m\} \ : \ g_{it} =1, g_{jt} =1 \}|$ is even.
Hence we see that for all $i,j \in \{1,\ldots,k\}$ we have
\[ \bg_i^X \bg_j^Z = (-1)^{b_{ij}} \bg_j^Z \bg_i^X = \bg_j^Z \bg_i^X \Rightarrow \left[ \bg_i^X, \bg_j^Z \right] = \boldsymbol{0} . \]
Thus we see that the above defined set of $2k$ operators commute with each other and clearly do not generate $-I_N$.  \hfill\IEEEQEDhere
%\end{IEEEproof}

\subsection{Proof of Theorem~\ref{thm:css_stabilizer}}
\label{proof:css_stabilizer}

%\begin{IEEEproof}
First we observe that since $X$ is a bit-flip operator satisfying $X \ket{0} = \ket{0 + 1} = \ket{1}, X\ket{1} = \ket{1 + 1} = \ket{0}$, the operator $\bg_i^X$ satisfies
\[ \bg_i^X \ket{\vecnot{u}} = \ket{\vecnot{u} + \vecnot{g}_i} \]
for any vector $\vecnot{u} \in \{0,1\}^m$, where $\vecnot{g}_i \in \MCCd$ is the row of $H_{\MCC}$ used to define $\bg_i^X$ in~\eqref{eq:css_generators}.
Since $\vecnot{c} + \vecnot{g}_i \in \MCCd$ for all $\vecnot{c} \in \MCCd$ we have
\[ \bg_i^X \ket{\psi_v} = \frac{1}{\sqrt{|\MCCd|}} \sum_{\vecnot{c} \in \MCCd} \bg_i^X \ket{\vecnot{c} + \vecnot{v}} = \frac{1}{\sqrt{|\MCCd|}} \sum_{\vecnot{c} \in \MCCd} \ket{(\vecnot{c} + \vecnot{g}_i) + \vecnot{v}} = \frac{1}{\sqrt{|\MCCd|}} \sum_{\vecnot{c} \in \MCCd} \ket{\vecnot{c} + \vecnot{v}} = \ket{\psi_v} . \]

Similarly, since $Z$ is a phase-flip operator satisfying $Z \ket{0} = \ket{0}, Z \ket{1} = -\ket{1}$, the operator $\bg_i^Z$ satisfies
\[ \bg_i^Z \ket{\vecnot{u}} = (-1)^{\vecnot{g}_i \cdot \vecnot{u}} \ket{\vecnot{u}} \]
for any vector $\vecnot{u} \in \{0,1\}^m$.
In each term of the superposition in the CSS state $\ket{\psi_v}$, we observe that $\vecnot{c} + \vecnot{v} \in \MCC$.
As $\vecnot{g}_i$ is a row of the parity-check matrix of $\MCC$ it automatically satisfies $\vecnot{g}_i \cdot (\vecnot{c} + \vecnot{v}) = 0$.
Therefore we have
\[ \bg_i^Z \ket{\psi_v} = \frac{1}{\sqrt{|\MCCd|}} \sum_{\vecnot{c} \in \MCCd} \bg_i^Z \ket{\vecnot{c} + \vecnot{v}} = \frac{1}{\sqrt{|\MCCd|}} \sum_{\vecnot{c} \in \MCCd} (-1)^{\vecnot{g}_i \cdot (\vecnot{c} + \vecnot{v})} \ket{\vecnot{c} + \vecnot{v}} = \frac{1}{\sqrt{|\MCCd|}} \sum_{\vecnot{c} \in \MCCd} \ket{\vecnot{c} + \vecnot{v}} = \ket{\psi_v} . \]
Thus we have shown that all the $2k$ operators $\bg_i^X, \bg_i^Z$ defined in~\eqref{eq:css_generators} stabilize the states $\ket{\psi_v} \in \MCQ$.
Also, the dimension of the space $V(S)$ stabilized by the group generated by $\{ \bg_i^X, \bg_i^Z \ ; \ i\in [k] \}$ is $2^{m-2k}$, which is exactly the dimension of $\MCQ$ too.
Therefore $V(S) = \MCQ$.   \hfill\IEEEQEDhere
%\end{IEEEproof}

\subsection{Proof of Lemma~\ref{lem:logical_Paulis_commute}}
\label{proof:logical_Paulis_commute}

Assume $G_{\MCC/\MCCd}^X \left( G_{\MCC/\MCCd}^Z \right)^T = I_{m-2k}$.
This implies $\vecnot{h}_i \cdot \vecnot{h}_j' = 1$ if $i=j$ and $\vecnot{h}_i \cdot \vecnot{h}_j' = 0$ if $i \neq j$.
Then using the property $(A \otimes B) (C \otimes D) = AC \otimes BD$ of Kronecker products we have
\begin{align*}
\lX_i \lZ_j = \bigotimes_{t=1}^{m} X^{h_{it}} Z^{h_{it}'} & = \bigotimes_{t=1}^{m} (-1)^{h_{it} h_{it}'} Z^{h_{it}'} X^{h_{it}} \\
  & = (-1)^{\vecnot{h}_i \cdot \vecnot{h}_j'} \bigotimes_{t=1}^{m} Z^{h_{it}'} X^{h_{it}} \\
  & = 
\begin{cases}
- \lZ_j \lX_i & {\rm if} \ i=j, \\
\lZ_j \lX_i & {\rm if} \ i \neq j 
\end{cases} . 
\end{align*}
Conversely, it is easy to see that the last equality above requires $G_{\MCC/\MCCd}^X \left( G_{\MCC/\MCCd}^Z \right)^T = I_{m-2k}$.  \hfill\IEEEQEDhere

\subsection{Proof of Theorem~\ref{thm:logical_Paulis}}
\label{proof:logical_Paulis}

As observed in the proof of Theorem~\ref{thm:css_stabilizer}, we have $\lX_i \ket{\vecnot{u}} = \ket{\vecnot{u} + \vecnot{h}_i}$ for any vector $\vecnot{u} \in \{0,1\}^m$.
Recall that the CSS state for $\ket{\vecnot{x}}_L$ is defined as
\[ \ket{\psi_x} \triangleq \frac{1}{\sqrt{|\MCCd|}} \sum_{\vecnot{c} \in \MCCd} \ket{\vecnot{c} + \vecnot{x} \cdot G_{\MCC/\MCCd}} = \frac{1}{\sqrt{|\MCCd|}} \sum_{\vecnot{c} \in \MCCd} \ket{\vecnot{c} + \sum_{j=1}^{m-2k} x_j \vecnot{h}_j} .
 \]
Therefore we have
\begin{align*}
\lX_i \ket{\psi_x} & = \frac{1}{\sqrt{|\MCCd|}} \sum_{\vecnot{c} \in \MCCd} \lX_i \ket{\vecnot{c} + \sum_{j=1}^{m-2k} x_j \vecnot{h}_j} \\
  & = \frac{1}{\sqrt{|\MCCd|}} \sum_{\vecnot{c} \in \MCCd} \ket{\vecnot{c} + \sum_{j=1, j \neq i}^{m-2k} x_j \vecnot{h}_j + (x_i\oplus 1) \vecnot{h}_i} \\
  & = \ket{\psi_{x'}} . 
\end{align*}
Similarly we have $\lZ_i \ket{\vecnot{u}} = (-1)^{\vecnot{h}_i' \cdot \vecnot{u}} \ket{\vecnot{u}}$.
For convenience we rewrite the CSS state $\ket{\psi_x}$ as
\begin{align*}
\ket{\psi_x} & = \frac{1}{\sqrt{|\MCCd|}} \sum_{\vecnot{c} \in \MCCd} \ket{\vecnot{c} + \sum_{j=1}^{m-2k} x_j \vecnot{h}_j} \\
  & = \frac{1}{\sqrt{|\MCCd|}} \sum_{\vecnot{c} \in \MCCd} \prod_{j=1}^{m-2k} \lX_j^{x_j} \ket{\vecnot{c}} \\
  & = \prod_{j=1}^{m-2k} \lX_j^{x_j} \frac{1}{\sqrt{|\MCCd|}} \sum_{\vecnot{c} \in \MCCd} \ket{\vecnot{c}} . 
\end{align*}
Using the commutation relations above we have $\lZ_i \lX_i = - \lX_i \lZ_i$ and $\lZ_i \lX_j = \lX_j \lZ_i$ for $j \neq i$.
Also, since $\vecnot{h}_i' \in \MCC$ it satisfies $\vecnot{h}_i' \cdot \vecnot{c} = 0$ for all $\vecnot{c} \in \MCCd$.
This implies
\begin{IEEEeqnarray*}{rCl+x*}
\lZ_i \ket{\psi_x} & = & \lZ_i \prod_{j=1}^{m-2k} \lX_j^{x_j} \frac{1}{\sqrt{|\MCCd|}} \sum_{\vecnot{c} \in \MCCd} \ket{\vecnot{c}} \\
  & = & (-1)^{x_i} \prod_{j=1}^{m-2k} \lX_j^{x_j} \frac{1}{\sqrt{|\MCCd|}} \sum_{\vecnot{c} \in \MCCd} \lZ_i \ket{\vecnot{c}} \\ 
  & = & (-1)^{x_i} \prod_{j=1}^{m-2k} \lX_j^{x_j} \frac{1}{\sqrt{|\MCCd|}} \sum_{\vecnot{c} \in \MCCd} (-1)^{\vecnot{h}_i' \cdot \vecnot{c}} \ket{\vecnot{c}} \\ 
  & = & (-1)^{x_i} \prod_{j=1}^{m-2k} \lX_j^{x_j} \frac{1}{\sqrt{|\MCCd|}} \sum_{\vecnot{c} \in \MCCd} \ket{\vecnot{c}} \\ 
  & = & (-1)^{x_i} \ket{\psi_x} . & \IEEEQEDhere
%  \\* &&& \IEEEQEDhere
\end{IEEEeqnarray*}

\newpage

\section{Circuit Identities}
\label{sec:circuit_ids}

Let $\iota \triangleq \sqrt{-1}$.
The definitions of the single-qubit gates (operators) appearing in the following table are:
\begin{align*}
X \triangleq 
\begin{bmatrix}
0 & 1 \\
1 & 0
\end{bmatrix} , \ 
Z \triangleq 
\begin{bmatrix}
1 & 0 \\
0 & -1
\end{bmatrix} , \  
Y \triangleq \iota XZ = 
\begin{bmatrix}
0 & -\iota \\
\iota & 0
\end{bmatrix} , \ 
H \triangleq \frac{1}{\sqrt{2}}
\begin{bmatrix}
1 & 1 \\
1 & -1
\end{bmatrix} , \ 
P \triangleq 
\begin{bmatrix}
1 & 0 \\
0 & \iota
\end{bmatrix} .
\end{align*}

\vspace{-0.2cm}

\begin{center}
\begin{longtable}{ccc|cc}

(a1) & Controlled-Z Gate (CZ) & & & Controlled-NOT Gate (CNOT) \\
 & & & & \\
 &
\begin{tikzpicture}

\node[draw,rectangle] (Z10) at (1,0) {$Z$};

\path[draw] (0,0) -- (Z10) -- (2,0);
\path[draw] (0,0.75) -- (2,0.75);

\draw[fill=black] (1,0.75) circle (0.1);
\path[draw] (1,0.75) -- (Z10);

\node[align=center] at (3,0.375) {$\equiv$};

\path[draw] (4,0) -- (6,0);
\path[draw] (4,0.75) -- (6,0.75);

\draw[fill=black] (5,0.75) circle (0.1);
\draw[fill=black] (5,0) circle (0.1);
\path[draw] (5,0.75) -- (5,0);

\end{tikzpicture}
& \qquad & \qquad &
\begin{tikzpicture}

\node[draw,rectangle] (X10) at (1,0) {$X$};

\path[draw] (0,0) -- (X10) -- (2,0);
\path[draw] (0,0.75) -- (2,0.75);

\draw[fill=black] (1,0.75) circle (0.1);
\path[draw] (1,0.75) -- (X10);

\node[align=center] at (3,0.375) {$\equiv$};

\path[draw] (4,0) -- (6,0);
\path[draw] (4,0.75) -- (6,0.75);

\draw[fill=black] (5,0.75) circle (0.1);
\draw (5,0) circle (0.15);
\path[draw] (5,0.75) -- (5,-0.15);

\end{tikzpicture}

\\
 & & & &
\\

(a2) & Controlled-Z Gate (CZ) & & & Controlled-NOT Gate (CNOT) \\
 & & & & \\
 &
\begin{tikzpicture}

\node[draw,rectangle] (Z10) at (1,0) {$Z$};

\path[draw] (0,0) -- (Z10) -- (2,0);
\path[draw] (0,-0.75) -- (2,-0.75);

\draw[fill=black] (1,-0.75) circle (0.1);
\path[draw] (1,-0.75) -- (Z10);

\node[align=center] at (3,-0.375) {$\equiv$};

\path[draw] (4,0) -- (6,0);
\path[draw] (4,-0.75) -- (6,-0.75);

\draw[fill=black] (5,-0.75) circle (0.1);
\draw[fill=black] (5,0) circle (0.1);
\path[draw] (5,-0.75) -- (5,0);

\end{tikzpicture}
& \qquad & \qquad &
\begin{tikzpicture}

\node[draw,rectangle] (X10) at (1,0) {$X$};

\path[draw] (0,0) -- (X10) -- (2,0);
\path[draw] (0,-0.75) -- (2,-0.75);

\draw[fill=black] (1,-0.75) circle (0.1);
\path[draw] (1,-0.75) -- (X10);

\node[align=center] at (3,-0.375) {$\equiv$};

\path[draw] (4,0) -- (6,0);
\path[draw] (4,-0.75) -- (6,-0.75);

\draw[fill=black] (5,-0.75) circle (0.1);
\draw (5,0) circle (0.15);
\path[draw] (5,-0.75) -- (5,0.15);

\end{tikzpicture}

\\
 & & & &
\\

(b) & & & \\
 &
\begin{tikzpicture}

\path[draw] (0,0) -- (2,0);
\path[draw] (0,0.75) -- (2,0.75);

\draw[fill=black] (1,0.75) circle (0.1);
\draw[fill=black] (1,0) circle (0.1);
\path[draw] (1,0.75) -- (1,0);

\node[align=center] at (2.5,0.375) {$=$};

\node[draw,rectangle] (H40) at (4,0) {$H$};
\node[draw,rectangle] (H60) at (6,0) {$H$};

\path[draw] (3,0) -- (H40) -- (H60) -- (7,0);
\path[draw] (3,0.75) -- (7,0.75);

\draw[fill=black] (5,0.75) circle (0.1);
\path[draw] (5,0.75) -- (5,-0.15);
\draw (5,0) circle (0.15);

\end{tikzpicture}
& \qquad & \qquad &
\begin{tikzpicture}

\path[draw] (0,0) -- (2,0);
\path[draw] (0,0.75) -- (2,0.75);

\draw[fill=black] (1,0.75) circle (0.1);
\draw (1,0) circle (0.15);
\path[draw] (1,0.75) -- (1,-0.15);

\node[align=center] at (2.5,0.375) {$=$};

\node[draw,rectangle] (H40) at (4,0) {$H$};
\node[draw,rectangle] (H60) at (6,0) {$H$};

\path[draw] (3,0) -- (H40) -- (H60) -- (7,0);
\path[draw] (3,0.75) -- (7,0.75);

\draw[fill=black] (5,0.75) circle (0.1);
\path[draw] (5,0.75) -- (5,0);
\draw[fill=black] (5,0) circle (0.1);

\end{tikzpicture}

\\
 & & & &
\\

(c) & & & \\
 &
\begin{tikzpicture}

\node at (-0.5,0.75) {$X$};

\path[draw] (0,0) -- (2,0);
\path[draw] (0,0.75) -- (2,0.75);

\draw[fill=black] (1,0.75) circle (0.1);
\draw[fill=black] (1,0) circle (0.1);
\path[draw] (1,0.75) -- (1,0);

\node[align=center] at (2.5,0.375) {$=$};

\path[draw] (3,0) -- (5,0);
\path[draw] (3,0.75) -- (5,0.75);

\draw[fill=black] (4,0.75) circle (0.1);
\draw[fill=black] (4,0) circle (0.1);
\path[draw] (4,0.75) -- (4,0);

\node at (5.5,0.75) {$X$};
\node at (5.5,0) {$Z$};

\end{tikzpicture}
& \qquad & \qquad &
\begin{tikzpicture}

\node at (-0.5,0.75) {$X$};

\path[draw] (0,0) -- (2,0);
\path[draw] (0,0.75) -- (2,0.75);

\draw[fill=black] (1,0.75) circle (0.1);
\draw (1,0) circle (0.15);
\path[draw] (1,0.75) -- (1,-0.15);

\node[align=center] at (2.5,0.375) {$=$};

\path[draw] (3,0) -- (5,0);
\path[draw] (3,0.75) -- (5,0.75);

\draw[fill=black] (4,0.75) circle (0.1);
\draw (4,0) circle (0.15);
\path[draw] (4,0.75) -- (4,-0.15);

\node at (5.5,0.75) {$X$};
\node at (5.5,0) {$X$};

\end{tikzpicture}

\\
 & & & &
\\

(d) & & & \\
 &
\begin{tikzpicture}

\node at (-0.5,0.75) {$Z$};

\path[draw] (0,0) -- (2,0);
\path[draw] (0,0.75) -- (2,0.75);

\draw[fill=black] (1,0.75) circle (0.1);
\draw[fill=black] (1,0) circle (0.1);
\path[draw] (1,0.75) -- (1,0);

\node[align=center] at (2.5,0.375) {$=$};

\path[draw] (3,0) -- (5,0);
\path[draw] (3,0.75) -- (5,0.75);

\draw[fill=black] (4,0.75) circle (0.1);
\draw[fill=black] (4,0) circle (0.1);
\path[draw] (4,0.75) -- (4,0);

\node at (5.5,0.75) {$Z$};

\end{tikzpicture}
& \qquad & \qquad &
\begin{tikzpicture}

\node at (-0.5,0.75) {$Z$};

\path[draw] (0,0) -- (2,0);
\path[draw] (0,0.75) -- (2,0.75);

\draw[fill=black] (1,0.75) circle (0.1);
\draw (1,0) circle (0.15);
\path[draw] (1,0.75) -- (1,-0.15);

\node[align=center] at (2.5,0.375) {$=$};

\path[draw] (3,0) -- (5,0);
\path[draw] (3,0.75) -- (5,0.75);

\draw[fill=black] (4,0.75) circle (0.1);
\draw (4,0) circle (0.15);
\path[draw] (4,0.75) -- (4,-0.15);

\node at (5.5,0.75) {$Z$};

\end{tikzpicture}

\\
 & & & &
\\

(e) & & & \\
 &
\begin{tikzpicture}

\node at (-0.5,0) {$Z$};

\path[draw] (0,0) -- (2,0);
\path[draw] (0,0.75) -- (2,0.75);

\draw[fill=black] (1,0.75) circle (0.1);
\draw[fill=black] (1,0) circle (0.1);
\path[draw] (1,0.75) -- (1,0);

\node[align=center] at (2.5,0.375) {$=$};

\path[draw] (3,0) -- (5,0);
\path[draw] (3,0.75) -- (5,0.75);

\draw[fill=black] (4,0.75) circle (0.1);
\draw[fill=black] (4,0) circle (0.1);
\path[draw] (4,0.75) -- (4,0);

\node at (5.5,0) {$Z$};

\end{tikzpicture}
& \qquad & \qquad &
\begin{tikzpicture}

\node at (-0.5,0) {$X$};

\path[draw] (0,0) -- (2,0);
\path[draw] (0,0.75) -- (2,0.75);

\draw[fill=black] (1,0.75) circle (0.1);
\draw (1,0) circle (0.15);
\path[draw] (1,0.75) -- (1,-0.15);

\node[align=center] at (2.5,0.375) {$=$};

\path[draw] (3,0) -- (5,0);
\path[draw] (3,0.75) -- (5,0.75);

\draw[fill=black] (4,0.75) circle (0.1);
\draw (4,0) circle (0.15);
\path[draw] (4,0.75) -- (4,-0.15);

\node at (5.5,0) {$X$};

\end{tikzpicture}

\\
 & & & &
\\

(f) & & & \\
 &
\begin{tikzpicture}

\node at (-0.5,0) {$X$};

\path[draw] (0,0) -- (2,0);
\path[draw] (0,0.75) -- (2,0.75);

\draw[fill=black] (1,0.75) circle (0.1);
\draw[fill=black] (1,0) circle (0.1);
\path[draw] (1,0.75) -- (1,0);

\node[align=center] at (2.5,0.375) {$=$};

\path[draw] (3,0) -- (5,0);
\path[draw] (3,0.75) -- (5,0.75);

\draw[fill=black] (4,0.75) circle (0.1);
\draw[fill=black] (4,0) circle (0.1);
\path[draw] (4,0.75) -- (4,0);

\node at (5.5,0.75) {$Z$};
\node at (5.5,0) {$X$};

\end{tikzpicture}
& \qquad & \qquad &
\begin{tikzpicture}

\node at (-0.5,0) {$Z$};

\path[draw] (0,0) -- (2,0);
\path[draw] (0,0.75) -- (2,0.75);

\draw[fill=black] (1,0.75) circle (0.1);
\draw (1,0) circle (0.15);
\path[draw] (1,0.75) -- (1,-0.15);

\node[align=center] at (2.5,0.375) {$=$};

\path[draw] (3,0) -- (5,0);
\path[draw] (3,0.75) -- (5,0.75);

\draw[fill=black] (4,0.75) circle (0.1);
\draw (4,0) circle (0.15);
\path[draw] (4,0.75) -- (4,-0.15);

\node at (5.5,0.75) {$Z$};
\node at (5.5,0) {$Z$};

\end{tikzpicture}

\\
 & & & &
\\

(g) & & & \\
 &
\begin{tikzpicture}

\node at (-0.5,0.75) {$X$};
\node at (-0.5,0) {$X$};

\path[draw] (0,0) -- (2,0);
\path[draw] (0,0.75) -- (2,0.75);

\draw[fill=black] (1,0.75) circle (0.1);
\draw[fill=black] (1,0) circle (0.1);
\path[draw] (1,0.75) -- (1,0);

\node[align=center] at (2.5,0.375) {$=$};

\path[draw] (3,0) -- (5,0);
\path[draw] (3,0.75) -- (5,0.75);

\draw[fill=black] (4,0.75) circle (0.1);
\draw[fill=black] (4,0) circle (0.1);
\path[draw] (4,0.75) -- (4,0);

\node at (5.5,0.75) {$Y$};
\node at (5.5,0) {$Y$};

\end{tikzpicture}
& \qquad & \qquad &
\begin{tikzpicture}

\node at (-0.5,0.75) {$X$};
\node at (-0.5,0) {$Z$};

\path[draw] (0,0) -- (2,0);
\path[draw] (0,0.75) -- (2,0.75);

\draw[fill=black] (1,0.75) circle (0.1);
\draw (1,0) circle (0.15);
\path[draw] (1,0.75) -- (1,-0.15);

\node[align=center] at (2.5,0.375) {$=$};

\path[draw] (3,0) -- (5,0);
\path[draw] (3,0.75) -- (5,0.75);

\draw[fill=black] (4,0.75) circle (0.1);
\draw (4,0) circle (0.15);
\path[draw] (4,0.75) -- (4,-0.15);

\node at (5.5,0.75) {$-Y$};
\node at (5.5,0) {$Y$};

\end{tikzpicture}

\\
 & & & &
\\

(h) & Hadamard Gate ($H$) & & & Hadamard Gate ($H$) \\
 & & & & \\
 &
\begin{tikzpicture}

\node[draw,rectangle] (H10) at (1,0) {$H$};

\node at (-0.5,0) {$Z$};
\path[draw] (0,0) -- (H10) -- (2,0);

\node[align=center] at (2.5,0) {$=$};

\node[draw,rectangle] (H40) at (4,0) {$H$};

\path[draw] (3,0) -- (H40) -- (5,0);
\node at (5.5,0) {$X$};

\end{tikzpicture}
& \qquad & \qquad & 
\begin{tikzpicture}

\node[draw,rectangle] (H10) at (1,0) {$H$};

\node at (-0.5,0) {$X$};
\path[draw] (0,0) -- (H10) -- (2,0);

\node[align=center] at (2.5,0) {$=$};

\node[draw,rectangle] (H40) at (4,0) {$H$};

\path[draw] (3,0) -- (H40) -- (5,0);
\node at (5.5,0) {$Z$};

\end{tikzpicture}

\\
 & & & &
\\

(i) & Phase Gate ($P$) & & & Phase Gate ($P$) \\
 & & & & \\
 &
\begin{tikzpicture}

\node[draw,rectangle] (S10) at (1,0) {$P$};

\node at (-0.75,0) {$Z$};
\path[draw] (0,0) -- (S10) -- (2,0);

\node[align=center] at (2.5,0) {$=$};

\node[draw,rectangle] (S40) at (4,0) {$P$};

\path[draw] (3,0) -- (S40) -- (5,0);
\node at (5.5,0) {$Z$};

\end{tikzpicture}
& \qquad & \qquad & 
\begin{tikzpicture}

\node[draw,rectangle] (S10) at (1,0) {$P$};

\node at (-0.75,0) {$X$};
\path[draw] (0,0) -- (S10) -- (2,0);

\node[align=center] at (2.5,0) {$=$};

\node[draw,rectangle] (S40) at (4,0) {$P$};

\path[draw] (3,0) -- (S40) -- (5,0);
\node at (5.5,0) {$Y$};

\end{tikzpicture}
%
%\\
% & & & & \\
%\hline
\end{longtable}
\end{center}

\vspace{-0.75cm}

\subsection*{Calculating Conjugations}

Formally, the controlled-NOT (CNOT) and controlled-Z (CZ) gates are defined as
\begin{align*}
\cnot{1}{2} & \triangleq \ketbra{0} \otimes I + \ketbra{1} \otimes X = 
\begin{bmatrix}
1 & 0 & 0 & 0 \\
0 & 1 & 0 & 0 \\
0 & 0 & 0 & 1 \\
0 & 0 & 1 & 0
\end{bmatrix} , \\
{\rm CZ}_{12} & \triangleq \ketbra{0} \otimes I + \ketbra{1} \otimes Z = 
\begin{bmatrix}
1 & 0 & 0 & 0 \\
0 & 1 & 0 & 0 \\
0 & 0 & 1 & 0 \\
0 & 0 & 0 & -1
\end{bmatrix} = I \otimes \ketbra{0} + Z \otimes \ketbra{1} , 
\end{align*}
where $I$ is the $2 \times 2$ identity matrix, the subscript ``$1 \rightarrow 2$'' implies qubit $1$ is the control and qubit $2$ is the target.
We see that the CZ gate is symmetric about its inputs whereas the CNOT is not, i.e., ``$1 \rightarrow 2$'' and ``$2 \rightarrow 1$'' are distinct operators.

Let us see two (interesting) examples for calculating the transformation that these operators induce on their input under conjugation.

\begin{enumerate}

\item Let the input to $\cnot{1}{2}$ be $X \otimes Z$.
Then we have
\begin{align*}
\cnot{1}{2} \left( X \otimes Z \right) \cnot{1}{2}^{\dagger} & = \left( \ketbra{0} \otimes I + \ketbra{1} \otimes X \right) \left( X \otimes Z \right) \left( \ketbra{0} \otimes I + \ketbra{1} \otimes X \right) \\
& = \left( \ketbra{0}{1} \otimes Z + \ketbra{1}{0} \otimes XZ \right) \left( \ketbra{0} \otimes I + \ketbra{1} \otimes X \right) \\
& = \ketbra{0}{1} \otimes ZX + \ketbra{1}{0} \otimes XZ \\
& = \left( \ketbra{1}{0} - \ketbra{0}{1} \right) \otimes XZ \\
& = -Y \otimes Y .
\end{align*}

\item Let the input to ${\rm CZ}_{12}$ be $X \otimes X$.
Then we have
\begin{align*}
{\rm CZ}_{12} \left( X \otimes X \right) {\rm CZ}_{12}^{\dagger} & = \left( \ketbra{0} \otimes I + \ketbra{1} \otimes Z \right) \left( X \otimes X \right) \left( \ketbra{0} \otimes I + \ketbra{1} \otimes Z \right) \\
& = \left( \ketbra{0}{1} \otimes X + \ketbra{1}{0} \otimes ZX \right) \left( \ketbra{0} \otimes I + \ketbra{1} \otimes Z \right) \\
& = \ketbra{0}{1} \otimes XZ + \ketbra{1}{0} \otimes ZX \\
& = \left( \ketbra{0}{1} - \ketbra{1}{0} \right) \otimes XZ \\
& = Y \otimes Y .
\end{align*}

\end{enumerate}
These are the two identities appearing in part (g) in the above table.
The other identities can be derived in a similar fashion.

\fi

\end{document}